\documentclass[sigconf,authorversion,nonacm,dvipsnames]{acmart}

\usepackage{graphicx}
\usepackage{caption}
\usepackage{subfig}
\usepackage{amsmath}
\usepackage[vlined,commentsnumbered,ruled,linesnumbered]{algorithm2e}
\usepackage{algpseudocode}
\usepackage{verbatimbox}
\usepackage{float}
\usepackage{tcolorbox}
\usepackage{listings}
\usepackage[capitalize]{cleveref}
\usepackage{booktabs}
\usepackage{multirow}
\usepackage{lineno}
\usepackage{xspace}
\usepackage{colortbl}
\usepackage{mathtools}
\usepackage{enumitem}
\usepackage{balance}
\usepackage{standalone} % for the revision letter
\usepackage{etoolbox}
\providetoggle{long}

\usepackage{makecell}

\settoggle{long}{true}
\iftoggle{long}{
\newcommand{\ag}[1]{\reminderc{\bf (AG)~#1}{\typeout{#1}}}
\newcommand{\shunit}[1]{\textcolor{violet}{\bf (SA)~[#1]}{\typeout{#1}}}
\newcommand{\brit}[1]{\textcolor{magenta}{\bf (Brit)~[#1]}{\typeout{#1}}}
\def\benny#1{{\color{brown}BK: #1}}
}
{\newcommand{\ag}[1]{}
\newcommand{\shunit}[1]{}
\newcommand{\brit}[1]{}
\def\benny#1{}}

\newcommand\vldbdoi{XX.XX/XXX.XX}

\newcommand\vldbavailabilityurl{URL_TO_YOUR_ARTIFACTS}
\newcommand\vldbpagestyle{plain} 
\newcommand{\reminderc}[1]{\textcolor{blue}{[[[#1]]]}}

\newcommand\red[1]{\textcolor{red}{#1}}

\newcommand{\Aagg}{A_{agg}}
\newcommand{\Agb}{A_{gb}}
\newcommand{\attrset}{\mathord{\mathrm{Att}(D)}}
\newcommand{\splitset}{\mathrm{SAtt}(D)}

\newcommand{\view}{V}

\newcommand{\aggfunction}[1]{\mathsf{#1}}
\newcommand{\CNT}{\aggfunction{Count}}
\newcommand{\SUM}{\aggfunction{Sum}}
\newcommand{\AVG}{\aggfunction{Average}}
\newcommand{\MED}{\aggfunction{Median}}
\newcommand{\STDDEV}{\aggfunction{StdDev}}
\newcommand{\MIN}{\aggfunction{Min}}
\newcommand{\MAX}{\aggfunction{Max}}

\newcommand{\group}{g}
\newcommand{\val}{v}
\newcommand{\claim}{\kappa}
\newcommand{\nat}{\nu}
\newcommand{\predspace}{\ensuremath{\mathcal{P}}}
\newcommand{\pred}{\ensuremath{p}}
\newcommand{\bigo}{\ensuremath{\mathcal{O}}}

\newcommand{\cut}[1]{}
\newcommand{\eat}[1]{}

\def\exptitle#1{\vspace{1mm}\noindent\underline{#1}\,\,}
\newcommand{\oursolution}{{\textsc{Merged Top-$k$}}\xspace}
\newcommand{\oursolutionk}{{\textsc{Merged Top-$k$}}\xspace}
\newcommand{\oursolutionserial}{{\textsc{Serial Top-$k$}}\xspace}
\newcommand{\oursolutionsample}{{\textsc{$x$\% Sample}}\xspace}
\newcommand{\oursolutionsampleone}{{\textsc{$1$\%~Sample}}\xspace}
\newcommand{\oursolutionsamplefive}{{\textsc{$5$\%~Sample}}\xspace}
\newcommand{\oursolutionsampleten}{{\textsc{$10$\%~Sample}}\xspace}
\def\paratitle#1{\vspace{1mm}\noindent\underline{#1}\,\,}

\newtheorem{theorem}{Theorem}[section]

\newtheorem{example}[theorem]{Example}

\newtheorem{definition}[theorem]{Definition}

\definecolor{shadecolor}{rgb}{0.01,0.199,0.1}
\newcolumntype{g}{>{\columncolor{shadecolor}}c}

\def\coverage{\mathit{Coverage}}
\def\statsig{\mathit{StatSig}}
\def\anova{\mathit{ANOVA}}
\def\embsim{\mathit{EmbSim}}
\def\mi{\mathit{MI}}

\newcommand{\defeq}{\vcentcolon=}

\newcommand{\reva}[1]{{\leavevmode\color{black}{#1}}}
\newcommand{\revb}[1]{{\leavevmode\color{black}{#1}}}
\newcommand{\revc}[1]{{\leavevmode\color{black}{#1}}}
\newcommand{\common}[1]{{\leavevmode\color{black}{#1}}}

\def\benny#1{{\color{brown}BK: #1}}

\def\OREOS{OREO$^\star$\xspace}

\begin{document}
\setlength{\abovedisplayskip}{2.5pt}
\setlength{\belowdisplayskip}{2.5pt}

%\input{00-Revision-letter}

% \title{Make your data tell you what you want to hear: Automated Cherrypicking}
\title{Finding Convincing Views to Endorse a Claim}

%%
%% The "author" command and its associated commands are used to define the authors and their affiliations.
\settopmatter{authorsperrow=3}

\author{Shunit Agmon}
\orcid{0000-0001-9605-4131}
\affiliation{%
  % \institution{Technion -- Israel Institution of Technology}
  \institution{Technion}
  \streetaddress{}
  % \city{Haifa}
  % \state{Israel}
  % \postcode{3200003}
}
\email{shunita@campus.technion.ac.il}

\author{Amir Gilad}
\orcid{0000-0002-3764-1958}
\affiliation{%
  \institution{Hebrew University}
  % \city{Jerusalem}
  % \country{Israel}
}
\email{amirg@cs.huji.ac.il}

\author{Brit Youngmann}
\orcid{0000-0002-0031-5550}
\affiliation{%
  % \institution{Technion -- Israel Institution of Technology}
  \institution{Technion}
  % \city{Haifa}
  % \country{Israel}
  % \postcode{3200003}
}
\email{brity@technion.ac.il}

\author{Shahar Zoarets}
\affiliation{%
  % \institution{Technion -- Israel Institution of Technology}
  \institution{Technion}
  \streetaddress{}
  % \city{Haifa}
  % \state{Israel}
  % \postcode{3200003}
}
\email{shahar.zo@campus.technion.ac.il}

\author{Benny Kimelfeld}
\orcid{0000-0002-7156-1572}
\affiliation{%
  % \institution{Technion -- Israel Institution of Technology}
  \institution{Technion}
  % \streetaddress{}
  % \city{Haifa}
  % \country{Israel}
}
\email{bennyk@cs.technion.ac.il}

\begin{abstract}
Recent studies investigated the challenge of assessing the strength of a given claim extracted from a dataset, particularly the claim's potential of being misleading and cherry-picked. We focus on claims that compare answers to an aggregate query posed on a view that selects tuples. 
%Here, 
The strength of a claim amounts to the question of how likely it is that the view is carefully chosen to support the claim, whereas less careful choices would lead to contradictory claims. We embark on the study of the reverse task that offers a complementary angle in the critical assessment of data-based claims: given a claim, find useful supporting views. The goal of this task is twofold. On the one hand, we aim to assist users in finding significant evidence of phenomena of interest. On the other hand, we wish to provide them with machinery to criticize or counter given claims by extracting evidence of opposing statements. 

To be effective, the supporting sub-population should be significant and defined by a ``natural'' view. We discuss several measures of naturalness and propose ways of extracting the best views under each measure (and combinations thereof). The main challenge is the computational cost, as na\"ive search is infeasible. We devise anytime algorithms that deploy two main steps: \emph{(1)} a preliminary construction of a ranked list of attribute combinations that are assessed using fast-to-compute features, \emph{and (2)} an efficient search for the actual views based on each attribute combination. We present a thorough experimental study that shows the effectiveness of our algorithms in terms of quality and execution cost. \common{We also present a user study to assess the usefulness of the naturalness measures.}
\end{abstract}

\maketitle

\pagestyle{\vldbpagestyle}
%\begingroup\small\noindent\raggedright\textbf{PVLDB Reference Format:}\\
%\vldbauthors. \vldbtitle. PVLDB, \vldbvolume(\vldbissue): \vldbpages, \vldbyear.\\
%\href{https://doi.org/\vldbdoi}{doi:\vldbdoi}
%\endgroup
%\begingroup
%\renewcommand\thefootnote{}\footnote{\noindent
%This work is licensed under the Creative Commons BY-NC-ND 4.0 International License. Visit \url{https://creativecommons.org/licenses/by-nc-nd/4.0/} to view a copy of this license. For any use beyond those covered by this license, obtain permission by emailing \href{mailto:info@vldb.org}{info@vldb.org}. Copyright is held by the owner/author(s). Publication rights licensed to the VLDB Endowment. \\
%\raggedright Proceedings of the VLDB Endowment, Vol. \vldbvolume, No. \vldbissue\ %
%ISSN 2150-8097. \\
%\href{https://doi.org/\vldbdoi}{doi:\vldbdoi} \\
%}\addtocounter{footnote}{-1}\endgroup

%\ifdefempty{\vldbavailabilityurl}{}{
%\vspace{.3cm}
%\begingroup\small\noindent\raggedright\textbf{PVLDB Artifact Availability:}\\
%The source code, data, and/or other artifacts have been made available at \url{\vldbavailabilityurl}.
%\endgroup

\pagenumbering{arabic}

\section{Introduction}
\label{sec:intro}

Data collected rigorously can be used to identify important phenomena in an arguably objective manner. Hence, evidence from data is often used to qualify stated claims. Yet, for this reason, data-based arguments involve considerable risks. Cherrypicking, for instance, refers to basing the claim on a query with specific components that may seem natural, but are nevertheless crucial for supporting the claim rather than its opposite. The risk also goes in the other direction: basing a claim on the general population may give the false impression that the claim holds in important subpopulations (an extreme manifestation of this falsehood is the Simpson's Paradox).
Past work proposed ways of assessing whether a given claim, commonly phrased as an aggregate query over a database, is cherry-picked~\cite{lin2021detecting,asudeh2020detecting,asudeh2021perturbation,lin2022oreo}. This can be seen as a branch in the more general area of computational fact-checking~\cite{guo2022survey,karagiannis2020scrutinizer,zhou2020survey,farinha2018towards,hassan2017claimbuster,wu2014toward,wu2017computational}. Relevant techniques involve machine learning~\cite{lin2021detecting}, query perturbation~\cite{wu2014toward,wu2017computational,asudeh2020detecting, asudeh2021perturbation}, and Natural Language Processing (NLP)~\cite{farinha2018towards,hassan2017claimbuster}.

In this work, we study this task in the reverse direction: given a claim that is false in the database, we aim to find natural views where the claim holds. We refer to this problem as \emph{claim endorsement}. The objective is to enrich people's machinery with a complementary tool for the critical assessment of claims. In particular, effective claim endorsement can help users better understand the mechanism of cherry-picking and its involved risks. It can also allow users to look for queries that weaken or invalidate a stated claim; for example, a user can look for alternative phrasings that seem just as natural, yet support the opposite claim on the same database. Such an exercise can further help users assess how amenable a particular dataset is to support contradictory claims, and how suspicious we should be, in general, regarding allegations drawn from this dataset. \reva{Such abilities may be useful in many domains. For example, they may be used to closely inspect the claims of political candidates, or explore the extent of social issues.
They could also be applied in journalism, academic research, and various fields where critical analysis of data-driven claims is crucial.}

\begin{example}\label{ex:motivating1}\em
Consider the Stack Overflow Developers Survey dataset\footnote{\url{https://survey.stackoverflow.co/2022}} that includes information about Hi-Tech workers such as  backgrounds, demographics, and annual salaries. In this dataset, the average salary of people with a master's degree is \emph{lower} than that of people with only a bachelor's degree. Alex, a social scientist, aims to challenge this observation. She may do so to understand the potential of cherry-picking by a malicious party, to explore the extent  of the phenomenon, or to find out whether the dataset can still support the master's degree. Hence, she seeks support for the claim that \emph{individuals with a master's degree earn more, on average, than those with a bachelor's degree}. She finds that the claim holds for people in the field of Data Science and Machine Learning (Subpopulation~I). It also holds for people in Germany. In addition, she finds less compelling subpopulations, like 
people 
%who expressed a desire to use the ``Uno'' platform
who use Zoom for office  communication
(Subpopulation~II), and people who do not know the size of their organization.
\qed
\end{example}

The example illustrates that to be effective, the views that endorse the claim should be as significant and ``natural'' to support for the claim. 
As the notion of ``natural'' subpopulations is subjective, we incorporate various measures of naturalness. These measures are drawn from different fields such as information theory, NLP, and statistics. For instance, drawing from past proposals~\cite{wu2013scorpion,TaoGMR22}, one simple measure evaluates the subpopulation's size, with a larger size indicating broader applicability of the claim. Another measure assesses the claim's strength within a subpopulation, determined through tests of statistical significance~\cite{wasserman2004all}. Additionally, naturalness can be inferred based on the linguistic relationship between the target attribute and the predicate that defines the view, using language models such as sentence transformers~\cite{sentence-bert2019} (as we adopt in our implementation).
Our proposed framework is designed to identify and extract the highest-scoring views under each measure, as well as combinations of measures, for a given user claim.

%We illustrate the usability of the \prob problem for the task of finding a \emph{natural} supporting subpopulation via the following example.
\eat{
\brit{Yuval's comments: }
\begin{itemize}
    \item the example is coming very late in the introduction. it is hard to understand the problem before this. consider moving it earlier
    \item Alex's motivation is not clear. what exactly is her challenge? why existing tools are not enough? how exactly does our proposed tool answer her need?
    \item is it actually a different name for cherrypicking? what exactly is the connection to cherry picking? is she trying to do cherry-picking? 
\end{itemize}
}

\begin{example}\label{ex:motivating1-cont}\em
We continue \Cref{ex:motivating1}, and focus on Subpopulations I 
%(Data Science and ML people) 
and II.% (users of Zoom for office communication). 
The definition of 
Subpopulation II is, arguably,  more artificial and arbitrary, hence less convincing; Subpopulation I appears more natural and supportive of the claim. This intuition is captured by various measures of naturalness. First, Subpopulation~I is larger 
(1446 individuals vs.~743).
%(1,287 individuals vs.~20).
Second, the criterion of ``office synchronous communication tool'' is quite unexpected due to lack of connection to salaries and/or academic degrees; we aim to capture this difference through measures from statistics and NLP.
%Second, its defining attribute \red{(specific job title) is}
%(experience and country) are 
%related more closely to the salary, compared to Subpopulation II's defining attribute \red{(a desire to work with a specific platform)}.
%(knowledge of the organization size).
Third, within Subpopulation I, master's people earn considerably more than bachelor's (average of
%\$126K,
\$183K
vs.~\$144K),
%\$104K; 
whereas the advantage in Subpopulation II is  modest (\$117K compared to \$112K).
%\$71K compared to \$70K 
These considerations are indeed reflected by lower scores for Subpopulation II% compared to Subpopulation I, 
according to the naturalness measures that we consider in this paper. 
See \Cref{sec:exp:case_studies} for additional examples.
\qed
\end{example}

In more formal terms, we define (in \Cref{sec:problem}) the problem of \textit{claim endorsement}  as follows. 
We are given a SQL query $Q$ with grouping and aggregation (e.g., Average) over a database (relation) $D$, along with a claim stating that the aggregate value of one group is higher than that of another group. We seek useful \emph{refinements} that restrict $Q$ by adding predicates $\pred$ to the selection condition of $Q$, so that the claim holds true in the refined query. The goal is to compute the top-$k$ refinements according to a function $\nat$ that quantifies the naturalness of each refinement (in the context of $Q$ and $D$). As aforesaid, we discuss a collection of basic measures of naturalness that cover different aspects of what ``natural'' can be. 
% Hence, an implementation may solve multiple instances of the problem by considering different measures (as we present in our implementation).
% \red{More generally, our framework can be used with any naturalness measure and combinations thereof, as we demonstrate in \Cref{sec:compute}.}
Hence, a system can solve multiple instances of the problem with different measures (as we present in our implementation).

%\brit{refinement instead of adjustment? I would say this is not the focus of our work - instead, you can "our framework enables, beyond selecting a natural view that endorses the claim, critical inspection of the give claim...."}
\reva{% comment R1-D1
The goal in claim endorsement is to find a natural sub-population supporting the claim, which will convince a critical listener of the claim's truth. In effect, we do so by refining the query representing the claim with the condition of the sub-population. However, our focus is on the original claim, and finding the views that support it.}

The main technical challenge that we address (\Cref{sec:compute}) is the high computational cost of claim endorsement: refinements can be made out of many attributes, attribute combinations, and value assignments; moreover, each candidate refinement may require costly computation to verify its correctness and measure its naturalness. Efficiency is critical when claim endorsement is used within data exploration, where the user may interactively react to the results of some claims to formulate new ones; response times of days or even hours render the tool ineffective. In particular, listing all possible refinements before ranking is too costly.
Similar challenges have been encountered in the search for drill-down and roll-up operators to find the most interesting data parts for exploration~\cite{agarwal1999cube,sathe2001intelligent,joglekar2017interactive,DBLP:journals/pvldb/YoungmannAP22}, the discovery of intriguing data visualizations and explanations~\cite{vartak2015seedb,deutch2022fedex}, and the 
explanation of outliers in aggregate queries~\cite{wu2013scorpion,li2021putting}. 
% However, we do not see any way of incorporating in the algorithms of that body of work the requirement to satisfy the claim and, at the same time, be high on naturalness measures. 
However, the algorithms in that body of work cannot be applied in our setting where the requirement is to satisfy the claim and, at the same time, achieve high scores of naturalness measures, as they satisfy neither requirement (see \Cref{sec:related}).

Instead of materializing all possible candidates, we devise a framework for \emph{anytime algorithms} that target the incremental generation of high-quality refinements from the very beginning. We instantiate the framework on the aforementioned measures of naturalness. More technically, the algorithms enumerate refinements in a ranked fashion, where the ranking function is, intuitively, well correlated with the naturalness measure, yet efficient to handle. Our framework deploys two main steps: \emph{(1)} produce a ranked list of attribute combinations according to an easy-to-compute scoring function. \emph{(2)} for each attribute combination $C$ (in ranked order), compute all value assignments for the corresponding refinements.
%rather than iterating on all assignments to each attribute combination. 
For each step, we develop several solutions and optimizations that we evaluate in the experiments. 
% (next). 

%\brit{existing baselines instead if related work methods}
We present a thorough experimental study (\Cref{sec:experiments}) on three datasets: American Community Survey, Stack Overflow, and a flight-delays dataset. These datasets differ in the number of attributes (from tens to hundreds) and number of tuples (from tens of thousands to millions). \revb{We show that our anytime algorithms typically achieve 95\% of the quality of the true top-$k$ refinements faster than existing baselines by one-to-two orders of magnitudes.} We also present case studies on the three datasets, similarly to \Cref{ex:motivating1}.

\common{
Additionally, we report a user study that we have conducted to assess how well our proposed naturalness measures align with intuitive concepts, and to compare them to existing solutions. Our study indicates that these measures effectively capture aspects of intuitive naturalness and, furthermore,  that they align better with intuition than measures adapted from previous work (e.g., \cite{salimi2018hypdb}).
}
%In addition to the experimental study, we also conducted a user study, where we presented participants with statements created from refinements that endorse a given claim. The participants were asked to rate the statements based on how useful they would be in a text supporting the claim. We created plausible scenarios for each of the datasets and their claims. The statements were obtained through different measures: some from this work and some from previous works. The two measures most frequently rated at the top were coverage and statistical significance. The user study showed that measures defined in this work correctly capture the user's intuition of what is a natural statement.

%To evaluate the real world applicability of the problem and the naturalness measures (detailed in Section~\ref{sec:naturalness-measures}), we conducted a user survey, where we collected the viewpoints of 50 external users .... \shunit{what more to say about this here in the introduction?}}

In summary, our contributions are as follows: %\begin{enumerate}
%    \item 
(a) We introduce the claim-endorsement problem.
    %\item 
(b) \reva{We introduced the concept of naturalness measures for claim endorsement, and proposed several concrete measures, adjusted to the context of this problem.}
    %\item 
(c) We devise an anytime framework for claim endorsement.
    %\item 
(d) We instantiate the framework with algorithms for the proposed measures of naturalness. 
    %\item 
(e) \reva{We present an experimental study that shows the effectiveness of our solutions and analyzes several case studies, and a user study to assess the effectiveness of the naturalness measures.}
%\end{enumerate}

% \input{02-problem}
\section{Formal Framework}
\label{sec:problem}
In this section, we describe our formal model and define the problem of claim endorsement. We begin with preliminary definitions.
\vspace{-0.5em}

% double column version
% \begin{table*}[ht]
%     \centering \small
%      \caption{Sample of the ACS dataset~\cite{folktables2021} used as our running example in \Cref{sec:problem}.}
%     \label{tab:DB_example}
%     \vspace{-4mm}
%     \begin{tabular}{cclcllcc}
%         \hline
%         \rowcolor{lightgray}ID & Sex & OccupationGroup & Age (binned) & EducationLevel & MaritalStatus & QuarterOfBirth & TotalIncome (K)\\
%         \hline
%         1 & F & CS\&Math & 30-35 & Bachelor's degree & Never Married & 1 & 72 \\
%         2 & F & CS\&Math & 40-45 & Master's degree & Divorced & 3 & 95 \\
%         3 & F & Education & 40-45 & Master's degree & Married & 2 & 43 \\
%         4 & F & Sales & 40-45 & High School Diploma & Married & 1 & 35 \\
%         5 & F & Sales & 30-35 & Bachelor's degree & Never Married & 4 & 100 \\
%         6 & M & CS\&Math & 30-35 & Bachelor's degree & Never Married & 4 & 80 \\
%         7 & M & CS\&Math & 40-45 & Master's degree & Married & 3 & 90 \\
%         8 & M & Education & 40-45 & Master's degree & Married & 2 & 62 \\
%         9 & M & Sales & 40-45 & Bachelor's degree & Divorced & 1 & 70 \\
%         10 & M & Sales & 40-45 & High School Diploma & Never Married & 3 & 65 \\
%         \bottomrule
%     \end{tabular}
% \end{table*}

% Single column version
\begin{table}[t]
    \centering \scriptsize
     \caption{Records from ACS~\cite{folktables2021} used as a running example.}
    \label{tab:DB_example}
    \vspace{-4mm}
    \begin{tabular}{ccllcc}
        \hline
        \rowcolor{lightgray}ID & Sex & Occupation & EducationLevel & QoB & Income (K)\\
        \hline
        1 & F & CS\&Math & Bachelor's degree & 1 & 72 \\
        2 & F & CS\&Math & Master's degree & 3 & 95 \\
        3 & F & Education & Master's degree & 2 & 43 \\
        4 & F & Sales & High School Diploma & 1 & 35 \\
        5 & F & Sales & Bachelor's degree & 4 & 100 \\
        6 & M & CS\&Math & Bachelor's degree & 4 & 80 \\
        7 & M & CS\&Math & Master's degree & 3 & 90 \\
        8 & M & Education & Master's degree & 2 & 62 \\
        9 & M & Sales & Bachelor's degree & 1 & 70 \\
        10 & M & Sales & High School Diploma & 3 & 65 \\
        \bottomrule
    \end{tabular}
   \vspace{-2em}
\end{table}

\subsection{Databases and Queries}
% \subsection{Definitions}
%\paratitle{Databases and queries}
% Throughout this section, 
% \brit{add a comment about multi-relation datasets}
We consider an input database $D$ that consists of a single relation with the relation name $R$ and the attribute set $\attrset{=}\{A_1, {\ldots}, A_q\}$. Note that this relation can be the result of a query that joins multiple source relations that are outside of the model (as is indeed the case in our experiments with the Flights dataset; see \Cref{sec:experiments}).
We denote by $|D|$ the number of tuples of $D$.
By $D[A_{i_1},{\dots},A_{i_\ell}]$ we denote the (set-semantics) projection of $D$ to the attributes $A_{i_1},{\dots},A_{i_\ell}$. By a slight abuse of notation, for a single attribute $A_i$ we may view $D[A_i]$ as the set of \emph{values} rather than single-value \emph{tuples}.

We assume an aggregate SQL query $Q$ of the following form.
\begin{equation}\label{eq:query}
\textsf{SELECT $\Agb$, $\alpha(\Aagg)$ FROM D WHERE $\phi$ GROUP BY $\Agb$}
\end{equation}
where $\Agb\in\attrset$ is the group-by attribute, $\Aagg\in\attrset$ is the aggregate attribute, and $\alpha$ is an aggregate function among  $\CNT$, $\SUM$, $\AVG$, $\MED$, $\MIN$ and $\MAX$.
We will refer to a query of this form as a \emph{group-aggregate} query.
The result of $Q$ on $D$, denoted $Q(D)$, is a set of tuples of the form $(\group, \val)$, where $\group \in D[\Agb]$ and 
$$\val = \alpha(\{t.\Aagg \mid t \in D \land \phi(t) \land t.\Agb = \group\})
\,.$$
%\shunit{possibly define $\alpha(g)$ as $\val$}

\begin{example}\em
% \em
\label{ex:running_example}
% In our running example, we will use the sample dataset in \Cref{tab:DB_example}, based on average values from the ACS dataset~\cite{folktables2021}. 
Consider the sample of the ACS dataset~\cite{folktables2021} shown in \Cref{tab:DB_example}. An analyst may be interested in justifying a Master's degree, so they will first issue the query $Q$:
% \ag{Fix formatting:}
% \begin{small}
% \begin{equation*}
% % \begin{split}
% \textsf{SELECT EducationLevel, $\AVG$(TotalIncome) FROM R GROUP BY} \textsf{EducationLevel}
% % \end{split}
% \end{equation*}
% \end{small}
% \begin{center}

\def\br{\smallskip\noindent}

\br\hrule\br
\textsf{%
SELECT EducationLevel, Average(Income)\\
FROM D\\
GROUP BY EducationLevel;
}
\br\hrule\br
% \end{center}
% in \Cref{eq:query} where $\Agb$ is $\textsf{EducationLevel}$,  $\Aagg$ is $\textsf{TotalIncome}$, $\\alpha=\AVG$, and $\phi=\textsf{true}$ (i.e., no filter). \qed 
In $Q(D)$, we find that the average income for people with a Master's degree (\$72.5K) is lower than that of Bachelor's degree (\$80.5K).\qed
\end{example}

\subsection{Claims and Refinements}\label{sec:claims_and_refinements}
Similarly to previous work in the context of query result explanations \cite{li2021putting,abs-2207-12718,RoyS14,TaoGMR22},
we study the case where the analyst restricts attention to the relationship between two groups of interest, $\group_1$ and $\group_2$, in the result $Q(D)$. For these, the analyst may be interested in endorsing a specific claim. We define it formally as follows:
\begin{definition}[Claim]\label{def:claim}
%Let $Q$ be a group-aggregate query (\Cref{eq:query}), and
Consider two tuples $(\group_1, \val_1)$ and $(\group_2,\val_2)$ in $Q(D)$. We refer to the tuple $\claim = (\group_1, \group_2, >)$ as a \emph{claim}.  
A group-aggregate query $Q'$ 
%\emph{satisfies} 
\emph{endorses} $\kappa$ (on $D$), denoted 
 $Q'(D) \models \kappa$,  if there are numbers $\val_1'$ and $\val_2'$ such that
 $(\group_1,\val_1')\in Q'(D)$ and $(\group_2,\val_2') \in Q'(D)$ and $\val_1' > \val_2'$.
\end{definition}
%Intuitively, the meaning of the claim is `I am interested in showing that the aggregate value of $\group_1$ for attribute $\Aagg$ is larger than that of $\group_2$'. 
We consider the situation where $Q$ violates $(\group_1, \group_2, >)$, and seek a \emph{refinement} $Q'$ that satisfies it. We focus on refinements that add predicates to the WHERE clause (as done previously, e.g., \cite{lin2021detecting}).

\begin{example}\label{example:violated-by-Q}
\em
% Following the result of $Q$ in \Cref{ex:running_example},
Following \Cref{ex:running_example},
the analyst wishes to compare the average income for different degree holders, with the initial assumption that a higher degree implies a higher salary. Yet, she finds that the average income for people with a Master's degree (\$72.5K) is actually \emph{lower} than that of people with a Bachelor's degree (\$80.5K). In our formalism, the analyst is interested in the relationship between $\group_1=\mathsf{Master's}$ and $\group_2=\mathsf{Bachelor's}$ and their corresponding values in $Q(D)$, namely $\val_1=72.5$ and $\val_2=80.5$. Hence, the claim  $(\mathsf{Master's}, \mathsf{Bachelor's}, >)$ is violated by $Q$.
\qed
\end{example}

In this paper, we study the problem of searching for query refinements. %\footnote{We note that query refinement has recently been used for changing the representation in the query result~\cite{MishraK09,KoudasLTV06,VelezWSG97,Ortega-BinderbergerCM02,LiMSJ23,SalimiGS18}.\benny{Representation of what? Also, this comment seems way out of context here. Maybe it better fits in the Introduction? Or related work? I do not see how this sentence is part of the story, and it seems like we're just saying stuff so that people won't say that we ignore related work. We should tell how it is related, in the proper place.}\shunit{I've added to the related work - if it looks good there, we can remove this footnote from here.}} that endorse the claim. 
To that end, we assume a space $\predspace$ of predicates that can be used to refine the query $Q$. 
We consider equality predicates of the form $A {=} \val$, where $\val {\in} D[A]$, and conjunctions of up to $m$ such predicates, where $m$ is a parameter. This practice has been commonly used in prior work about query answer explanations~\cite{lin2021detecting,roy2014formal,wu2013scorpion,el2014interpretable,miao2019going}. 
More formally, we are given a set of \emph{split attributes} from $\attrset$ that does not include $\Agb$ and $\Aagg$, and we denote this set by $\splitset$.\footnote{This is similar to prior work on cherry-picking detection~\cite{lin2021detecting} and other work on counterfactual explanations, where a subset of attributes cannot be used in the explanation as it is non-actionable~\cite{GalhotraPS21,KarimiSV21,KarimiBSV23}.}
Then $\predspace$ consists of all predicates of the form
$A_1=\val_1\land\dots\land A_\ell=\val_\ell$ such that 
$\ell\leq m$, $A_i\in\splitset$ and $\val_i\in D[A_i]$ for all $i=1,\dots,\ell$. An {\em atom} or atomic predicate is a predicate of the form $A_i=\val_i$. 
Given a predicate $\pred$ of this form, denote by $Att(\pred) = (A_1,\dots,A_\ell)$ the set of attributes used to define $\pred$, sorted lexicographically.

\begin{definition}[Refinement]\label{def:refinment}
Let $Q$ be a group-aggregate query as in \Cref{eq:query}, and let $\pred {\in} \predspace$ be a predicate. The \emph{refinement of $Q$ by $p$} is the query $Q_{\pred}$ where $\phi$ is replaced by $\phi' {=} \phi \land \pred$. 
\end{definition}

Note that, by \Cref{def:claim}, a refinement $Q_\pred$ endorses the claim $\kappa$ if $Q_{\pred} \models \kappa$, that is, $\pred$ selects a subset of $D$ that satisfies $\alpha(g_1)>\alpha(g_2)$.

\begin{example}\label{ex:predicates}
\em
Recall that in \Cref{example:violated-by-Q}, $Q$ violates the claim $(\mathsf{Master's}, \mathsf{Bachelor's}, >)$. 
Now, consider the predicate $\pred$ given by the expression $\textsf{Occupation}=\textsf{CS\&Math}$. 
In contrast to $Q$, the refinement $Q_p$ satisfies the claim: the average income for Master's degree holders is \$92.5K, yet only \$76K for Bachelor's. Another possible refinement is defined by the predicate $\textsf{QoB}{=}3$ (where QoB stands for Quarter of Birth), where $v_1{=}92.5$ and $v_2{=}65$. \qed
%The analyst may decide to exclude in advance the ID attribute: $\exclusion=\{\textsf{ID}\}$, since it will not yield meaningful subgroups.
\end{example}

%The attribute set $\attrset$ can be split into a set of {\em exclusion attributes} $\exclusion \subseteq \attrset$ such that it cannot be used in predicates in \predspace. This is similar to prior work on bias detection \cite{lin2021detecting} and other works about counterfactual explanations, where a subset of attributes cannot be used in the explanation as it is non-actionable \cite{GalhotraPS21,KarimiSV21,KarimiBSV23}.
%Attributes in $\splitset = \attrset \setminus (\exclusion \cup \{\Agb, \Aagg\})$, which can be used in \predspace, are called \textit{split attributes}, and we denote this set of attributes by $\splitset$. 
% \brit{similar to what was done in \cite{lin2021detecting}} 
% Finally, the user may specify a minimal size of a group in a view to be considered, which we denote by $M$. Views where $|V\cup P_i|<M$ should not be returned.

%\shunit{new paragraph here}
%Each refinement can be used to form a refined claim, which is supported by the dataset. In the previous example, the analyst can say, ``In occupations in the fields of CS and Math, the average income of Master's graduates is higher than for Bachelor's.''

%\brit{We can also note that a similar approach has been adopted in query results explanation work, where conjunctions of equality predicates are considered intuitive and easy-to-understand explanations \cite{el2014interpretable,roy2015explaining}. use a paragraph* for remark, this does not need to be numbered}

%\paragraph*{The choice of predicate space.} %comment R1-D3, R1-D4
\reva{Note that we restrict the discussion to refinement in the form of conjunctions of equality predicates. We leave for future work the consideration of a richer class of refinements, including, for example, inequality predicates and disjunction, since the search space of conjunctions is already large and computationally challenging. The choice of conjunctions of equality predicates is also in line with past work on explanations, where such conditions are considered intuitive and easy-to-understand~\cite{el2014interpretable,roy2015explaining}. %Enriching the predicate language is an interesting direction, left for future work.
% The search space of conjunctions is large and provides a computational challenge. Expanding the predicate space to disjunctions and ineqalities is an interesting future work direction. 
% It is also possible to consider relaxations of the query conditions, but our focus is on the case where the claim does not hold without restrictions. \shunit{I don't like the flow of the last sentence here - but not sure how to improve this.}
}

\subsection{Naturalness}
Supporting the analyst claim can be performed by finding a certain refinement where the claim holds. However, this refinement should be \emph{natural} in the sense that it should not be overly specific and restricted. For example, the following refinement from the ACS dataset:  ``people who depart for work between 11:50 and 13:30 and have served in the US military after Sept. 2001'' is, arguably, overly specific and can hardly serve as significant support for the claim. 

\begin{definition}[Naturalness Measure]
%Let $\queryset$ denote the space of group-aggregate queries (in the form of \Cref{eq:query}).
Let $Q$ be a group-aggregate query, and
$\claim$ a claim. 
%, a dataset $D$, and a set of views $\viewset$ created by $\pred$-refinements of a query $Q$: $Q_\pred \in \queryset_\predspace$,
A \emph{naturalness measure} (for $Q$ and $\claim$) is a function $\nat$ that maps pairs $(Q_p,D)$, where $Q_p$ is a refinement and $D$ is a database, to a numerical score $\nat(Q_p,D)$.
  \label{def:nat_measure}
\end{definition}
Semantically, $\nat(Q_{\pred},D){>}\nat(Q_{\pred'},D)$ means that we consider $Q_p$ as more natural than $Q_{p'}$ for the database $D$ (and specifically the results $Q_p(D)$ and $Q_{p'}(D)$). 
We are not aware of any relevant formalization of naturalness. Intuitively, $\nat$ aims to quantify (or be well correlated with) the likelihood of a critical listener accepting the claim if it is presented with this refinement. 

%\benny{Shunit - please review the definitions up to now, and if we accept them then please change the remaining text of this section accordingly. (In particular, we do not use the term "view" any more but rather stick to the terms that we already have).}

An example of a naturalness measure is the coverage of $Q_{\pred}$: the fraction of database tuples covered by $\phi {\wedge} \pred$. 
% Another possible measure is the \emph{strength} of the claim, that is, the result of a statistical significance test for the difference between the groups discussed in the claim. In the contra-positive, this measure captures the randomness of satisfying the claim: \emph{how likely it is that the claim holds over this refinement simply due to noise?}
% %The statistical test result is affected by the group sizes and the difference between aggregate values.
% Another naturalness measure is based on the linguistic connection between the predicate $\pred$ and the target attribute $\Aagg$, and we can capture it by the similarity of their descriptions according to a language model. Such a measure is based solely on the predicate $\pred$ and not on the database $D$.
We provide additional examples of measures of naturalness in \Cref{sec:naturalness_metrics} and implement them as part of our framework and empirical study. 
%The user may also define custom measures of naturalness.

\eat{
\begin{example}
\label{ex:coverage}
\em
For the two predicates of~\Cref{ex:predicates}, the coverage of $\pred_1=(\textsf{Occupation}{=}\textsf{CS\&Math})$ is: 
% $\coverage(Q_{\pred_1}, D) {=}0.4$
$\coverage(Q_{\pred_1}, D) = |\sigma_{\pred_1}(D)|/|D|=4/10=0.4$, 
and $0.3$ for $(\textsf{QoB}=3)$. \qed
\end{example}
}

\subsection{Problem Definition}
Naturalness is a subjective notion that is unlikely to be expressed precisely by a mathematical formula. In the remainder of the paper, we propose and study several examples of measures that capture what we deem as intuitive aspects of naturalness. 
Additionally, to broaden the scope of candidates, we consider an output of $k$ refinements instead of a single one, as users are more likely to find the desirable refinements this way even if the measure $\nat$ does not completely capture their idea of naturalness. 
% among the first $k$ refinements by the measure $\nat$, rather than in the very top. 
Furthermore, good refinements may be found in the \emph{union} of the top-$k$ refinements of \emph{multiple} measures. Hence, we define our main problem as follows.

\begin{definition}[Claim Endorsement]
Fix a predicate space \predspace and a naturalness measure $\nat$. Claim endorsement is the following problem:
given a database $D$, a group-aggregate query $Q$,
a claim $\claim = (\group_1, \group_2, >)$ and a natural number $k$, find $k$ refinements $Q_p$ of $Q$ that endorse $\claim$ and have the highest $\nat$ scores.
\end{definition}

%\shunit{Please review - does this next paragraph clarify the point?}
This problem definition allows for defining an instance of the problem for each naturalness measure; often, the practical thing to do is to combine several different measures, that is, to retrieve the top-$k$ refinements according to each naturalness measure and to present all of them to the user to select from. Furthermore, the system can define a naturalness measure that combines the measures (e.g., via weighted sum). 
In our implementation, we also allow for a hyper-parameter $M$ that restricts the refinements to be such that each subpopulation ($g_1$ and $g_2$) is of size at least $M$, so that we avoid refinements that apply to tiny groups.
In \Cref{sec:compute}, we devise methods to prioritize the search by a combination of all naturalness measures, and we evaluate these methods empirically in \Cref{sec:experiments}.

\begin{example}
\em
% Over the database $D$ in \Cref{tab:DB_example}, the user computes the average income for each education level, as shown by the query in \Cref{ex:running_example}. 
% \smallskip\noindent%
% \textsf{%
% SELECT EducationLevel, Average(TotalIncome)\\
% FROM D\\
% GROUP BY EducationLevel;\\
% }
% Reconsider the ACS dataset~\cite{folktables2021}, a sample of which is depicted in \Cref{tab:DB_example} and the query computing the average income for each education level, as shown in \Cref{ex:running_example}. 
% An instance of \prob aims to find refinements of up to $m = 2$ atoms so that the aggregated value for $g_1=\textsf{Master's degree}$ is higher than the aggregated value for $g_2=\textsf{Bachelor's degree}$, and to retrieve the top-$k$ out of those, according to some natural measure of choice (e.g., coverage). 
% When running this query on the dataset, 
% % on the  ACS dataset~\cite{folktables2021}, 
% the first few ranked predicates according to the average of all naturalness measures we used are defined by the 
% person's health insurance and the last time they worked; in the top 20, we can also find $\textsf{Occupation}=\textsf{Budget Analysts}$ and $\textsf{Occupation}=\textsf{Financial Clerks}$. 
% These two can be phrased as simple and intuitive hypotheses: in some occupations, a master's degree contributes more to the person's ability to perform well, and to their salary.
% Reconsider the sample of the ACS dataset~\cite{folktables2021}, depicted in \Cref{tab:DB_example} 
Reconsider the dataset in \Cref{tab:DB_example} 
and the query computing the average income for each education level, as shown in \Cref{ex:running_example}. 
An example of an instance of claim endorsement aims to find refinements of up to $m {=} 2$ atoms so that the aggregated value for $g_1=\textsf{Master's}$ is higher than the aggregated value for $g_2=\textsf{Bachelor's}$ (as opposed to the trend on the entire data shown in \Cref{example:violated-by-Q}), and to retrieve the top-$20$ ($k=20$) out of those, according to its coverage as the natural measure of choice. Two
of the refinements are described in \Cref{ex:predicates}. 
% When running this query on the dataset, 
% A few predicates according to the average of all naturalness measures we used are defined by the 
% person's health insurance and the last time they worked; in the top 20, we can also find $\textsf{Occupation}=\textsf{Budget Analysts}$ and $\textsf{Occupation}=\textsf{Financial Clerks}$. 
% These two can be phrased as simple and intuitive hypotheses: in some occupations, a master's degree contributes more to the person's ability to perform well, and to their salary.
\qed
\end{example}

\section{Examples of Naturalness Measures}\label{sec:naturalness-measures}
\label{sec:naturalness_metrics}
We now present a collection of naturalness measures, adapted from prior work. 
% of a given query refinement and claim. 
%Naturalness is not a formally defined concept, but rather, an attempt to capture the intuition of a critical listener presented with the claim and refinement. Therefore, we supply examples of various measures, each capturing a different angle of the intuitive naturalness concept. 
The examples aim to cover varying intuitive aspects of true naturalness, and are sourced from various domains, including information theory, statistics, and natural language processing.
% and entail different computational challenges. 
%We expect that in the future, many other measures will be suggested.
% The first two measures (Coverage and Embedding similarity) are deterministic, while the other three measures are probabilistic.
%comment R3-W1
\revc{A user can define custom naturalness measures, by implementing $\nat(Q_\pred, D)$ and possibly a method to prioritize the search for faster discovery of refinements with high $\nat$ values. We further discuss custom measures in \Cref{sec:methods:prioritization}.}
\paragraph*{Coverage} This measure is the fraction of tuples covered by the refinement. % (as explained in~\Cref{ex:coverage}). 
More specifically, recall the condition $\varphi'{=}\varphi{\land} p$ from the definition of refinement (\Cref{def:refinment}). 
%In the following formula, we denote by $\sigma_{\varphi}(D)$ the set of tuples from $D$ that satisfy $\varphi$. 
Let $\sigma_{\varphi}(D)$ denote the set of tuples from $D$ that satisfy $\varphi$. 
We then define: $$\coverage(Q_\pred, D) {\defeq} |\sigma_{\varphi\land\pred}(D)|/|D|$$

% \begin{equation}\label{eq:coverage}
%  \coverage(Q_\pred, D) \defeq \frac{|\sigma_{\varphi\land\pred}(D)|}{|D|}    
% \end{equation}

\paragraph{Embedding similarity} 
Modern word embeddings well capture the semantics of textual data~\cite{caliskan2017}. The cosine similarity between the embedding of the predicate (treated as text) and that of the target attribute can measure their perceived relatedness. To compute this measure, a pre-defined mapping $T:A{\rightarrow} \Sigma^*$ from attribute identifiers to a textual representation is needed. This mapping can be manually defined or extracted from database documentation; in our implementation, we mapped each abbreviation to the full wording. For example, in the ACS dataset, we mapped ``MARHM'' to ``Married in the past 12 months,'' based on the ACS documentation.
An embedding model $E: \Sigma^*{\rightarrow} \mathbb{R}^d$ emits a $d$-dimensional vector of real numbers for a given textual input. The measure is given by: 
\begin{small}
\begin{equation}
% This equation is referenced in section 3, please do not remove its number.
\label{eq:embsim}
\begin{aligned}
   \embsim(Q_\pred, D) \defeq
   \mathit{CosSim} \big( E \big( \hspace{-0.5em}{\bigoplus_{(A_i{=}v_i) {\in} \pred}}\hspace{-0.5em}T (A_i ) {\oplus} T ( v_i) \big), E(T (\Aagg)) \big)
\end{aligned}
\end{equation}
\end{small}
Here, $\oplus$ represents string concatenation and $\mathit{CosSim}$ is the cosine similarity between two vectors.
This score aims to measure how natural it is to limit the records according to the predicate $A_i{=}v_i$, when asking a query about the target attribute. 
In our implementation, we used a general-purpose sentence-BERT\footnote{Specifically, \textsf{all-MiniLM-L6-v2}: \url{https://huggingface.co/sentence-transformers/all-MiniLM-L6-v2}} model~\cite{sentence-bert2019}.

% This model was trained on a dataset of 1B sentence pairs, and it is a sentence-level (not word-level) model, which makes it suitable for embedding sequences of words (such as attribute names and values).}

\begin{example}
 \em
    %Since the strings in our example dataset are already represented as readable text, assume $T$ only separates the camel-case names to separate words. 
    For $\pred_1$ from \Cref{ex:predicates} we have: 
    \begin{small}
    \begin{equation*}
    \label{eq:embsim_example1}
    \begin{aligned}
      & \embsim\big(Q_{\textsf{Occupation}=\textsf{``CS\&Math''}}, D\big) = \\
      & \mathit{CosSim}(E(\textsf{``Occupation CS \& Math''}), E(\textsf{``Income''})) {=}0.19
    \end{aligned}
    \end{equation*}
    \end{small}
    For the predicate $\pred_2$, the  value 
    $\mathit{EmbSim}(Q_{\textsf{QoB}{=}3}, D)$ is \\
     $\mathit{CosSim}(E(\textsf{``Quarter\ Of\ Birth\ 3''}), E(\textsf{``Income''})) {=} 0.07$. 
    This is to be expected since a person's occupation is considered more related to their income than the quarter of the year of birth. 
    \qed
\end{example}

\paragraph*{Statistical significance}
For this measure, we adopt \emph{hypothesis testing}, which determines whether the difference between group values is significant and indicative of an actual phenomenon. The database is viewed as a sample of a real-world population. %The null hypothesis usually represents the case when the observed result from the database is random and does not represent the real difference, while the alternative hypothesis is that the phenomenon seen in the database is indeed representative.

%To compute this score, a test statistic is computed based on values in the database, such as average, standard deviation, count, and median. Then a p-value is computed, representing the probability of observing the value of the test statistic, given that the null hypothesis is correct. Small p-values mean that the null hypothesis should be rejected, meaning that the difference between the groups is significant.
%\benny{Shunit - we do not need a lecture on p-values and statistical significance. We just need things to be well defined. Also, no need for the figure. It is a paper, not lecture notes.} \shunit{I shortened the explanation. But I'm a little confused about what is the important information here for our readers. Last week you said you don't understand this and none of our readers will either, so I added explanations. Apparently I added too much?}
%\benny{The reader needs to know the details on what is the quantity. How do I implement it? What is the formula? Where it comes from can be attributed to references, but the reader needs to know the formulas. If there is a standard measure from literature, then we explain the reader the formula in the language of our framework, but need not give the full background on its essence (though it is good to briefly remind).}

This score has been previously used as a measure of interestingness in the context of insights in views~\cite{tang2017}. This measure is defined only when $\alpha$ is $\AVG$ (where we use the measure $\statsig_{\mathrm{avg}}$) or $\alpha=\MED$ (where we use $\statsig_{\mathrm{med}}$).\footnote{For the other aggregations, there is no closed-form statistical test. It is possible to use bootstrapping or permutation tests, which we leave for future work.}
When $\alpha=\AVG$, the appropriate test is the \emph{two-sided independent T-test}~\cite{kanji2006100}. 
To define the measure precisely, we denote: 
$$\alpha(g_i) := \alpha(\{t.\Aagg | t\in D \wedge t.\Agb = g_i\})$$
Then, the test statistic is given by
$$t^* \defeq \frac{\AVG(g_1)-\AVG(g_2)}{\sqrt{(\STDDEV(g_1)^2/\CNT(g_1)+(\STDDEV(g_2)^2/\CNT(g_2)}}\,.$$
Under the null hypothesis, the test statistic comes from a T-distribution with $\mathit{df}$ degrees of freedom, where the formula for $\mathit{df}$ is as follows. 
%$\mathit{df}$ is computed based on $\STDDEV(g_i)$ and $\CNT(g_i)$.
Let $n_i = \CNT(g_i)$ and $s_i=\STDDEV(g_i)$. Then,
$$\mathit{df} \defeq \frac{(s_1^2/n_1 + s_2^2/n_2)^2}{(s_1^2/n_1)^2/(n_1-1) + (s_2^2/n_2)^2/(n_2-1)}\,.$$
%$$\mathit{df} = \frac{(\STDDEV(g_1)^2/\CNT(g_1) + \STDDEV(g_2)^2/\CNT(g_2))^2}{(\STDDEV(g_1)^2/\CNT(g_1))^2/(\CNT(g_1)-1) + (\STDDEV(g_2)^2/\CNT(g_2)^2/(\CNT(g_2)-1)}$$
%\benny{I have absolutely no idea what ''df' means and how it is computed from stddev and cnt. Is there a reference to that? An intuitive explanation of what it means?}
%\shunit{I re added the formula.}

Then, the p-value for the two-sided T-test is the probability of observing $t^*$ or a more extreme value, under the null hypothesis (that $X\sim T\left( \mathit{df} \right)$). It is given by
$\mathrm{pvalue} \defeq 2\cdot P(X\geq t^*| X\sim T_{\left( \mathit{df} 
\right)}).$

The naturalness score is defined as the complement of the p-value.
We use the complement instead of the raw p-value so that higher scores will be associated with stronger claims.
$$\statsig_{\mathrm{avg}}(Q_\pred, D) \defeq 1-\mathrm{pvalue}$$

For median difference,
we use the median test~\cite{sprent2007}
with Yates correction for continuity~\cite{yates1934contingency}.
Denote $a_i = |\{t | t.\Agb = g_i \wedge t.\Aagg > \MED(g_i)\}|$ and 
$b_i = \CNT(g_i)-a_i$.
Also, let $\mathit{EA}_i = (a_1+a_2)\cdot(a_i+b_i)/(a_1+b_1+a_2+b_2)$ and
   $\mathit{EB}_i = (b_1+b_2)\cdot(a_i+b_i)/(a_1+b_1+a_2+b_2)$.
% $a_i = \CNT({t | t.\Agb = g_i \wedge t.\Aagg > \MED(g_i)})$\\
% $b_i = \CNT(g_i)-a_i$\\
% $EA_i = (a_1+a_2)*(a_i+b_i)/(a_1+b_1+a_2+b_2)$\\
% $EB_i = (b_1+b_2)*(a_i+b_i)/(a_1+b_1+a_2+b_2)$\\
The test statistic is given by:
\def\infrac#1#2{#1/#2}
\begin{equation}
  \begin{aligned}
  x^*\defeq &\infrac{(|a_1-\mathit{EA}_1|-0.5)^2}{\mathit{EA}_1} + \infrac{(|b_1-\mathit{EB}_1|-0.5)^2}{\mathit{EB}_1}+ \\
  &+\infrac{(|a_2-\mathit{EA}_2|-0.5)^2}{\mathit{EA}_2}+\infrac{(|b_2-\mathit{EB}_2|-0.5)^2}{\mathit{EB}_2}    
  \end{aligned}  
\end{equation}
Under the null hypothesis, the test statistic comes from a Chi-square distribution with 1 degree of freedom. Then the naturalness score is defined as the complement of the probability of observing $x^*$ or a more extreme value, under the null hypothesis (that $X\sim \chi^2_{\left( 1 \right)}$).
$$\statsig_{\mathrm{med}}(Q_\pred, D) \defeq 1-P\left(x\geq x^*| x\sim \chi^2_{\left( 1 
\right)}\right)$$
% \subsubsection*{Statistical significance.} 
% %The difference in the aggregated target attribute between the two given populations can be very small. A claim such as ``Among high school students, the average income for women is 2413 versus 2442 for men'' is a weak claim, as the difference is so small that it may be the result of noise in the sample, rather than an actual phenomenon. 
% The goal of \emph{statistical significance testing} is to estimate how well the difference between group values is significant and indicative of an actual phenomenon. This score has been previously used as a measure of interestingness when searching for insights in views~\cite{tang2017}. This measure is defined only for the $\AVG$ and $\MED$ aggregation functions.\footnote{For the other aggregations, there is no closed-form statistical test. It is possible to use bootstrapping or permutation tests, which we leave for future work.}

\paragraph*{Mutual information (MI)} 
Arguably, predicates that are deemed natural would be based on attributes with some correlation to the target attribute. A possible method to quantify this dependence is MI between attributes, which has been previously used to recommend views for anomaly detection~\cite{kandel2012}. 
Given a refinement $Q_\pred$, we use MI between the attribute list $Att(\pred) = (A_1,\dots,A_\ell)$ that defines $\pred$ 
and the target attribute $\Aagg$. 
%\benny{Distribution on what??? We've never defined any probability space here. What am I missing? The framework is deterministic.}\shunit{please see if this is clearer now.}
The definition of MI uses probability distributions over the values of a list of attributes, computed based on their normalized frequencies in the database, and it is given by:
%\benny{It's been a long time since we've seen the definition of the split attributes. Are we talking about what we previously called SAtt? So why don't we use the notation? The reader might think that this is a different thing (and I'm not sure myself). Please remind the reader and give a pointer to where the notation is defined. }
\begin{equation*}
  \begin{aligned}
  \mi(Q_\pred,D)&{\defeq} I((A_1,{\ldots},A_l), \Aagg) \\
   &=   \Delta_{\mathrm{KL}}(P_{((A_1,{\ldots},A_l),\Aagg)} {\mid} {\mid} P_{((A_1,{\ldots},A_l)\otimes P_T)})   
  \end{aligned}  
\end{equation*}
where $\Delta_{\mathrm{KL}}$ %\benny{Same as before, KL should be in mathrm}
%\benny{Please fix $KL$ all throughout the paper. Latex thinks that there are two variables $K$ and $L$ and $KL$ is their product. Use $D_{\mbox{KL}}$, probably with a macro. Can we avoid reusing $D$? $D$ is a database in our framework. Maybe $\Delta$?}
is the KL-divergence distance between probabilities, $P_{((A_1,...,A_l),\Aagg)}$ is the joint distribution of $A_1,...,A_l$ and $\Aagg$, and $P_{((A_1,...,A_l)\otimes P_{\Aagg})}$ is their product distribution.

Mutual information is non-negative and unbounded. To maintain the value between [0,1] we normalize all values by dividing them by the largest MI value of an attribute or attribute combination.
% \benny{Skipped this measure as well.}

% \benny{Got here.}
\paragraph{Analysis of variance ($\anova$)} %\benny{Please be consistent on the capitalization of titles... previously we use lowercase ("variance") and not first-cap ("Variance")}
Another method to quantify the dependence between an attribute combination $(A_1,\dots,A_\ell)$ and the aggregate attribute $\Aagg$ is $\anova$. 
Our use is similar to what was done by Wu et al.~\cite{wu2009promotion}
to assess the connection between a view dimension and an outcome dimension. 
% While both ANOVA and MI measure the dependence between variables, each highlights different attribute combinations, thus diversifying the top results.

As in the statistical significance score, the database is viewed again as a sample from a real world population.
Given a predicate $\pred$, consider the attribute combination used in it: $(A_1,\dots,A_\ell)$. %\benny{No need for a new notation, $Att(\pred)$. We already have lots of notation, and this one is not used. Moreover, it is again a word inside math...}
This attribute combination induces a partition of $\Aagg$ values into $\theta$ groups (bags) $\{x_i^1\}_{i=1}^{s_1},\dots,\{x_i^\theta\}_{i=1}^{s_\theta}$, where each group is associated with a value combination $(v_1,\dots,v_\ell)\in D[A_1,\dots,A_\ell]$: 
%Meaning,
$$\{x_i^j\}_{i=1}^{s_j} = \{t.\Aagg | t\in D \wedge \bigwedge_{p=1}^{\ell} t.A_p=v_p\}$$
Each group represents a sample from a normally distributed population.
The null hypothesis in $\anova$ is that each of the $\theta$ normal distributions has the same mean. The test statistic is based on the averages of each group of numbers in the partition. Let $N$ denote the total sample size: $N\defeq\sum_{j=1}^\theta s_j$. Let $\mu$ 
%\benny{You cannot write math-comma-math.}
denote the total average of $\Aagg$, and denote by $\mu_j$ the average of number group $j$.
$$\anova(Q_\pred,D) \defeq \frac
{\frac{1}{\theta - 1}\sum_{j=1}^\theta s_j\cdot\left(\mu_j - \mu\right)^2}
{\frac{1}{N-\theta} \sum_{j=1}^{\theta} \sum_{i=1}^{s_i} \left(x_i^j-\mu_j\right)^2}$$
%\benny{What is $s_j \left(\mu_j - \mu\right)$? Is $s_j$ a coefficient or a function? In the first case, please use $\cdot$.}
%The F-statistic can be used to calculate a corresponding p-value. However, we chose to use the F-statistic directly due to its wider range of values, which allows for a better differentiation of predicates.
%\benny{Is it F statistic (w/o hyphen) or F-statistic (with a hyphen?)}
The $\anova$ statistic is non-negative 
but unbound from above; as with $\mi$, we normalize it by dividing it by the largest value in $D$.

\paragraph*{Filtering by Generality}
%\subsection{Filtering by Generality}
\label{sec:gen_filter}
\revb{%R2-D1
In OREO~\cite{lin2022oreo}, the authors propose a method to retrieve counterarguments of a claim, and suggest to filter the ones contained in others, keeping only the most general. We adapt this concept as follows. 
Let $Q_{p_1}$ and $Q_{p_2}$ be two refinements. Recall that they are comprised of conjunctions of atomic predicates $A=v$. We say that $Q_{p_1}$ is \textit{more general} than $Q_{p_2}$ if the set of atoms comprising $p_1$ is a subset of the set of atoms comprising $p_2$. For example, if $p_1=(A_1=v_1)$ and $p_2=(A_1=v_1) \wedge (A_2=v_2)$, than $Q_{p_1}$ is more general than $Q_{p_2}$.
This filter can be composed over any measure of naturalness $\nat$, proposed in this section or otherwise. After retrieving the set of refinements and ranking them by $\nat$, we can filter only the most general refinements, and present the top-k out of those. 
We evaluate the effectiveness of this approach through the user study presented in \Cref{sec:user_study}.}
% \input{03-Methods}
% \section{Computing High-Quality Refinements}
\section{Computing Refinements}
\label{sec:compute}
%\shunit{note change in section title}
%\shunit{\textit{Retrieving} High-Quality Refinements?}
%\brit{the generality filtering should be presented first in this section, not at the user study for the first time. }

In this section, 
%we investigate the computational challenge of the claim-endorsement problem with the naturalness measures presented in \Cref{sec:naturalness-measures}. %We begin with a high-level and abstract overview of the algorithm, and then detail our concrete instantiations.
%\paragraph*{Solution Overview}
%\label{sec:alg_overview}
%Specifically, 
we present an \emph{anytime}
algorithmic framework to search for the top-scoring refinements. Recall that an \emph{anytime algorithm} can be asked for results at any moment, the quality of the results improves over time, and the goal is to reach close-to-optimal quality early in the computation~\cite{zilberstein1996anytime}.
%\shunit{Anytime algorithms have been used in the context of probabilistic databases~\cite{fink2013anytime} where the error guarantees improve over time, and top-k queries~\cite{arai2009anytime} where the confidence in the retrieved results is calculated at each time point.}
%where the given claim holds. 
\Cref{alg:guided} shows the general approach.
We first generate all possible attribute combinations up to a maximal number of $m$ atoms (\Cref{line:combs}). Next, we rank the combinations by some prioritization, based on pre-computation and heuristics (\Cref{line:sort}). We then go over the ranked attribute combinations (\Cref{line:for_loop}) and retrieve the refinements 
of each combination where the claim holds (\Cref{line:find_preds}). For each of these, we compute all naturalness measures (\Cref{line:compute_nat}) and print them (\Cref{line:print}).
In the remainder of this section, we describe the functions used in \Cref{alg:guided}.
 
% The algorithm

% For this algorithm to be fully specified, we need to implement four tasks:
% \begin{enumerate}
%     \item Materializing all attribute combinations, up to $m$ atoms.
%     \item Prioritizing the attribute combinations (so that more "promising" combinations are considered first.)
%     \item {For a given attribute combination, in the order defined in (2), finding all predicates that satisfy $\claim$.}
%     \item Computing the naturalness measures for all found predicates.
% \end{enumerate}

\begin{algorithm}[t]
\small
\DontPrintSemicolon
\SetKwInOut{Input}{Input}\SetKwInOut{Output}{Output}
\LinesNumbered
\newcommand\mycommfont[1]{\footnotesize\ttfamily\textcolor{blue}{#1}}
\SetCommentSty{mycommfont}
\SetKwFunction{FindPredicates}{Predicates}
\SetKwFunction{SortCombs}{Sort}
\SetKwFunction{ComputeNaturalnessMeasures}{NaturalnessMeasures}
\Input{Dataset $D$, query $Q$, user claim $\kappa$, natural numbers $m$ and $k$, a naturalness measure $\nat$.}
\Output{Set $M$ of refinements of $Q$.} 
$\mathit{Combs}\gets $ all combinations~from $\splitset$ of size $\leq m$ \label{line:combs} \\
%\tcp*[l]{\Cref{sec:methods:optimizations}}\label{line:combs}
% SortedCombs $\gets$ \Sortedcombs according to {\color{blue} \Cref{sec:methods:prioritization}}\label{line:sort} \\
$\mathit{CombsSorted}\gets \SortCombs(\mathit{Combs},k)$ \tcp*[l]{\Cref{sec:methods:prioritization}}\label{line:sort}
\For {$\mathit{comb}\in\mathit{CombsSorted}$} {
\label{line:for_loop}
$P\gets\FindPredicates(\mathit{comb},Q,\Aagg,\Agb,\claim)$ \tcp*[l]{\Cref{sec:methods:find_preds}}\label{line:find_preds}
$M \gets\ComputeNaturalnessMeasures(P)$ \label{line:compute_nat}

%\tcp*[l]{\Cref{sec:methods:optimizations}}\label{line:compute_nat}
$\mathsf{Print}(M)$\label{line:print}
}
\caption{Guided search for claim endorsement}
\label{alg:guided}
\end{algorithm}

% \ag{Shunit - please update the text to follow the pseudo code.}
% We next describe how to find predicates based on a given attribute combination (\Cref{line:find_preds}) in \Cref{sec:methods:find_preds} and how to prioritize the attribute combinations (\Cref{line:sort}) in \Cref{sec:methods:prioritization}. Generating all attribute combinations (\Cref{line:combs}) and computing the naturalness measures (\Cref{line:compute_nat}), along with some optimizations for them, are covered in \Cref{sec:methods:optimizations}.

\subsection{Finding Predicates for Given Attributes}
\label{sec:methods:find_preds}
We first describe \Cref{line:find_preds} in \Cref{alg:guided}: generating all refinements of a specific attribute combination.
%In the next section, we discuss prioritization strategies. 
Given an attribute combination $(A_1,...A_l)$, we wish to retrieve all predicates $\pred$ to endorse the claim, that is, $\pred=\bigwedge_{i=1}^{l}(A_i=v_i)$ such that $Q_\pred(D)\models \claim$.
A naive way is to run the following 
%\red{\emph{predicate-level}} \benny{I have no clue on what this addition contributes to the reader. What does it mean, actually?}\shunit{The purpose was to define what it means to search at the predicate level, iterating over value assignments (as opposed to iterating over attribute combinations). We refer to it from the experiments section when we discuss that OREO searches per-predicate and it's less efficient + from section~\ref{sec:ablation}.} 
%\benny{Not a legitimate answer. You explain things that the reader cannot know when she reads it, like the internal meaning you have in mind, and that you are using a term for later reference. You can say "that we refer to as \emph{predicate level} since..."}
query, that we refer to as \emph{predicate level}, since it is run 
for each combination of attribute values $v_1,...,v_l$:
\smallskip
\hrule\smallskip
\noindent\textsf{SELECT} $\Agb, \alpha(\Aagg)$
    \textsf{ FROM $D$}\\
    \textsf{ WHERE } $\Agb$ \textsf{IN (} $g_1,g_2$ \textsf{) AND} $A_1=v_1$ \textsf{AND ... AND} $A_l=v_l$ \\
    \textsf{GROUP BY} $\Agb$  
\hrule
\smallskip
\noindent 
We then verify that $Q_\pred(D) \models \claim$, that is, $\alpha(g_1) > \alpha(g2)$.
\eat{
\begin{figure}[t]
\centering
\begin{minipage}[b]{1.0\linewidth}
\begin{tcolorbox}[colback=white]
\vspace{-2mm}
% \begin{center}
\small
    \begin{tabular}{l}
    \textsf{(1) SELECT} $\Agb, \alpha(\Aagg)$\\
    \textsf{(2) FROM D}\\
    \textsf{(3) WHERE } $\Agb$ \textsf{IN (} $g_1,g_2$ \textsf{) AND} $A_1=v_1$ \textsf{AND ... AND} $A_l=v_l$ \\
    \textsf{(4) GROUP BY} $\Agb$    
    \end{tabular}
% \end{center}
\vspace{-2mm}
\end{tcolorbox}
\end{minipage}%%
\vspace{-4mm}
\caption{Naive predicate level query a given predicate $\pred=\bigwedge_{i=1}^n \left(A_i=v_i\right)$.}
\label{fig:naive_pred_level_query}
\end{figure}
}
Nevertheless, that would require the execution of a large number of queries---one for each $B\in P(\splitset)$ up to size $m$, for each combination of values of in the database. 
% In what follows, assume that $\splitset = Att(D)\setminus \{\Agb, \Aagg\}$ and $m=q-2$, meaning that all attributes can be used in the refinement except that aggregate and group-by attributes, and the predicates can be up to size $q-2$.
%In the worst case, when $\splitset = A$ and $m = n =|A|$, this totals in 
% Then the naive predicate-level approach would total in
% $\bigo(2^q \cdot q \cdot p)$ queries,
%$$\sum_{i=1}^{n}\sum_{\substack{B\in P(S) \\  \wedge|B|=i}}\sum_{\mathbf{v}\in \Pi_B(D)}1$$
%totaling in $\bigo(2^n \cdot n \cdot p)$, where $n = |A|$
% , $m$ is the number of allowed predicate conjunctions, 
% where $p = \prod_{A_i \in \splitset} |\dom(A_i)|$\footnote{In some cases, it is possible to reduce the number of queries, using a pre-processing step to determine the existing value combinations in the database and limiting the predicates by that information. However, in the worst case, all value combinations exist in the database.}. 
%
Instead, we execute one query for each attribute combination. 
%The query is run for each attribute combination, which results in fewer query runs. 
% totaling in $\bigo(2^q \cdot q)$. 
% $$\sum_{i=1}^{n}\sum_{\substack{B\in P(S) \\  \wedge|B|=i}}1$$
%For each attribute combination, 
The query returns all value combinations that define refinements that endorse the claim. %\shunit{"proper" sounds like something that should be defined mathematically, but maybe here it just means "refinements that endorse the claim"? In that case, maybe "claim-endorsing" refinements? If we do change the phrase, there is one more instance of it to change (in 3.2.2).}

The query depends on the choice of aggregate function; we provide an example for the query for $\alpha=\AVG$: %in the case of average: % (\Cref{fig:mean_gb_query}). 
% and median (\Cref{fig:median_gb_query}).

\hrule\smallskip\noindent
%\parbox{\textwidth}{
\textsf{%
SELECT $A_1, ..., A_\ell$\\
\-\quad AVG(CASE WHEN $\Agb=\group_1$ THEN $\Aagg$ END) as m1,\\
\-\quad AVG(CASE WHEN $\Agb=\group_2$ THEN $\Aagg$ END) as m2,\\
FROM D\\
GROUP BY $A_1, ..., A_\ell$\\
HAVING AVG(CASE WHEN $\Agb=\group_1$ THEN $\Aagg$ END)<\\
\-\quad\quad AVG(CASE WHEN $\Agb=\group_2$ THEN $\Aagg$ END)\\ 
\-\quad AND COUNT(CASE WHEN $\Agb=\group_1$ THEN $\Aagg$ END)> $M$ \\
\-\quad AND COUNT(CASE WHEN $\Agb=\group_2$ THEN $\Aagg$ END)> $M$}
%}
\hrule\smallskip

%The structure of these queries is similar. Both perform 
%The query performs a group-by for each attribute combination (Line (5)). 
For each group, we define the two subgroups $\group_1$ and $\group_2$ using the CASE WHEN statements. We found that the use of CASE WHEN yields faster executions than a simple join between a view for each population (see \Cref{sec:case_when} for an experimental evaluation). The HAVING statement verifies that $\alpha(\group_1)>\alpha(\group_2$). 
We also incorporate (in HAVING) the user-defined threshold on group sizes so that the refinements define sets of size at least $M$.

%\red{TODO add example of runtimes for the single atom case on ACS.} For example, on the ACS dataset with $n=1$ the predicate-level solution takes \red{X} hours to run on \red{Y} attributes, while the attribute level solution takes only \red{Z} minutes.

%\begin{example}\em
On Stack Overflow with
$m=1$ and $\AVG$, predicate-level search takes 10x longer than the attribute-level search (48.5s  vs.~4.5s). Predicate-level search uses 842 queries versus only 47 for attribute-level search. See additional comparisons in \Cref{sec:ablation}.
%\qed
%\end{example}

%\subsubsection*{Additional implementation choices.}
As an optimization, we apply pruning based on the group sizes. We pre-compute the maximal group size in a group-by query according to each attribute combination. If that size is at most $M$, we remove the attribute combination from the generated set, since the query above returns an empty result due to the violation of the HAVING statement. 
Additionally, we preprocess the dataset to count the distinct values of each attribute. We prune attributes with a single distinct value from the generated combinations, since a query refinement using this attribute will have no effect, as all tuples have the same value in this attribute.
%grouping by them is like not grouping at all.
%\benny{Not clear. You're saying that the group is of size one, so no grouping is required. Then why is it bad? You just found a way to avoid grouping. I don't see what reason we have to avoid a combination just because its computation is easier. What am I missing?}\shunit{If all tuples belong to the same group then there is no hope to find a refinement that endorses the claim based on this attribute. I rephrased. Clearer now?}
Lastly, we extend the query above to incorporate the computation of some of the naturalness measures, particularly $\coverage$ and $\statsig$. This is done by adding aggregates to the SELECT clause: standard deviation, count, and counts of values above/below the median.

\subsection{Prioritization of Attribute Combinations}
\label{sec:methods:prioritization}

%\subsubsection{Classification of Naturalness Measures}
Next, we propose ways to prioritize attribute combinations in
\Cref{line:sort}.
%of \Cref{alg:guided}.

% \subsubsection{Precomputation.}
\subsubsection{Prioritization per measure.}
\label{sec:methods:precomp_details}

We differentiate between two types of measures.
% \textbf{Attribute-level Naturalness Measures.} 
\emph{Attribute-level measures}
are naturalness measures that are based only on the connection between the attributes $A_1,...,A_l$ defining the refinement and the aggregate attribute $\Aagg$. In our case, these are $\mi$ and $\anova$. For these, the full computation can be done in advance, before knowing the exact query $Q$, the claim $\claim$, or the groups $g_1$ and $g_2$.
The computation may be costly, but it is done offline before the analyst issues the query, and can be stored and read when needed.
%
% \textbf{View-level Naturalness Measures.} 
In contrast, the \emph{predicate-level measures}
are naturalness measures that are based on the specific predicates that define the refinement, and possibly on the results $\alpha(g_1)$ and $\alpha(g_2)$. Therefore, they cannot be precomputed accurately. In this work, these are the measures $\coverage$, $\statsig$, and $\embsim$.
We will nevertheless propose heuristic methods to estimate these measures based on attribute-level information.

% Next, we describe the heuristics and pre-computations for each naturalness measure. 
%Next, we describe our prioritization strategy (accurate computation or heuristic) for each of the naturalness measures defined in~\Cref{sec:naturalness_metrics}.
% \ag{Complexity of each strategy?}

% \textbf{ANOVA and MI: exact pre-computations.} 
\exptitle{$\anova$ and $\mi$.}
For these, the prioritization is straightforward: we calculate in advance the exact values for each attribute combination, and order by decreasing value.
%In addition to guiding the search, this strategy also reduces the runtime, since computing these scores for all attribute combinations can be costly. %For each attribute combination, ANOVA can be calculated using a group by operation and another pass on the data to compute the square difference of each point from the total and group averages. 
% totaling in $\bigo (2^q |D|^2 log(|D|))$. \shunit{for MI?}
%\red{Attribute combinations are prioritized by decreasing MI/ANOVA scores.} 

% \textbf{Embedding similarity: heuristic.}
% comment R3-W2
\exptitle{$\embsim$.}
Here, we prioritize combinations by a simplified version of \Cref{eq:embsim}. Define $\textit{EmbSimSimple}(Q_\pred, D)$ to be \\
% \begin{equation}
% \label{eq:embsim_simple}
% EmbSim(A_i, T) = CosSim(E(Trans(A_i)), E(Trans(T)))
% \end{equation}
%\begin{equation}
%\label{eq:embsim_simple}
%\begin{aligned}
%  & \textit{EmbSimSimple}(Q_\pred, D) =\\
   $\textit{CosSim} \left( E \left( \bigoplus_{(A_i=v_i) \in \pred} T \left(A_i \right)  \right), E \left( T \left( \Aagg \right) \right) \right)$. This version compares only the attributes that define the predicate (without the values) against the representations of the aggregate attributes. \revc{Intuitively, the first string is a part of the full string defining the predicate, so we expect $\textit{EmbSimSimple}$ to be close to $\embsim$.}
%\end{aligned}
%\end{equation}
%The complexity of inference for a transformers embedding model as the one we used is quadratic in the length of the input text~\cite{devlin2018bert}. 
% Denote by $w$ the maximal length of an attribute name in the database, then the complexity of computing this heuristic is $\bigo(2^q (q\cdot w)^2)$ in the case where $m=q-2$.
%\shunit{maybe we need to say something about the quality of the approximation, like how far it is, on average, from the cosine similarity with the value. Or maybe it's enough to show in the main quality experiments.}

% \textbf{Coverage: heuristic.} 
\exptitle{$\coverage$.}
This one prioritizes by (the reverse of) the number of values with over $K$ occurrences for some $K$ (100 in our experiments). \revc{We expect attribute combinations that define many large groups to yield large groups where the claim holds.}
%The attribute combinations are ranked descending according to that number. 
% The complexity of this method is $\bigo(2^q\cdot |D| log (|D|))$ as it requires a group-by-count query for each attribute combination.

% \textbf{Statistical significance: heuristic.}
\exptitle{$\statsig$.}
%Since the statistical significance score cannot be computed in advance, we employ a heuristic method to find the predicates where the claim would hold most strongly, i.e., where $\alpha(g_1)-\alpha(g_2)$ is largest.
\revc{To find predicates with a high $\statsig$ score, we wish to quickly identify the attribute combinations that cause the most polarization between the two groups, i.e., where $\alpha(g_1)-\alpha(g_2)$ is largest.}
As a heuristic precomputation, we train two linear-regression models, $M_1$ and $M_2$, on  $g_1$ and $g_2$, respectively, to predict the (numeric) value of the attribute $\Aagg$. 
%In linear regression models, the target is modeled as a linear combination of the inputs. 
For the split attributes $\splitset = \{A_1,...A_q\}$, 
each model has the form 
$M(A_1,...A_q) = \sum_{i=1}^{q}w_i A_i + c$.
%The weights $w_1,...,w_q$ are tuned to minimize the sum of squares error $\sum_{i=1}^n (M(A_1^i,...,A_q^i)-\Aagg^i)^2$. 
The weight $w_i$ is indicative of 
$A_i$'s contribution to predicting $\Aagg$. 
Hence, for each $A_i$, we define 
$\mathit{RegScore}(A_i) = w_i^1-w_i^2$
where $w_i^j$ is the weight for feature $i$ in regression model $M_j$.
%Attributes with a high $\mathit{RegScore}$ grow more with $\Aagg$ in $g_1$ and grow less, or are reduced as $\Aagg$ grows in $g_2$. 
The score for a combination of attributes is given by
$\mathit{RegScore}(A_1,...,A_l) = \sum_{i=1}^l \mathit{RegScore}(A_i)$.
We rank combinations by decreasing $\mathit{RegScore}$.

\subsubsection{Prioritization for combining all measures}\label{sec:prioritization}
In many cases, a single naturalness measure does not capture all the intuitively natural refinements. Furthermore, the analyst may not know in advance which naturalness measure is best for her query. For these reasons, we
next describe prioritization methods that 
combine single-measure rankings into a single prioritization strategy.
%\benny{Why do we do so? What is the goal? What is the point? As far as the reader is concerned, she always has exactly one naturalness measure in mind, so why caring about the others?} 
For each naturalness measure,
%\benny{What is precomputable here? is it the attribute-level vs predicate-level?} \shunit{looks like a relic. removed it.}
we either compute it (or the appropriate heuristic estimate) in advance, or use a fast computation as a preprocessing step, as detailed in \Cref{sec:methods:precomp_details}. We rank the list of attribute combinations according to each computed score. Then, we combine the rankings using one of the following three variations. 

\paratitle{Serial:} for each individual measure, solve claim endorsement independently and stop when $k$ refinements have been found. If the deadline of \Cref{alg:guided} has not passed, the search continues over the remaining attribute combinations, in their original order.

\paratitle{Merged:} merge the ranked lists in an interleaving fashion. More precisely, we take the first element from each ranking, then the second element from each ranking, and so on. Then we perform the search according to the single merged ranking. 

\paratitle{Sampling.}
% \label{sec:methods:sampling}
%\brit{we expermently find that not to perform well in practice. Maybe we can just move this to the experiments as a baseline method? since the best method to-k-merged actually doesn't use it all}
A general prioritization heuristic can be done by sampling a portion of the database and finding all refinements that endorse the claim via brute-force search. Note that the time for executing each group-by query on the sample is faster than the full database, so this search is generally faster. (It may also be lossy, since some combinations may be missing.) 
%For each combination, we compute its priority by calculating 
For each refinement, we compute all measures, relying on either the precomputation for the attribute-level measures ($\anova$ and $\mi$) or the sample-based results for the predicate-level measures. 
%We also compute the average of all measures. 
%\benny{And what do we do with this average??}
As a heuristic, we define the priority of a combination as the maximum of the average naturalness, over all refinements of the combination.
%Next, we run another claim endorsement search, this time on the full database, where the attribute combinations are prioritized by the results from the search on the sample. We rank the attribute combinations according to the maximal average score given to a refinement returned from each combination. Other rankings may also be considered, e.g., by the number of refinements returned from each attribute combination, or by the maximal score in a single naturalness measure.
%The reason for the second search (on the full database) is that the sample-based results may not be accurate in terms of the coverage and statistical significance, and they may not even include the full set of \red{claim endorsing} refinements. % where the claim holds. 
%Therefore, we run another cherrypicking search on the full database, where the attribute combinations are prioritized by the results from the search on the sample. We rank the attribute combinations according to the maximal average score given to a predicate returned from each combination.

\revc{%comment R3-W1
The advantage of sampling is its independence from the choice of a specific measure. For example, for a custom naturalness measure, if there is a prioritization method for the new measure (like the regression weight for $\statsig$, or the simplified embedding similarity for $\embsim$), it can be combined with the other rankings. But even without a prioritization method, sampling can still be used for the custom measure, since the process of sampling and searching the sample for predicates for each attribute combination (Section~\ref{sec:methods:find_preds}) is independent from the choice of measures.
}

\section{Experimental Evaluation}
\label{sec:experiments}
We present experiments to evaluate our framework. We aim to address two main research questions. \textbf{($Q_1$)} What is the effectiveness of our framework for claim endorsement? Specifically, what insights are revealed by our framework, and how do they differ from prior work intended for similar (yet different) goals? \textbf{($Q_2$)} To what extent do the optimizations contribute to the performance?

%$Q2$: 
%How does each phase of \algoName contributes to its ability to find an explanation summary that satisfies our optimization goal and constraints? 
\paragraph*{Summary}
Before we delve into the details of the experiments, we briefly summarize our findings. 
We observed that our framework generates intuitive and understandable refinements that endorse %support
the claim across various settings. Our approach can generate more convincing refinements than prior work~\cite{lin2022oreo, salimi2018hypdb}. 
We found \oursolution to be the best-performing prioritization.
%, reaching a score recall of $95\%$ in an average time of \ag{X} compared to \ag{Y} for the second-best variant. 
%\ag{Two more sentences}
\revb{Over all naturalness measures and  datasets, oursolutionk was on average 13.4 (and up to 78.3) times faster than the
random-order baseline, and on average 16.2 (up to 82.5) times faster than HypDB
%in terms of the time until high-quality results,
%than the original order and random order baselines (which serve as alternative prioritization methods). 
(as alternative prioritization).}
Furthermore, \oursolution method is almost indifferent to the number of tuples and %to the number of 
attributes in the database, and to the number of top results ($k$) and maximal number of atoms ($m$).
% While our approach quantifies the naturalness of a view to aid users in discovering views supporting their claim, OREO does not address this ranking aspect, which could result in users having to navigate through an overwhelming number of candidate views. 
% While covariate attributes found by HypDB are useful in explaining the gap in the outcome between two groups, such attributes do not necessarily yield natural views that support the claim. 

\subsection{Experimental Setup}\label{sec:exp_setup}

Our code is in Python and publicly available online.\footnote{\url{https://github.com/shunita/claimendorse}}
%\shunit{will change soon to a better name}} 
We used the Pandas library for accessing raw dataset files and SQLAlchemy for accessing a PostgreSQL database. We also used 
%scikit-learn and scipy for statistical models, and
sentence transformers~\cite{sentence-bert2019} \footnote{\url{https://huggingface.co/sentence-transformers/all-MiniLM-L6-v2}} for the word-embedding model. The experiments were executed on a PC with a
$2.50$GHz CPU, and $512$GB memory.

\begin{table}
	\centering
%	\captionsetup{justification=centering}	
	\small
	\caption{Examined datasets.}
        \vspace{-1em}
	\label{tab:datasets}
			% \vspace{-4mm}
	\begin{tabular}{cccc}
 % {|p{15mm}|p{7mm}|p{7mm}|p{18mm}|p{18mm}|}
		\toprule
	    \textbf{Dataset} & \textbf{\#Tuples}& \textbf{\#Atts}& \textbf{Max vals per att} \\
		\midrule
        ACS
        &1,420,652 & 288 & 723\\
        Stack Overflow & 73,268 & 78 & 180\\
        Flights & 5,819,079 & 43 & 6952\\
		\bottomrule
	\end{tabular}
 \vspace{-1.5em}
% 	\vspace{-1mm}
\end{table}

\subsubsection*{Datasets, queries, and claims}
Similarly to previous work on similar topics~\cite{lin2021detecting, youngmann2023explaining, salimi2018},
we examined multiple common datasets
and devised queries and claims that are inspired by real-world resources 
and statistical studies~\cite{vandenbroucke2018married,borrego2018pursuing}.

\exptitle{ACS.} We accessed the American Census Survey (ACS) data through the Folktables library~\cite{folktables2021}. We use the 2018 data that includes the seven largest US states: CA, TX, FL, NY, PA, IL, OH. This resulted in a dataset of 1,420,652 rows and 288 columns. Attribute names are encoded strings and attribute values are encoded numbers or strings. We mapped them to human readable strings using the ACS PUMS data dictionary.\footnote{\url{https://www.census.gov/programs-surveys/acs/microdata/documentation.html}} We also added two attributes based on groupings of the OCCP and NAICSP fields---the readable strings for these fields begin with a 3-letter encoding of the occupation field, which we extracted to a new attribute. The query is the average total income grouped by gender, and we search for refinements where women's income is higher than men's.

%\footnote{We recognize the complexity of sex and gender, and we know that it is not a binary concept. However, due to the limitations of the ACS dataset, where the gender field has only two possible values, we are limited to this binary representation. Our work is easily applicable to future versions of the data, should they include additional possible values for gender.}

   \exptitle{Stack Overflow.} We used the Stack Overflow developer survey from 2022.\footnote{\url{https://survey.stackoverflow.co/2022}} The dataset contains 73,268 responses and 78 attributes, covering demographic and professional information, as well as the yearly compensation of the participant. In some attributes, some rows contain multiple values, separated by a delimiter; in these, we kept only the first value. We focused on the median yearly compensation, and compared bachelor's degree graduates to master's degree graduates. Overall, the median salary of master's graduates was slightly higher, so we searched for refinements where the median salary of bachelor's graduates is higher.
    
    \exptitle{Flight Delays.} The flight delays dataset, available on kaggle,\footnote{\url{https://www.kaggle.com/datasets/usdot/flight-delays}} contains 5,819,079 flights from 2015, over various carriers and airports. We joined the airports and flights relations to create a 43-attribute relation, which includes scheduled and actual departure and arrival times, as well as carrier, airplane number, etc. We compared two weekdays, Monday and Saturday, and counted the flights with departure delays exceeding 10 minutes. On Mondays, there were more delays than on Saturdays (192,219 versus 134,681).
    %two carriers, American Airlines (AA) and United Airlines (UA). Overall, UA had a higher average arrival delay (5.43 hours versus 3.45 hours for AA). 
    We aim to endorse the claim of the reverse direction.
%\end{itemize}
%\textbf{Chicago Crime.}\\
  % \\
%\textbf{Car Accidents?}\\

\subsubsection*{Data preprocessing}
For numeric attributes other than the target attribute $\Aagg$, we perform binning similarly to prior work~\cite{deutch2022fedex,PradhanZGS22} %\shunit{in FEDEX they used equal frequency binning, not equi-width, does that matter?}. 
The interval size is chosen according to the order of magnitude of the range of values in the attribute. The start and end points of the bins are rounded to create natural-sounding ranges, (e.g., ``10-20 years of coding'' is more natural than ``8-19 years of coding'').
% \red{TODO cite some works that also used equi-width binning}

% \paratitle{Claims}
% The examined claims were inspired by real-world resources
% % , ranging from Stack Overflow user reports \cite{} to 
% and statistical studies~\cite{vandenbroucke2018married,borrego2018pursuing}. They are used to generate insights about job opportunities and technology trends or expose problems of contemporary society. \brit{rewrite the above according to the examined claims and add cites}. Similar approaches were taken in \cite{lin2021detecting, youngmann2023explaining, salimi2018}.

\subsubsection*{Algorithm variants}
We examined the following prioritization methods for attribute combinations   (\Cref{line:sort} in \Cref{alg:guided}).
%, as described in \Cref{sec:methods:prioritization}.

    \exptitle{\oursolutionk:} Interleaving merge combination of naturalness measure heuristics and pre-computations.
    % , as described in Section~\ref{sec:methods:prioritization}.
    
    \exptitle{\oursolutionserial:} Serial combination of naturalness measure heuristics and pre-computations.
    % , as described in Section~\ref{sec:methods:prioritization}. 
    We iterate over the metrics from the most accurately precomputed to the least accurate heuristic: $\anova$, $\mi, \embsim, \statsig,$ and $\coverage$.
    
   \exptitle{\oursolutionsample:} The search is guided by a sample size of $x$\% of the number of tuples in the database. We experimented with sample sizes of 1\%, 5\% and 10\%. 
    % As described in Section~\ref{sec:methods:prioritization}, 
    We first perform a search on the sample, and use the results to prioritize the attribute combinations in the search over the full database. In the experiments below, the presented results for \oursolutionsample methods represent an average of 3 runs.

\vspace{-0.5em}
\subsubsection*{Baselines}
As baselines, we used approaches from the literature and some naive ones. Since the claim-endorsing problem is new, there are no existing solutions for this exact problem. However, we compare our methods to previous approaches to reminiscent problems through case studies (\Cref{sec:exp:case_studies}).

% The naive baselines are described next.
%\exptitle{Original order:} This baseline iterates over attribute combinations according to their original order in the database, instead of prioritizing by their (potential) naturalness as in our proposed algorithm.

\exptitle{Random order:} This baseline iterates over the attribute combinations in a random order. In the experiments below, the presented results for this baseline represent an average of 3 runs.

\exptitle{Single naturalness measure:} This baseline prioritizes the attribute combinations according to a single naturalness measure. Each measure has a different prioritization method (detailed in \Cref{sec:methods:precomp_details}).

\exptitle{OREO~\cite{lin2021detecting,lin2022oreo}:} This system is designed to identify cherry-picked generalization statements.  
The approach includes a scoring function that considers all supporting subpopulations, weighted by size, where a higher score indicates stronger support. 
% They also explore identifying counterarguments with an algorithm that finds {\em all} maximal query refinements (in terms of generalization) that oppose the claim.
Additionally, OREO employs an algorithm for discovering counter-examples to the examined statement in the form of query refinements.
As opposed to our approach, OREO provides \emph{all} \revb{of the most general query refinements (as described in \Cref{sec:gen_filter})}. %, signifying that by removing a predicate, the refinement no longer serves as a counterargument to the statement. 
This is compounded by the fact that OREO uses a predicate-level search instead of an attribute-level search (see \Cref{sec:methods:find_preds}), limiting scalability to high-dimensional datasets (up to 12 attributes in the examined datasets).
%For example, if the predicate $(A_i {=} v_i)$ was returned, all predicates of the form $(A_i {=} v_i) \wedge (A_j {=} v_j)$ will not be returned. 
Also, the refinements are \emph{not ranked}.

\revb{For our case studies and user study, we use this method to find all maximal refinements. In our quantitative evaluation below, we run an enhanced version of OREO with some differences: (1) Iterating over combinations of up to 2 attributes instead of all possible combinations (to handle large datasets); (2) Disabling generality filtering (\Cref{sec:gen_filter}) to retrieve a larger set of predicates; and (3) Anytime-style output - output the predicates as they are found instead of at the end of the run. 
We further give an unfair advantage to OREO by pre-selecting only the attributes responsible for top k predicates (found using other baselines). To avoid confusion, we name this version of OREO with the advantage as \OREOS.
}
% Closest to our work, \cite{lin2021detecting} introduces OREO, which assesses the level of support for a user claim using data. They define a scoring function that considers all supporting subpopulations, weighted by size, where a higher score indicates stronger support from the data. They also explore identifying counterarguments by developing an algorithm to find all maximal query refinements (in terms of generalization) that oppose the claim.
% However, OREO has two main limitations. Firstly, it does not rank the refinements (counterarguments). As shown in our experiments (Section \ref{sec:exp:case_studies}), the number of maximal counterarguments can be extensive, making it hard to find meaningful ones. Secondly, it uses a predicate-level search instead of an attribute-level search, limiting scalability to high-dimensional datasets (up to 12 attributes in the examined datasets).
% In contrast, our approach ranks refinements based on naturalness and can scale efficiently to high-dimensional datasets with hundreds of attributes using an efficient attribute-level search strategy.

\exptitle{HypDB \cite{SalimiGS18,salimi2018hypdb}:} This system detects bias in average-group-by SQL queries. Given a query, HypDB finds a set of \emph{covariates} (attributes with uneven distributions among the groups) that serve as an explanation for the query results, based on causal analysis. Each selected attribute is associated with a responsibility score. Although this approach does not directly offer query refinements supporting the user claim, it highlights attributes with uneven distributions among the groups defined by the grouping attribute. However, we demonstrate that attributes relevant to defining natural refinements that support the user's claim are not necessarily covariates. \revb{In our effectiveness evaluation, we use the responsibility score to prioritize the attribute combinations, where the score of an attribute combination is the sum of responsibility scores of each attribute.}

\subsubsection*{Evaluation metric}
We report the \textit{score recall} of each examined baseline. 
%\brit{what is the score recall?}. 
For a generated set $r$ of refinements, and naturalness measure $\nat$, let $\nat_i^{r}$ denote the top $i$'th $\nat$ score found in $r$. %\benny{What is $v$? Why is $t$ given if $r$ is known?} 
Define
$S_{r} \defeq \sum_{i=1}^{k}\nat_i^{r}$.
The \emph{score recall} of $r$ is given by:  $\infrac{S_{r}}{S_{\textsf{full}}}$ where $S_{\textsf{full}}$ is sum of top-$k$ scores of an exhaustive search.
%This score recall was computed by considering the full run of the algorithm and the obtained recall of each evaluated prioritization method at various time points during the search. We compute it for each proposed naturalness measure, and for the average naturalness.
%\brit{is this correct? what is the average score recall in Fig 3(a)}
%\red{discuss other possible recall variations and why they are not a good fit.}
While there are other metrics to evaluate retrieval results, they are not suitable for our task. Classic recall (based on exact matches), Normalized Discounted Cumulative Gain (NDCG)~\cite{jarvelin2002ndcg}, and Kendall's-$\tau$~\cite{kendall1938new}, are all based on the existence of an underlying ground truth ranking. However, this is not the case here, as there can be many results of similar quality. 

\subsubsection*{Default configuration}
We use a default of $m=2$ maximal number of atoms in each combination. 
%limit the number of attributes in each combination to $m \leq 2$, 
%resulting in predicates of the forms $(A_i{=}v_i)$ or $(A_i{=}v_i) \wedge (A_j{=}v_j)$. 
Even with this limitation, predicate search is computationally intensive. For example, in the ACS dataset, there are 120 split attributes, and 7140 2-attribute combinations, resulting in a total of 7260 pairwise attribute combinations. 
%With single attribute predicates, the search takes 2.1 minutes to go over the 120 split attributes; 
Processing all 7260 takes over 1.5 hours, thus creating an experimental setup where the differences between methods can be evaluated. %\shunit{For flights predicate examples I used single atom, because it was hard to find good examples from the 2-atom search, since most of the attributes are dependent. But for the run time experiments I used the 2-atom configuration, as for all the others. Where should we mention this?} 

%Throughout the experiments, 
As default we use $k = 100$. 
%We view such $k$ represents a proper balance between being effective to process by a user (i.e., manually examining $k$ refinements) and being large enough to allow for discovering high-quality refinements. 
In addition, we evaluate the performance of our algorithm with varying values of $k$ (see \Cref{sec:k_sensitivity}).

\subsection{Case Studies}
\label{sec:exp:case_studies}

\begin{table}[t]
    \centering
    \caption{Example refinements found through \oursolution.}
    \vspace{-0.5em}
    \label{tab:example_preds}
    \footnotesize
    \begin{tabular}{r|p{3.2cm}|ccc} \hline
        \textbf{Dataset} &\textbf{Predicate} & \textbf{Avg.~Nat} & \textbf{Rank} & \textbf{Time (s)} \\
        \hline
        \multirow{3}{1cm}{Stack Overflow} & 
    \makecell{\textsf{OpSysProfUse}=\textsf{Linux-based} \\ $\wedge$ \textsf{YearsCode}=\mbox{0-10}}& 0.44 & 32 & 2 \\ \cline{2-5}
        & \makecell{$(\textsf{Employment}=\textsf{Full-time})$ \\ $ \wedge (\textsf{Org.~size}=\textsf{``1K-5K''})$} &  0.10 & 59 ($\mathit{Cov.}$) & 13 \\
        % & \makecell{\textsf{Country}=\textsf{Germany} $\wedge$ \\ \textsf{Branch}=\textsf{``dev.~by profession''}} & 0.35 & 133 & 12 \\
        %& \red{\makecell{$\textsf{Country}=\textsf{Germany}$}} & 0.34 & 139 & 8 \\ 
        \hline
        \multirow{6}{1cm}{ACS}& \makecell{\textsf{GradeLevelAttending}=12 $\wedge$ \\ \textsf{Occupation}=\textsf{Customer Service}} & 0.40& 73 & 197\\ \cline{2-5}
        & \makecell{\textsf{Occupation}=\textsf{Cooks}
        $\wedge$ \\ \textsf{HoursPerWeek}=\textsf{10-20}} & 0.41 & 20 & 70\\ \cline{2-5}
        &\makecell{ \textsf{MaritalStatus}=\textsf{NeverMarried} $\wedge$ \\ \textsf{WhenLastWorked}=\textsf{>5y ago}} & 0.39 & 56 & 475\\
        \hline
        \multirow{2}{1cm}{Flights}& 
        % using single atom search:
        %$\textsf{DepartureTime}=\textsf{7:00-8:00}$ & 0.52 & 28 & 46 \\
        %\cline{2-5}
        %& $\textsf{OriginAirportCode}=\textsf{MIA}$ & 0.34 & 54 & 61 \\
        % using 2-atom search:
        \makecell{\textsf{Airline}=\textsf{Hawaiian Airlines Inc.}} & 0.36 & 1 & 36 \\
        \cline{2-5}
        & \makecell{\textsf{ScheduledDeparture}=\textsf{3:00-4:00}} & 0.30 & 2 & 40 \\
        %& $\textsf{DepartureDelay}=\textsf{0-100}$ &0.68 & 2 & 40 \\
        \hline
    \end{tabular}
    \vskip-1.5em
\end{table}

We use \oursolutionk prioritization, which achieved the best results among our methods in terms  (see \Cref{sec:main_quality_exp,sec:sensitivity} in the sequel).
We note that the choice of prioritization method affects the time it takes to find the best refinements, but if given unlimited time, all prioritization methods retrieve the same set of refinements; the prioritization only affects the order in which the attribute combinations are searched, but eventually, we iterate over all of them. 
In this section, we allow the search to finish and compare the final set of retrieved refinements to those achieved by existing solutions, since existing solutions have no time limit. Nevertheless, our approach retrieves convincing refinements that endorse the claim in reasonable time (\Cref{tab:example_preds}).
% \ag{This begs the question: why do we have an anytime algorithm if we need to run it without deadline to find good predicates? Can we answer that?}. \shunit{we find the good predicates first but we want to compare the set of all the found predicates to other solutions, since they don't have a time limit. I've added that the good predicate examples below are found in 12 seconds.}
%utilize the Merged Top K prioritization method, which has demonstrated the best performance, as discussed in Sections~\ref{sec:main_quality_exp}.
We compare \oursolution  with existing solutions on three example scenarios, to showcase the distinctions in our results. Specifically, we consider OREO and HypDB.

\subsubsection*{Stack Overflow}
We consider the claim ``the average salary of people with M.Sc.~is higher than that of people with B.Sc.''\footnote{We used Average since HypDB is suitable only for average aggregation.}
% Two of the top scoring refinements (and their average naturalness score) found by \oursolution are $\textsf{OpSysProfessional use}=\textsf{Linux-based} \wedge \textsf{YearsCode}=0-10$ (0.44, ranked 32) and $\textsf{Country}=\textsf{Germany} \wedge \textsf{MainBranch}=\textsf{"I am a developer by profession"}$ (0.35, ranked 133), both found in under 12 seconds of the search. \shunit{I'm not sure what to do about this - I wanted to show two different predicates but all the top 100 are using yearscode, so to find a predicate not using this attribute I had to go beyond 100.}
\oursolution discovered meaningful and comprehensible refinements that endorse the claim, as shown in \Cref{tab:example_preds}.
One of the top scoring refinements is $(\textsf{OpSysProfessional use}=\textsf{Linux-based}) \land (\textsf{YearsCode}=0-10)$ (Avg.~nat.=0.44, ranked 32). The rationale is that Hi-Tech workers who use Linux are typically those in technical positions, and for those with little experience, a master's degree may well contribute to a higher salary.
%As all 100 predicates found by \oursolution, in this case, contained the \textsf{YearsCode} attribute, and for the sake of the example, we increased $k$ to $150$ to increase diversity.In this case, \oursolution found the predicate $(\textsf{Country}=\textsf{Germany}) \wedge (\textsf{MainBranch}=\textsf{``Dev.~by profession''})$ (0.35, ranked 133). Both predicates were found in under 12 seconds. 
Ranking by different measures, we get additional results. With $\coverage$, a top refinement is $(\textsf{Employment}=\textsf{Full-time}) \wedge (\textsf{Org.~size}=\textsf{``1K-5K''})$ that covers over 3000 people. Both predicates were found in under 13 seconds.
%$\textsf{DevType}=\textsf{Data scientist or machine learning specialist}$ (0.22).

Due to the small size of this dataset, and the limitation of at most two predicate atoms, OREO returned a manageable set of 282 maximal predicates. Some are quite convincing, like $\textsf{DevType}=\textsf{``data scientist or machine learning specialist''}$. In this situation, it may be beneficial to combine the two approaches, by incorporating OREO's output in the refinements that we rank by scores of naturalness. 
Nevertheless, the results also show that being a maximal predicate may fall short of capturing our objective. \revb{Maximal refinements may be 
% comment R2-D3
reflecting a side effect of the survey design, and not necessarily a characteristic of the group, like $\textsf{Ethnicity}{=}\textsf{``I don't know''}$.} % (people who have selected not to fill out their ethnicity).
Other maximal predicates seem convincing at first but turn out to be weak. An example is $\textsf{Country}{=}\textsf{India}$, where the difference in salaries is tiny (\$51.7K vs.~\$52.5K). Furthermore, the set of maximal predicates may omit convincing refinements. For example, OREO did not return the above-mentioned $(\textsf{OpSysProfessional use}=\textsf{Linux-based}) \land (\textsf{YearsCode}=0-10)$, which is statistically significant and shows a considerable difference in salaries (\$120.1K vs. \$133.6K), simply because $(\textsf{OpSysProfessional use}{=}\textsf{Linux-based})$ satisfies the claim with a tiny difference (\$165.7K vs.~\$167.2K).

\begin{figure*}[t]
\centering
\subfloat[$\anova$]{
\label{fig:quality_main_anova_ACS}
  \includegraphics[width=27mm]{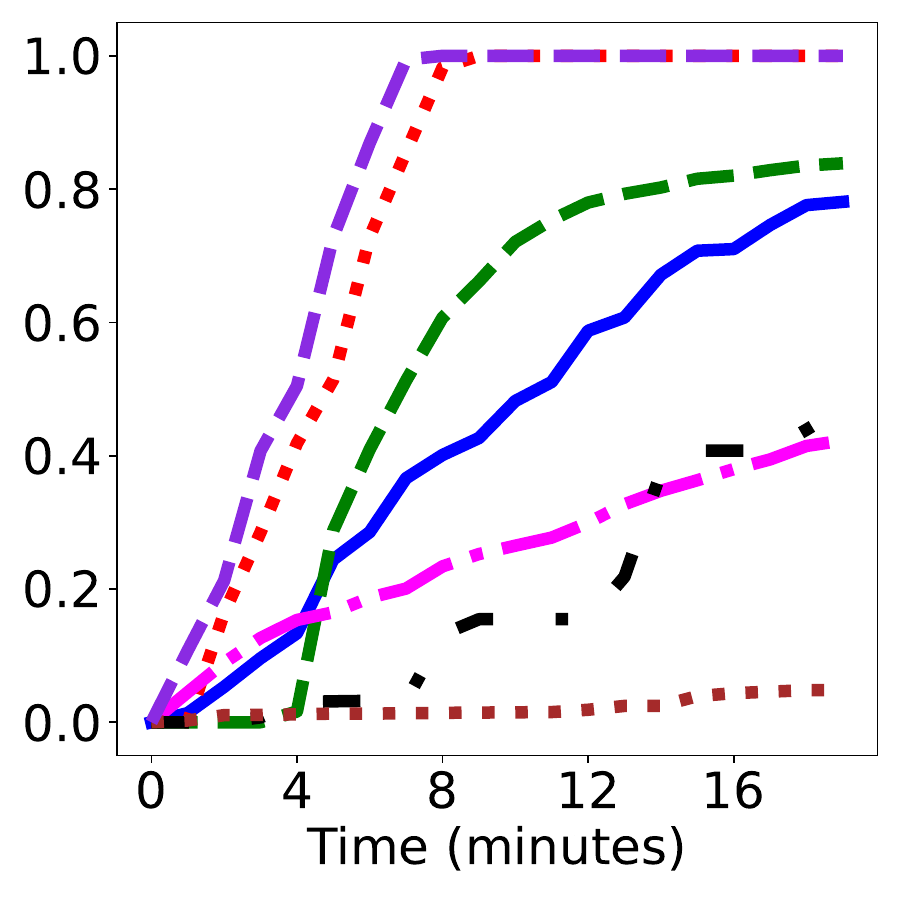}
}
\subfloat[$\mi$]{
\label{fig:quality_main_mi_ACS}
  \includegraphics[width=27mm]{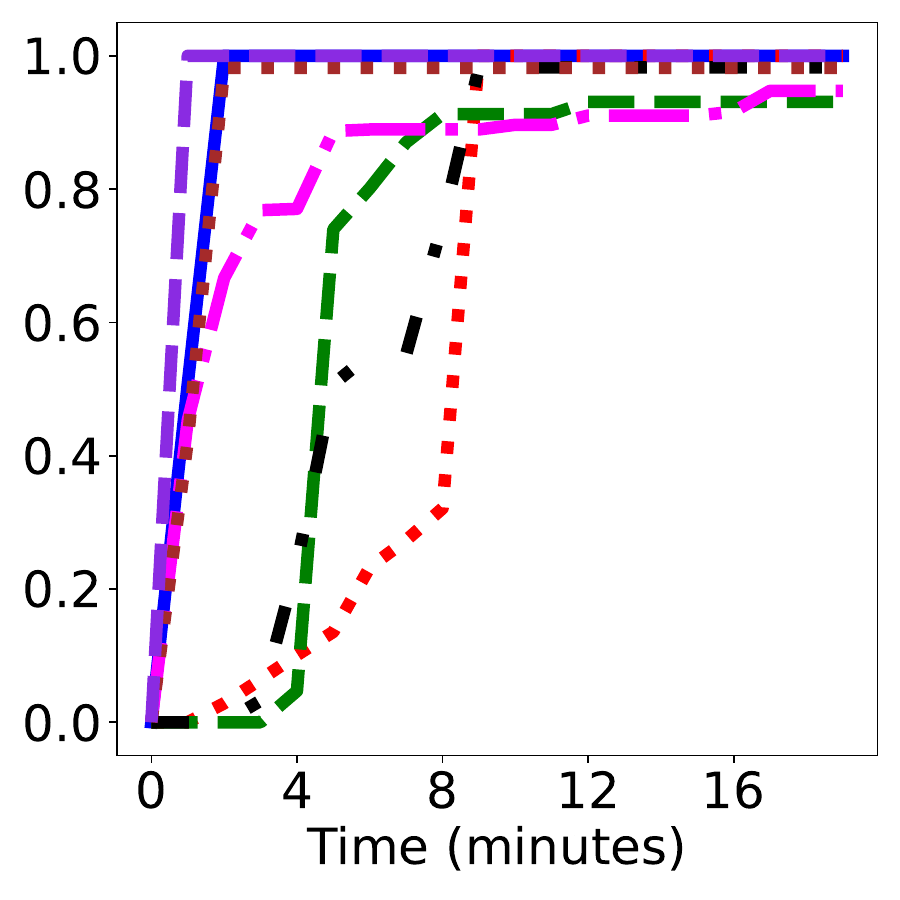}
}
% \hspace{0mm}
\subfloat[$\embsim$]{
\label{fig:quality_main_embsim_ACS}
  \includegraphics[width=27mm]{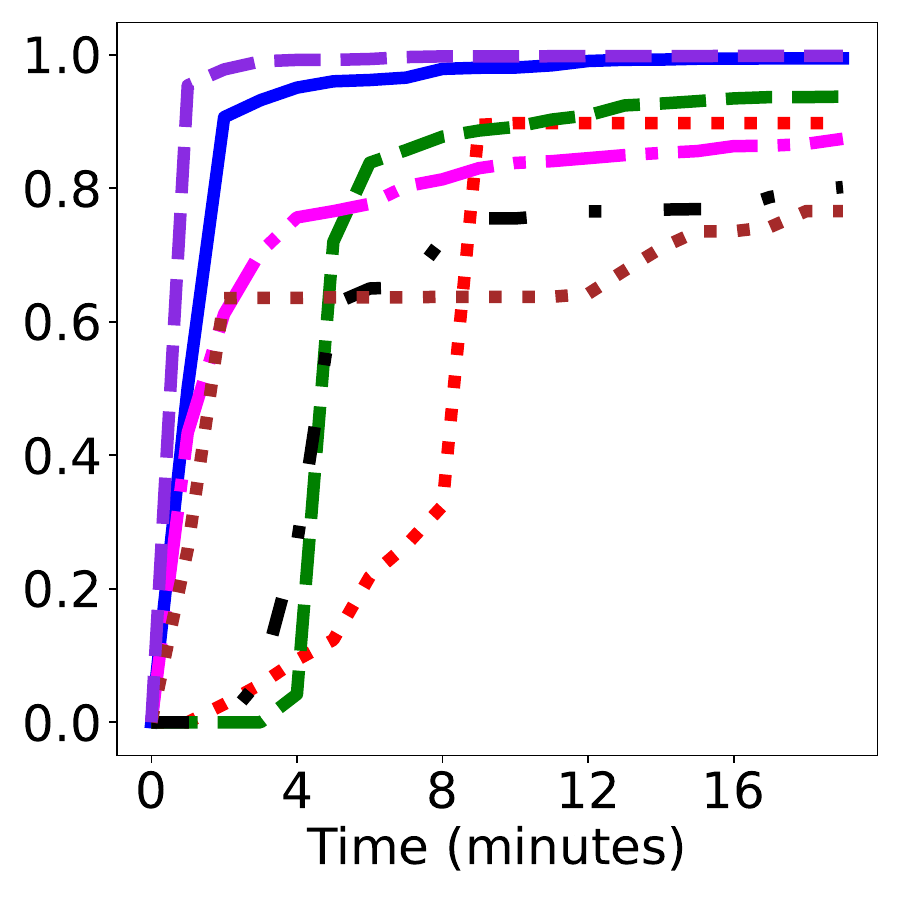}
}
\subfloat[$\coverage$]{
\label{fig:quality_main_coverage_ACS}
  \includegraphics[width=27mm]{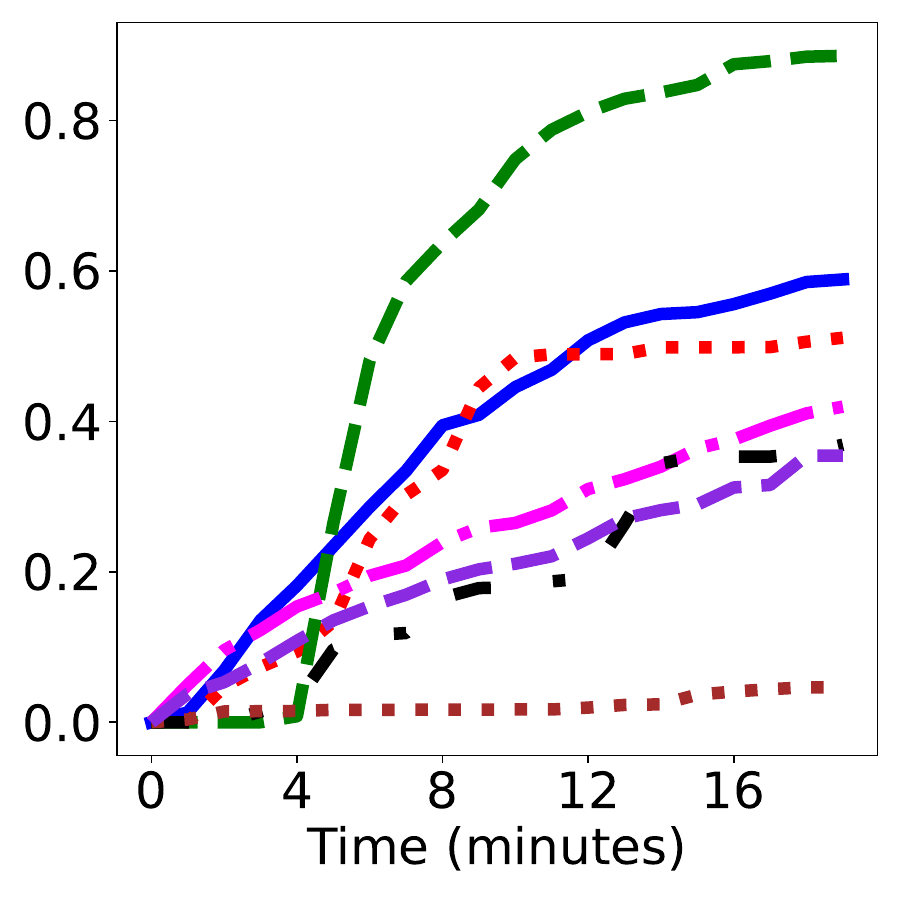}
}
\subfloat[$\statsig$]{
\label{fig:quality_main_statsig_ACS}
  \includegraphics[width=27mm]{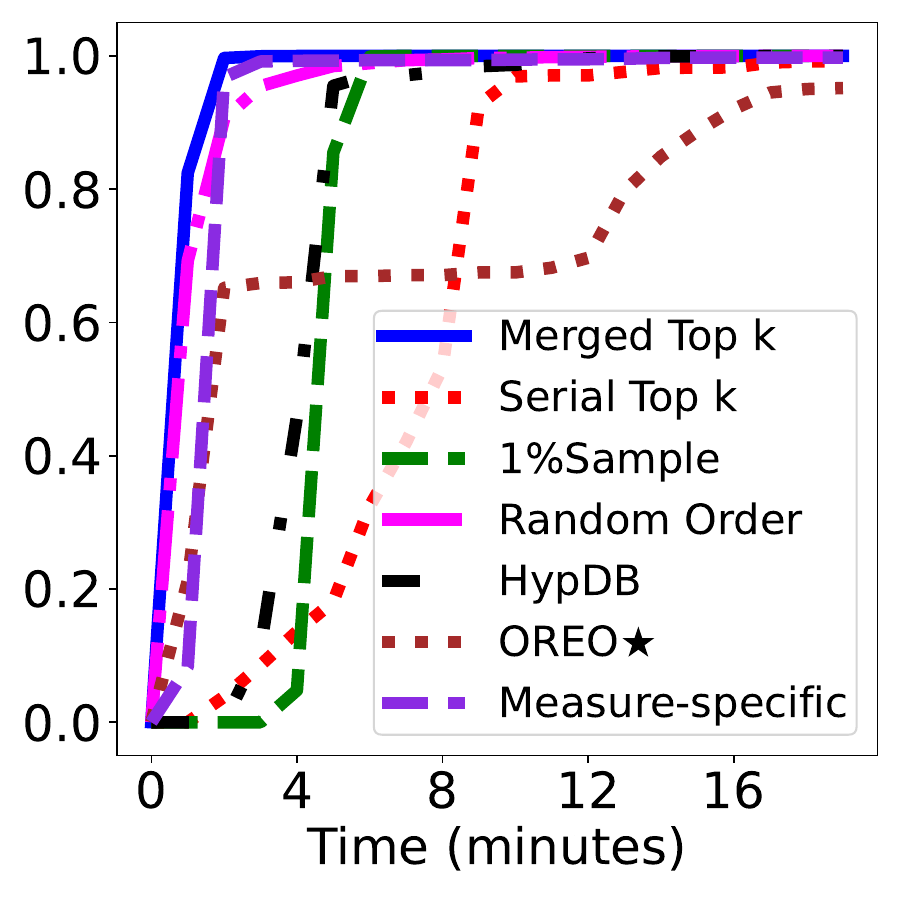}
}
\subfloat[Average of all]{
\label{fig:quality_main_avg_ACS}
  \includegraphics[width=27mm]{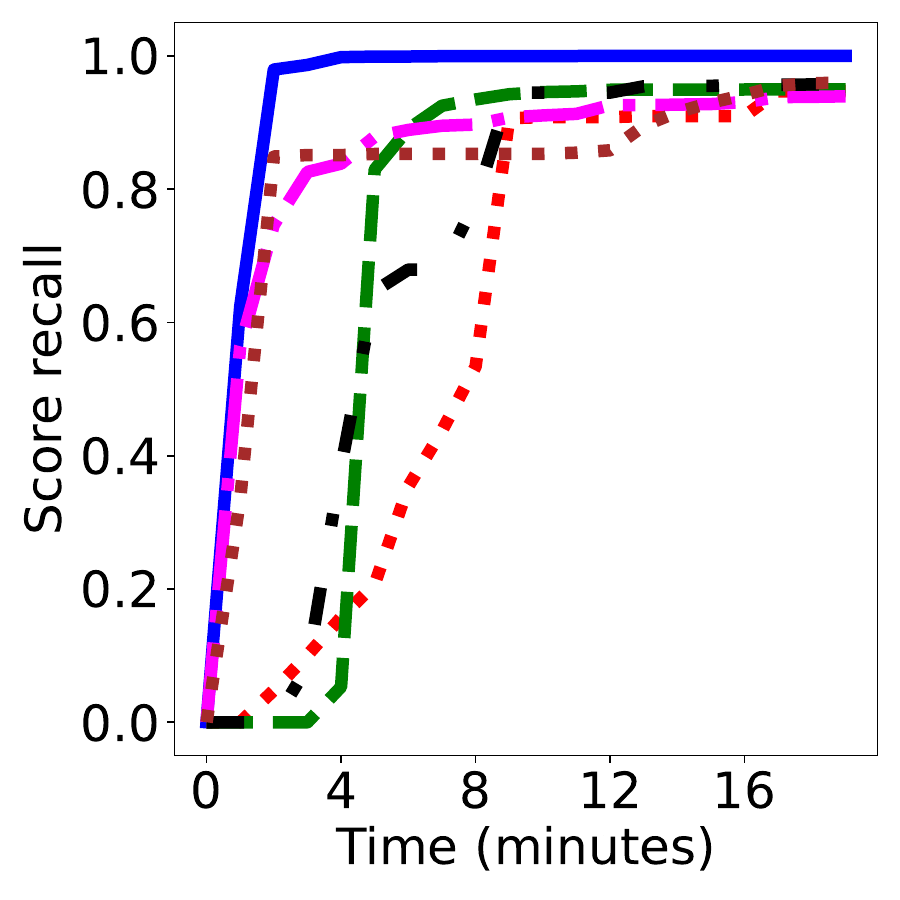}
}
% \vspace{-0.7em}
% % now sampling
% \subfloat[$\anova$]{
% \label{fig:quality_sampling_anova_ACS}
%   \includegraphics[width=27mm]{figures/sampling_ACS/ACS7_numeric_mean_2atoms_F_gt_M_sample_size_Normalized Anova F Stat_top100score_recall_over_time_multi.pdf}
% }
% \subfloat[$\mi$]{
% \label{fig:quality_sampling_mi_ACS}
%   \includegraphics[width=27mm]{figures/sampling_ACS/ACS7_numeric_mean_2atoms_F_gt_M_sample_size_Normalized_MI_top100score_recall_over_time_multi.pdf}
% }
% \subfloat[$\embsim$]{
% \label{fig:quality_sampling_embsim_ACS}
%   \includegraphics[width=27mm]{figures/sampling_ACS/ACS7_numeric_mean_2atoms_F_gt_M_sample_size_Cosine Similarity_top100score_recall_over_time_multi.pdf}
% }
% \subfloat[$\coverage$]{
% \label{fig:quality_sampling_coverage_ACS}
%   \includegraphics[width=27mm]{figures/sampling_ACS/ACS7_numeric_mean_2atoms_F_gt_M_sample_size_Coverage_top100score_recall_over_time_multi.pdf}
% }
% \subfloat[$\statsig$]{
% \label{fig:quality_sampling_statsig_ACS}
%   \includegraphics[width=27mm]{figures/sampling_ACS/ACS7_numeric_mean_2atoms_F_gt_M_sample_size_Inverted pvalue_top100score_recall_over_time_multi.pdf}
% }
% \subfloat[Average of all]{
% \label{fig:quality_sampling_avg_ACS}
%   \includegraphics[width=27mm]{figures/sampling_ACS/ACS7_numeric_mean_2atoms_F_gt_M_sample_size_Metrics Average_top100score_recall_over_time_multi.pdf}
% }
\vspace{-1em}
\caption{\revb{Top 100 score recall for each naturalness measure over time for our methods and external baselines for ACS. %(a-f) and specifically for various sampling sizes in sampling guided search (g-l) over the ACS dataset.
}}
\label{fig:quality_our_methods_ACS}
\vspace{-0.5em}
\end{figure*}

For explaining the impact of M.Sc.~versus B.Sc.~on the average salary, the covariates (and responsibility scores) found by HypDB are 
BuyNewTool (0.34), % 0.3422
LanguageWantToWorkWith (0.33), %0.3313
and RemoteWork (0.33). % 0.3265 
These attributes accounted for 8 out of the 224 single-atom refinements that satisfy the claim, while RemoteWork did not yield any predicates. 
The average naturalness score of these attributes (i.e., their corresponding predicates) ranges between $0.04-0.19$. Our viewpoint concurs as they are not very intuitive for supporting claims. Other predicates that satisfy the claim had a much higher average naturalness score. %\brit{what was the highest  naturallness score for hypDB?}. 

\subsubsection*{ACS}
We consider the
claim ``The average salary for women is higher than the
the average salary for men''. \oursolution found meaningful and understandable refinements that endorse the claim, as shown in \Cref{tab:example_preds}. For example, 
there are 27 occupations where women's average income is higher, e.g., medical transcriptionists and tutors. Other examples are high school students attending the 12th grade, and people living with a single working parent.  
% \ag{Shunit, be specific - what are the obtained predicates? What were their ranks among the 100 obtained predicates?}.
% The top 2 predicates (and their naturalness score) found by our method are: \brit{Shunit, please complete this}.

OREO identified 2,646 maximal predicates supporting this claim. This is a larger number of predicates than the user can usually consider. Among these, OREO prioritizes the 97 single-atom refinements but lacks a method to distinguish between predicates or determine their naturalness. For instance, OREO does not differentiate between the predicates:
``People who have Indian Health Service'' and ``People who have a doctorate degree''.

% To compare to OREO, we filtered the up to 2 atom predicates by generality. Out of 35,997 refinements where the claim holds, 2,646 most general predicates remained, which is still an overwhelming number of refinements for the user to consider. OREO ranks the 97 single atom refinements over the rest of the refinements (2-atom), but other than that, there is no way to differentiate between the predicates, and no concept of naturalness.
% %\shunit{TODO discuss if we want to mention 3-atom preds.}
% %We also evaluated the potential for 3-atom combinations:
% %out of 287,861 3-atom combinations, 87,941 were not filtered by generality and would be searched in the OREO method, but we note that as the number of atoms increases, the refinements become more narrow and less natural. Again, the only form of naturalness measured by the OREO method is generality. 
% For example, OREO does not differentiate between the predicates:
% "People who have Indian Health Service" and "People who have a doctorate degree".

% Using HypDB, we searched for covariates of the treatment ("SEX") and the outcome ("PINCP"). 

The top-3 covariates (and scores) by HypDB were ``Gave birth within past year'' (0.27),  %FER, 0.2706
``Veteran Service Disability Rating'' (0.27), and  %DRAT 0.2690
``Year of naturalization'' (0.25). %CITWP 0.2450
% "Ability to speak English" (0.2020), %ENG
% "Cognitive difficulty" (0.0135), %DREM
% and "Insurance through employer or union" ($1.31 \times 10^{-5}$). 
Out of the 6 covariants, only one (``Veteran Service Disability Rating'') yielded any legal refinements. The refinement is defined by ``Veteran Service Disability Rating''=``Not reported,'' which contains a small group of women (58) and a small difference between the average incomes of men and women. The found covariates are attributes that can explain income gaps between the genders. For example, giving birth in the past year may harm the yearly income, due to unpaid leave from work. However, as demonstrated, a covariate attribute does not necessarily define a subgroup where the opposite direction holds. %of the income gap holds.

% The cherrypicking search, on the other hand, yielded such attributes, which define meaningful views that satisfy the claim:
% there are 27 occupations where women's average income is higher, e.g., medical transcriptionists, tutors, and computer network architects. Other view examples are high school students attending the 12th grade, and people living with a single working parent.

% \ag{Why don't we have use cases for the flights dataset? This is a bit suspicious}
\subsubsection*{Flights}
We consider the claim ``Departure delays of more than 10 minutes are more common on Saturdays than Mondays.''
Here we use $m=1$. The difference between the approaches is similar to the previous cases.
Some refinements obtained from \oursolution are in \Cref{tab:example_preds}. 
%due to the many dependencies between attributes. For example, $(\textsf{ScheduledDeparture},\textsf{ActualDeparture})$ and $\Aagg=\textsf{DepartureDelay}$, are highly dependant.
%
\oursolution returned 2,300 refinements that endorse the claim. The refinement with the highest average naturalness score (0.36) was Hawaiian Airlines. Another refinement is flights scheduled to depart between 3AM and 4AM. However, for the opposite claim (departure delays are more common on Mondays), the algorithm returned many more refinements (6,488) and the highest average naturalness was 0.57, distinctively higher than the highest naturalness for the opposite claim (0.36). This case shows the merit of using our framework as a touchstone for verifying claims, in addition to endorsing a given claim through refinements.
The use cases demonstrate that claim endorsement cannot be directly solved by finding covariates of an intervention on an outcome, as covariate attributes do not necessarily yield successful refinements that endorse the claim. 
Furthermore, sorting the naturalness of refinements is one of the pillars of this work, but it is not discussed in HypDB and OREO. 
Finally, in OREO, the search is done on a predicate level, and as we show in \Cref{sec:pred_level}, our attribute level approach is faster by an order of magnitude. \revb{Nevertheless, we compare our approach to these baselines in the next experiments.}

\subsection{User Study}\label{sec:user_study}
\begin{figure}[t]
\includegraphics[width=\linewidth]{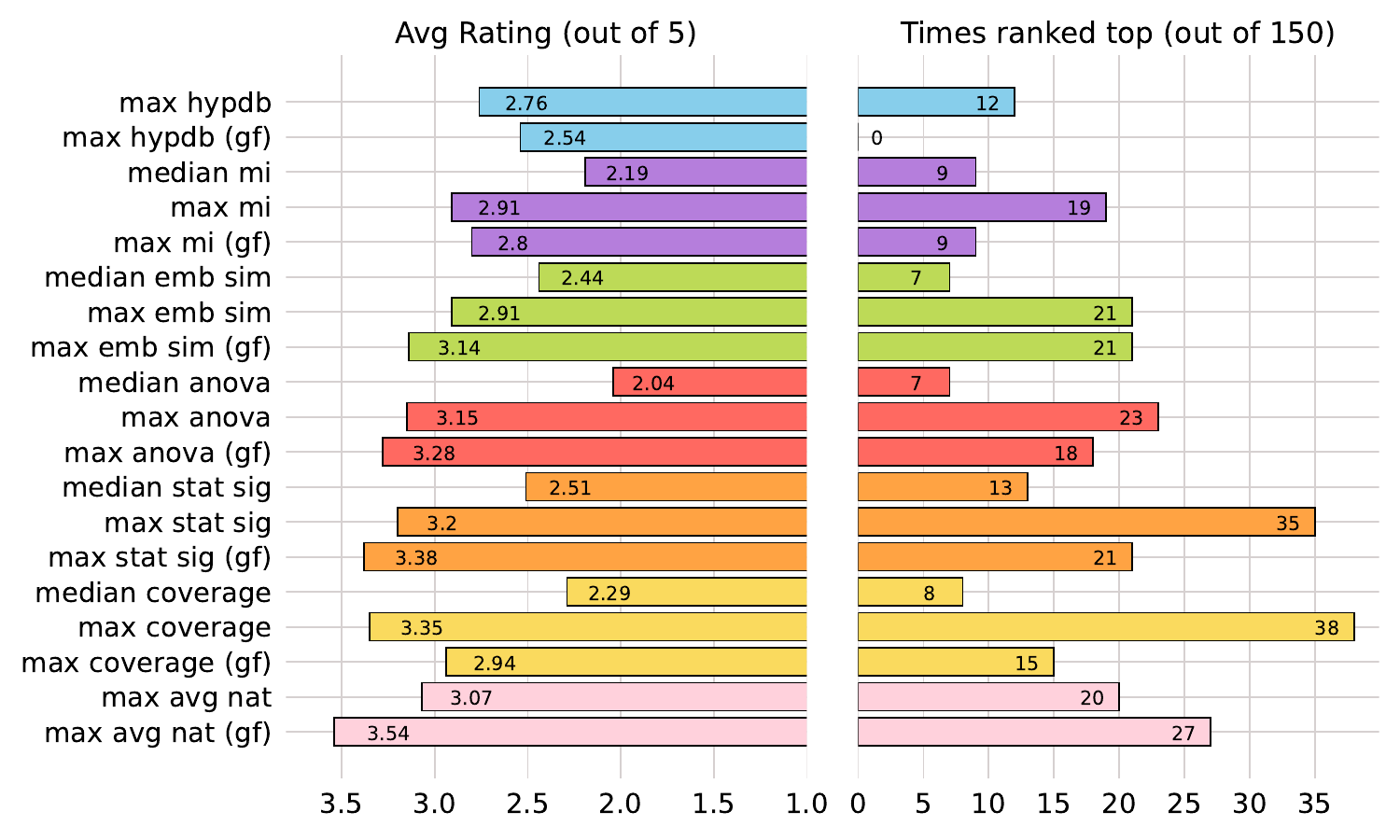}
\vspace{-2em}
\caption{\common{User study results: average rating and times ranked top for each method. (gf) stands for generality filter.}}
\vspace{-1em}
\label{fig:user_study}
\end{figure}

%comment R1-D2, R1-D5, R1-D8, R2-D1, R3-W1
\common{
We conducted a user study (1) to evaluate the extent to which the naturalness measures correspond to intuitive naturalness, (2) to find out which naturalness measures are preferred, (3) to inspect the effect of generality filtering adapted from OREO, and (4) to compare the measures to the HypDB responsibility score.

For each dataset (and corresponding claim), we generated several statements supporting the claim\footnote{The full set of survey claims and statements is included in \Cref{sec:survey_questions}.}.
%\footnote{For the full survey with all claims and statements please see \url{https://shorturl.at/eCLk0}.}. 
We presented the statements to 50 users through the Prolific Academic platform\footnote{www.prolific.com} and asked them to rate, on a scale of 1 to 5, how much they would recommend using each to an author of an article about the claim. 
% how we created the questions
We included the statements with the maximal score in each measure, %($\anova$, $\mi$, $\embsim$, $\coverage$, $\statsig$, and an average of all). 
%We also included 
along with the statement with the maximal HypDB responsibility. For each of these, we also applied a generality filter (\Cref{sec:gen_filter}) and chose the statement with the maximal score out of the most general statements, thus combining our measures with the generality property of counter arguments used in OREO~\cite{lin2022oreo}.
We also retrieved the statements with the median score in each measure.
For each choice method we computed the average rating, and the number of times its statements were ranked at the top of their group (\Cref{fig:user_study}).
}

% \begin{table}[tb]
% \centering
% \footnotesize
% \begin{tabular}{|l|c|c|}
% \hline
% \textbf{Method} & \textbf{Average rating} & \textbf{Times ranked top} \\ \hline
% max avg nat + gen filter & 3.54 & 27 \\ 
% max stat sig + gen filter & 3.38 & 21 \\
% \textbf{max coverage} & 3.35 & \textbf{38} \\
% max anova + gen filter & 3.28 & 18 \\
% \textbf{max stat sig} & 3.20 & \textbf{35} \\
% max anova & 3.15 & 23 \\
% max emb sim + gen filter & 3.14 & 21 \\
% max avg nat & 3.07 & 20 \\
% max coverage + gen filter & 2.94 & 15 \\
% max emb sim & 2.91 & 21 \\
% max mi & 2.91 & 19 \\
% max mi + gen filter & 2.80 & 9 \\
% max hypdb & 2.76 & 12 \\
% max hypdb + gen filter & 2.54 & 0 \\
% median stat sig & 2.51 & 13 \\
% median emb sim & 2.44 & 7 \\
% median coverage & 2.29 & 8 \\
% median mi & 2.19 & 9 \\
% median anova & 2.04 & 7 \\ \hline
% \end{tabular}
% \caption{User study results sorted by average rating. The two methods ranked the most times at the top are in bold. \brit{everything that is new should be colored - all tables figures etc}}
% \label{tab:user_study}
% \end{table}

%\brit{I would summarize this in bullets so the conclusions will be clear}
\common{
We conclude the following.
(1) The refinements in the medians of the measures had significantly lower ratings than their maximum counterparts, hence the naturalness measures coincide with naturalness as perceived by the participants. 
%This strengthens our hypothesis that our chosen measures represent aspects of the intuitive naturalness concept. 
(2) Maximal $\coverage$ was marked best more times than any other method, followed closely by maximal $\statsig$. The highest average rating belongs to maximal average naturalness with generality filtering.
(3) Generality filtering, adapted from OREO~\cite{lin2022oreo} and applied on each selection method, increased the rating of each method, and statistically significantly for $\statsig$, $\embsim$ and average naturalness.
(4) HypDB responsibility score was significantly lower than max $\coverage$, max $\statsig$ and max average naturalness, which coincides with our observations in our case studies, that HypDB solves a different problem.}

\subsection{Effectiveness Evaluation}
\label{sec:main_quality_exp}

We compare our methods and external baselines based on score recall over time. \Cref{tab:time_to_recall} shows the time till score recall 95\% for each prioritization method and naturalness measure. The score recall over time for each naturalness measure for ACS is shown in \Cref{fig:quality_our_methods_ACS}. We focus here on the ACS dataset since it has the largest number of attributes and over 1M rows, and therefore requires the longest run times. Similar trends were seen for the other datasets.

% We compare our methods and evaluate the score recall over time for each naturalness measure. 
% The results for ACS are shown in \Cref{fig:quality_our_methods_ACS}.
% %\Cref{fig:quality_main_anova_ACS,fig:quality_main_mi_ACS,fig:quality_main_embsim_ACS,fig:quality_main_coverage_ACS,fig:quality_main_statsig_ACS,fig:quality_main_avg_ACS}. 
% Similar trends were seen for the other datasets.
% %Figure~\ref{fig:quality_our_methods_SO} for the stack overflow dataset.
% The time until score recall above 0.95 for each prioritization method in each naturalness measure is shown in Table~\ref{tab:time_to_recall}.
% We focus here on the ACS dataset, which has the largest number of attributes and over 1M rows, and therefore requires the longest run times.

For most of the naturalness measures and datasets, the best method was \oursolutionk.
In ACS, this method was the first to reach 0.95 recall for $\mi,\embsim,\statsig$, and the average of all naturalness measures.
\revb{Over all naturalness measures and all datasets, \oursolutionk was on average 13.4 (and up to 78.3) times faster to reach 0.95 recall than the %original-order and 
random-order baseline, and on average 16.2 (up to 82.5) times faster than HypDB.}
For the Flights dataset, \oursolutionserial was the fastest to reach 0.95 recall for all measures except $\coverage$, with \oursolutionk as a close second. The best prioritization for $\anova$ was \oursolutionserial, simply because it is the first measure in the serial order.
%comment R2-D1
\revb{The external baselines yielded inferior results for all datasets and measures. Prioritizing by HypDB scores performed similarly to random. \OREOS was not applicable to the median query used in stack overflow (since median is not supported in OREO), but on ACS and Flights it was slower to reach 95\% recall, if it reached it at all. This is likely because even the enhanced \OREOS version we used operates on a predicate level, which is much more time consuming than our attribute level solution.}

%, and so the retrieval of top $k$ predicates by $\anova$ is the fastest possible among all methods. In the \oursolutionk method, the ranking by $\anova$ is merged with other naturalness rankings, thus slowing it down.

The $\coverage$ score recall was the slowest to reach $95\%$ prioritization methods (e.g., over 48 minutes for ACS). This is because $\coverage$ is a predicate-level measure, and the hardest to provide an accurate heuristic for. Our heuristic of prioritizing attribute combinations by their number of large groups hardly predicted large groups that satisfy the user. Still, in Stack Overflow, \oursolutionk\ was the fastest to reach 95\% score recall for $\coverage$.
%and Flights datasets. 
For the ACS dataset, the sampling method with 1\% was the fastest, despite the long preprocessing time. 
%This is because the sampling method is based on a predicate-level heuristic: it examines the retrieved refinements from the sample and searches for attribute combinations with the highest scoring predicates. 
We conclude that in the absence of an accurate heuristic for a given measure,  sampling can yield fast results, but for naturalness measures with a good predictor (heuristic or pre-computation),  \oursolutionserial and \oursolutionk prevail.

%\shunit{This is new and summarizes the sample size experiment instead of the subsection. Possibly shorten more?}
\common{For the \oursolutionsample method we experimented with sample sizes of 1\%, 5\%, and 10\%. The complete set of results is in \Cref{sec:sample_size_exp}. We observed a trade-off between the time until the first result (which grows with the sample size) and the time until a high score recall (which is shorter for larger samples, as the preliminary search yields more accurate results).
However, for most naturalness measures, 1\%, so we consider it a recommended sample size.
%does not fall short behind 5\% and 10\%. We concluded that 1\% is a good sample size for the tested naturalness measures. 
Similar trends were seen for the other datasets, as shown in \Cref{tab:time_to_recall}.}

Due to the results of \oursolutionsample on the $\coverage$ measure, one may consider to integrate the sampling method into the \oursolutionk or \oursolutionserial methods, as another ranking of the attribute combinations. However, the pre-processing of \oursolutionsample will increase the time until high recall is reached in all other measures.

The anytime nature of the algorithm is reflected in that the set of retrieved refinements grows, thus the sum of top-$k$ scores monotonically increases, as seen in \Cref{fig:quality_our_methods_ACS}. In the following experiments, we used this observation to optimize the experimentation time by stopping when the score recall reaches a critical threshold, computed with reference to a full run of the algorithm. When the score recall is not known,
one could design an early stopping strategy, based on the size of the change in the top $k$ sum of naturalness scores. For example, ``If the $c$ last attribute combinations did not improve the sum of top-$k$ $\nat$ scores by more than $\epsilon$, stop the search''. 

%\brit{add a discussion about the results. what is the best baseline? are the results consistent across different datasets? different naturalness measures? after how long is the recall above 0.9? }

% \iftoggle{long}{
% \begin{figure*}
% \centering
% \subfloat[Measures Average]{
%   \includegraphics[width=27mm]{figures/quality_main_SO/SO_median_2atoms_Bsc_gt_Msc_main_Metrics Average_top100score_recall_over_time_multi.png}
% }
% \subfloat[ANOVA]{
%   \includegraphics[width=27mm]{figures/quality_main_SO/SO_median_2atoms_Bsc_gt_Msc_main_Normalized Anova F Stat_top100score_recall_over_time_multi.png}
% }
% \subfloat[MI]{
%   \includegraphics[width=27mm]{figures/quality_main_SO/SO_median_2atoms_Bsc_gt_Msc_main_Normalized_MI_top100score_recall_over_time_multi.png}
% }
% % \hspace{0mm}
% \subfloat[Embedding Similarity]{
%   \includegraphics[width=27mm]{figures/quality_main_SO/SO_median_2atoms_Bsc_gt_Msc_main_Cosine Similarity_top100score_recall_over_time_multi.png}
% }
% \subfloat[Coverage]{
%   \includegraphics[width=27mm]{figures/quality_main_SO/SO_median_2atoms_Bsc_gt_Msc_main_Coverage_top100score_recall_over_time_multi.png}
% }
% \subfloat[Statistical Significance]{
%   \includegraphics[width=27mm]{figures/quality_main_SO/SO_median_2atoms_Bsc_gt_Msc_main_Inverted pvalue_top100score_recall_over_time_multi.png}
% }
% \caption{Top 100 score recall for each naturalness measure over time for our methods over the stack overflow dataset. \shunit{This figure will be removed before submission - but it's useful for writing the experiments section.}}
% \label{fig:quality_our_methods_SO}
% \end{figure*}
% }

\begin{table}[t]
\scriptsize
    %\centering
    \caption{\revb{Time (seconds) to 0.95 score recall and first result time (FRT) for different prioritizations and baselines (\OREOS and HypDB). \OREOS is inapplicable to Stack Overflow as it does not support median aggregation. Entries ``t/o'' mean that timeout is reached (2 hours for ACS7 and 5 hours for Flights).}\label{tab:time_to_recall}}
    \vspace{-1.2em}
    
    % \begin{tabular}{lcccccc}
    %     \toprule
    %     \multirow{2}{*}{Method/Metric} & EmbSim & Statistical significance & ANOVA & MI & Coverage & Average Naturalness \\
    %     \cmidrule(lr){2-7}
    %     & EmbSim & Statistical significance & ANOVA & MI & Coverage & Average Naturalness \\
    %     \midrule
    %     500Sample & 190 & 31 & 336 & 28 & 312 & 29 \\
    %     1KSample & 57 & 47 & 74 & 45 & 287 & 45 \\
    %     5KSample & 90 & 71 & 92 & 68 & 362 & 69 \\
    %     10KSample & 111 & 90 & 106 & 88 & 368 & 88 \\
    %     Original Order & 202 & 10 & 242 & 314 & 270 & 242 \\
    %     Random Order & 228 & 8 & 224 & 269 & 246 & 156 \\
    %     Serial Top K & 115 & 9 & 4 & 5 & 246 & 7 \\
    %     Merged Top K & 25 & 7 & 11 & 4 & 206 & 7 \\
    %     \bottomrule
    % \end{tabular}
    \begin{tabular}[b]{l@{\hskip.15cm}l@{\hskip.15cm}c@{\hskip.15cm}c@{\hskip.15cm}c@{\hskip.15cm}c@{\hskip.15cm}c@{\hskip.15cm}c|c}
\toprule
Dataset & Prioritization & $\embsim$ & $\statsig$ & $\anova$ & $\mi$ & $\coverage$ & Avg. & FRT \\
\multirow{7}{*}{\makecell[cl]{ACS \\ $\alpha=\mathsf{Avg}$}}
& \oursolutionsampleone (3) & 1303 & 300.7 & 3881 & 833.3 & \textbf{2886} & 401 & 191.0 \\ 
&\oursolutionsamplefive (3) & 734 & 598.7 & 5733.3 & 584.33 & 3720.7 & 588.3 & 582.4\\ 
&\oursolutionsampleten (3) & 1006.3 & 839.3 & 3812 & 830.3 & 3895.7 & 834 & 828.1\\ 
%&Original Order & 1896 & 501 & 6078 & 1879 & 5701 & 1941 & 15.8\\ 
&Rand.~Order (3) & 2914.3 & 177 & 5129 & 909.7 & 5518.7 & 1479 &10.2\\ 
&\oursolutionserial & 3415 & 554 & \textbf{469} & 503 & 4566 & 1218 & 84.3\\ 
&\oursolutionk & \textbf{231} & \textbf{71} & 1557 & \textbf{66} & 4886 & \textbf{76} & 49.9 \\ 
& \revb{HypDB} & 1992 & 290 & 6079 & 502 & 4272 & 722 & 139.9\\
%& \red{OREO (generality filtering)} & 7380 & 5505 & 7380 & 7380 & 7380 & 7380 & 38.49\\
%& \revb{OREO (no generality filtering)} & 5913 & 1051 & 7380 (c) & 98 & 7380 (c) & 982 & 38.49 \\ % c- cutoff. 95% recall not reached (we take the attrs responsible for the top 100 in avg. nat. so it's possible we missed some that are responsible for top 100 in other metrics.
& \revb{\OREOS} & 5913 & 1051 & t/o & 98 & t/o & 982 & 38.5 \\
\midrule
\multirow{7}{*}{\makecell[cl]{Stack O. \\ $\alpha=\mathsf{Med}$}} 
& \oursolutionsampleone (3) & 63.3 & 41.7 & 298 & 43.3 & 332.3 & 40.7 & 38.1 \\
& \oursolutionsamplefive (3) & 77.3 & 57.7 & 100.3 & 54.7 & 358 & 55 & 53.5 \\
& \oursolutionsampleten (3) & 94.3 & 71.7 & 99.3 & 70 & 362.7 & 70 & 68.7 \\
%& Original order & 202 & 10 & 242 & 314 & 270 & 242 & 1.2 \\
& Rand.~Order (3) & 206.3 & 9 & 207.3 & 313.3 & 290.7 & 200 & 1.1 \\
& \oursolutionserial & 115 & 9 & \textbf{4} & 5 & 246 & \textbf{7} & 1.8 \\
& \oursolutionk & \textbf{25} & \textbf{7} & 11 & \textbf{4} & \textbf{206} & \textbf{7} & 1.9 \\
& \revb{HypDB} & 280 & 28 & 313 &330 &261 & 203 & 9.7\\

% \midrule
% \multirow{7}{*}{Flights  (AA vs.~UA)}
% &\oursolutionsampleone (3) & 130 & 116.67 & 122.67 & 116.67 & 534.33 & 115 & 105.85 \\
% & \oursolutionsamplefive (3) & 223.67 & 213.67 & 219 & 213.67 & 607.33 & 210.33 & 200.76 \\
% &\oursolutionsampleten (3) & 308.33 & 298.33 & 303.67 & 298.33 & 675.67 & 295.33 & 285.57 \\
% &Original Order & 1160 & 43 & 1552 & 1166 & 1552 & 1166 & 24.74 \\
% &Random Order (3) & 486 & \textbf{34.67} & 1493.67 & 545 & 1493.67 & 1061.33 & 24.26 \\
% &\oursolutionserial & 63 & 50 & \textbf{48} & \textbf{59} & 1331 & \textbf{50} & 41 \\
% &\oursolutionk & \textbf{44} & 44 & 82 & 75 & \textbf{232} & 59 & 39.87 \\

\midrule
\multirow{7}{*}{\makecell[cl]{Flights \\ $\alpha=\mathsf{Cnt}$}} %(Mon. vs. Sat.) 
& \oursolutionsampleone (3) & 133.7 & - & 131.3 & 131.3 & 1022.3 & 131.3 & 129.8 \\
& \oursolutionsamplefive (3) & 261 & - & 259 & 259 & 1148.7 & 259 & 257.4 \\
& \oursolutionsampleten (3) & 344 & - & 342 & 342 & 1232.7 & 342 & 340.4 \\
%& Original Order & 100 & - & 50 & 50 & 881 & 50 & 20.4 \\
& Rand.~Order (3) & 111.7 & - & 512.7 & 381.3 & \textbf{793} & 435.3 & 20.9 \\ 
& \oursolutionserial & \textbf{37} & - & \textbf{35} & \textbf{35} & 895 & \textbf{35} & 34.5 \\
& \oursolutionk & 40 & - & 38 & 38 & 932 & 38 & 34.6 \\
& \revb{HypDB} & 239 & - & 979 & 979 & 928 & 979 & 89.9 \\ 
%& \revb{OREO} & 13050 & - & 20750(c) & 20750(c) & 12298 & 20750(c) & 12.71 \\ 
& \revb{\OREOS} & 13050 & - & t/o & t/o & 12298 & t/o & 12.7 \\% OREO was given the 26 attributes responsible for the top 100 results (w.r.t each metric). It also recieved a smaller version of the db, filtered using the WHERE clause (dp. delay >10). It did not reach 95% recall in any metric.

% small flights DB
% &1\% Sample(3) & 12 & 4.33 & 7 & 33 & 8.67 & 12 & 2.98 \\
% &5\% Sample(3) & 29.33 & 6.33 & 12.67 & 24 & 18.33 & 6 & 5.55 \\
% &10\% Sample(3) & 10 & 9.33 & 14.33 & 32 & 27 & 9.33 & 8.56 \\
% & Original Order & 5 & 2 & 9 & 24 & 21 & 5 & 0.99 \\
% & Random Order(3) & 24.67 & 2.33 & 31.33 & 23.33 & 34.33 & 5.33 & 1.13 \\
%&MI & 6 & 4 & 13 & 3 & 35 & 4 & 1.921123981 \\
%&ANOVA & 6 & 3 & 2 & 21 & 11 & 5 & 1.321282864 \\
%&EmbSim & 2 & 2 & 14 & 11 & 33 & 2 & 1.248801231 \\
%&Regression & 6 & 3 & 35 & 30 & 35 & 6 & 1.453836441 \\
%&Coverage100 & 25 & 2 & 19 & 28 & 21 & 10 & 1.133672953 \\
% &Serial Top k & 5 & 3 & \textbf{3} & \textbf{5} & \textbf{20} & 5 & 1.67 \\
% &Merged Top k & \textbf{2} & \textbf{2} & 11 & 9 & 27 & \textbf{2} & 1.89 \\
        \bottomrule
    \end{tabular}
%\fbox{
\vspace{-1em}
\end{table}

% \subsection{Precomputed metric guided search}
% We evaluate the top k set of predicates retrieved over time in each of the guiding methods. We compute three types of recall-based scores. Each of the scores is computed for each naturalness measure separately.
% For a given naturalness measure $NM$:
% \begin{itemize}
%     \item Recall$_{NM}$@T - the number of exact predicates retrieved by time $T$ that are in the top k predicates according to $NM$, out of the actual top k that were retrieved.
%     \item Score recall$_{NM}$@T - the sum of naturalness score values of the top k predicates retrieved by time $T$, divided by the sum of $NM$ values of the actual top k predicates. Since many predicates can have the same naturalness score (e.g., the same ANOVA score is given to all predicates created by a given attribute combination), classic recall may not provide the full picture.
%     \item First out over last in (FO/LI)$_{NM}$@T - the highest $NM$ score of a predicate not retrieved at time $T$, divided by the lowest $NM$ score of a predicate in the top k retrieved at time T. 
% \end{itemize}

% The results for ANOVA-, MI- and approximate cosine similarity- guided searches are shown in Figure~\ref{fig:precomp_metrics}.
% The most interesting thing to notice is that the naturalness measures are not independent; while the search is guided by one metric, the others rise with it, but slower. 
% \red{TODO: discuss results. The time until score recall 1 in the guiding metric vs. the other metrics. The score recall 1 vs. revall 0.3 in MI guided search. Etc.}

\subsection{Parameter Sensitivity}\label{sec:sensitivity}
\subsubsection*{Sensitivity to number of tuples and columns}
%We assess the sensitivity of our algorithm's runtime to the the number of tuples (\Cref{fig:num_tuples_ACS}), and to the number of columns in the database (\Cref{fig:num_columns_ACS}). 
For the number of tuples (\Cref{fig:num_tuples_ACS}), we sample varying amounts of 
%For the experimental evaluation, we save the subset as a materialized view and run all of the search queries on the same sample.
tuples out of the full ACS dataset. %\red{save the sample as a materialized view, and evaluate each method on it.} %and run the full cherrypicking search on the samples. 
For the number of columns (\Cref{fig:num_tuples_ACS}), we sample increasing sizes of subsets of $\splitset$ in the ACS dataset. For each number of tuples or columns, we show the average of three runs. 
We measure the
time until a 95\% score recall was achieved, for the naturalness measures average score, in each of our five main prioritization methods \revb{and in the external baselines - HypDB and \OREOS}. 
%We measure the time until the full search is done, and out of that, the total time for SQL queries run during the search. We also measure the time for setup (generating the attribute combinations and reading the precomputed scores and heuristics).
% The results on the ACS dataset are shown in Figure~\ref{fig:num_tuples_ACS} for the number of tuples and Figure~\ref{fig:num_columns_ACS} for the number of columns.
Similar trends were observed for the other two datasets.

% \begin{figure}[htbp]
% % \centering
% % \includegraphics[width=0.4\linewidth]{figures/scalability_by_tuples_SO_median.png}
% \subfloat[Sensitivity to number of tuples]{
%     \label{fig:num_tuples_SO}
%   \includegraphics[width=0.49\linewidth]{figures/SO_Bsc_gt_Msc_DB_size_sensitivity.png}
% }
% \subfloat[Sensitivity to number of columns]{
% \label{fig:num_columns_SO}
% % \includegraphics[width=0.49\linewidth]{figures/scalability_by_columns_SO_median.png}
% \includegraphics[width=0.49\linewidth]{figures/SO_Bsc_gt_Msc_DB_width_sensitivity.png}
% }
% \caption{Time for 0.95 score recall of avg. naturalness in cherry-picking run for 2 atoms, with varying number of tuples and columns, over stack overflow.}
% \end{figure}

For the naive prioritization of random order \revb{and for HypDB}, the time until 95\% recall steadily rises with the number of tuples or columns. This is also the case for \oursolutionsampleone, since the time to run the pre-processing step of the search on the 1\% sample naturally depends on the number of tuples in the database. \revb{For \OREOS, the time to reach 95\% recall was on average 7.8X longer than random order for the number of tuples experiment, and 84X longer for the number of columns experiment. This is due to the predicate level approach taken in \OREOS, which does not scale to large datasets.}
However, for \oursolution, the time until 95\% recall is almost constant. This is due to the prioritization of the search, and covering the most promising attribute combinations at the beginning. For \oursolutionserial, there is a rise in the time until 95\% score recall with the number of tuples or columns, although not as steep as the random order baseline. A possible explanation is that in \oursolutionserial, only after $k$ refinements for a specific naturalness measure are found, do we continue to the next measure. If there is only a small number of refinements for a specific measure, this may delay the recall for the other measures (and for their average).

\subsubsection*{Sensitivity to $k$ (number of top refinements)}\label{sec:k_sensitivity}

\begin{figure*}[t]
% \centering
% \includegraphics[width=0.4\linewidth]{figures/scalability_by_tuples_SO_median.png}
% \vspace{-0.5em}

\subfloat[Sensitivity to number of tuples]{
    \label{fig:num_tuples_ACS}
  \includegraphics[width=40mm]{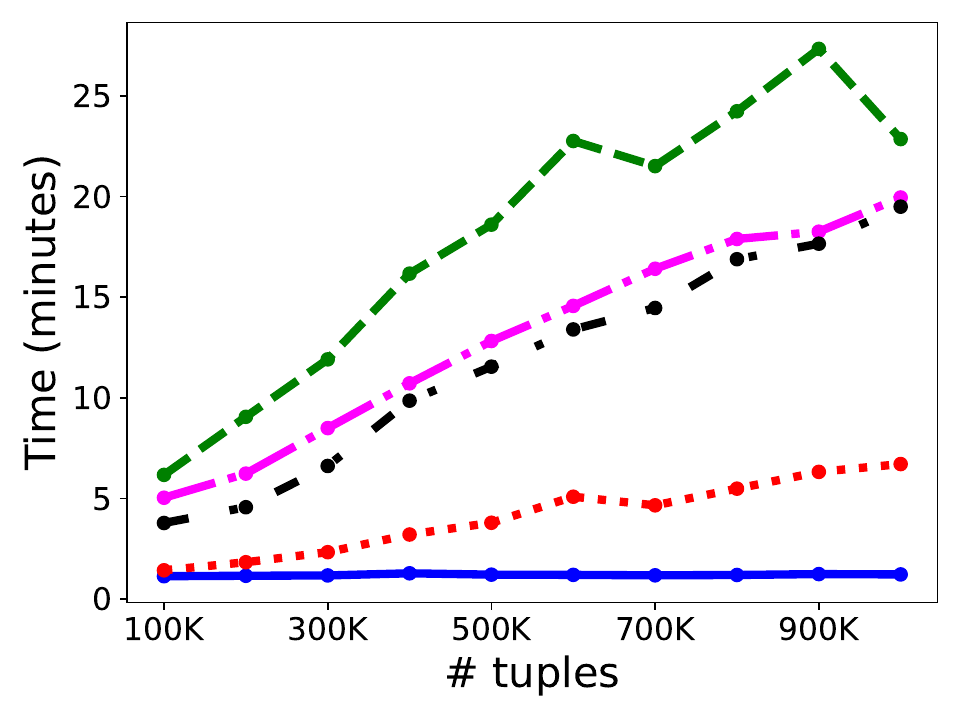} % generated with sensitivity experiments 5,6,7
}
\subfloat[Sensitivity to number of columns]{
\label{fig:num_columns_ACS}
\includegraphics[width=40mm]{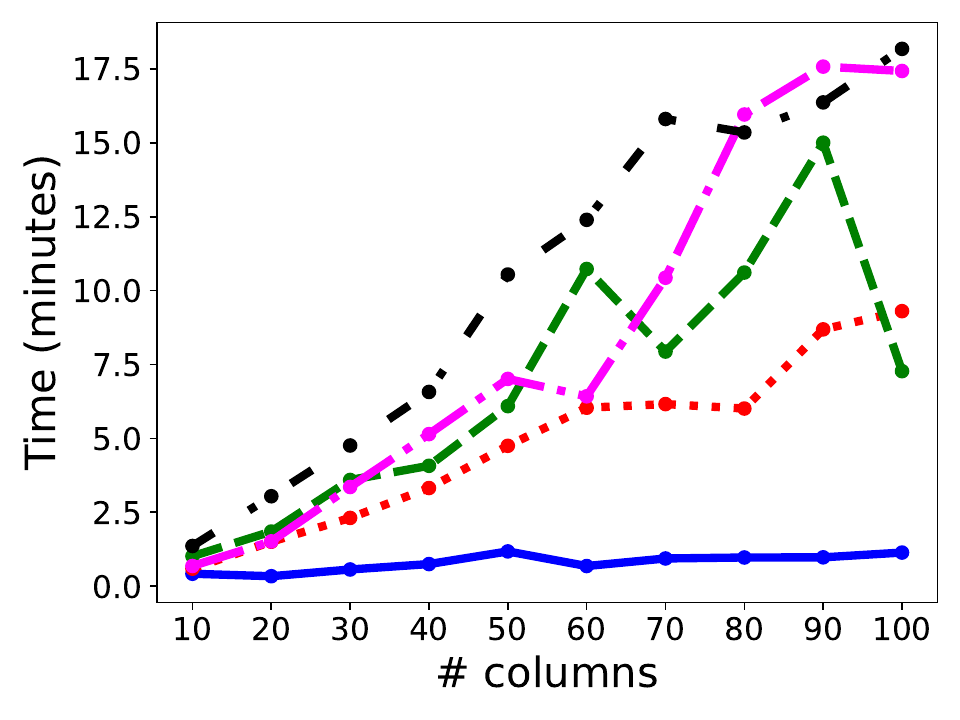}
}
%\hspace{0}
% \vspace{-0.5em}
% \begin{minipage}{\linewidth}
\subfloat[Sensitivity to $k$]{
\label{fig:k_sensitivity_ACS}
\includegraphics[width=40mm]
{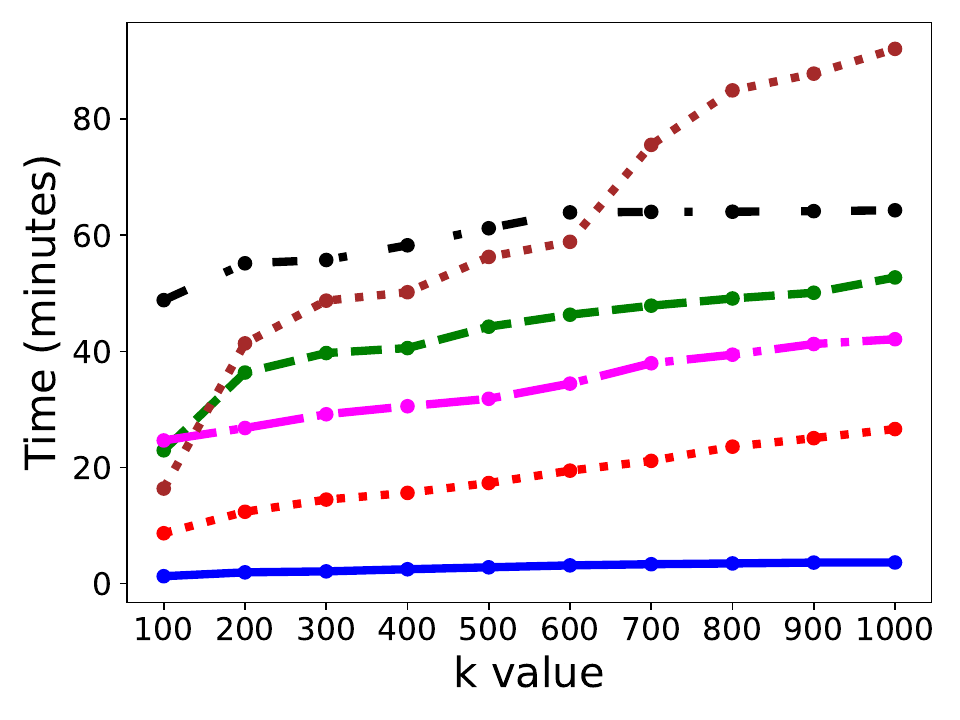}
}
%\hspace{0}
%\vspace{-0.5em}
\subfloat[Sensitivity to $m$]{
\label{fig:num_atoms_SO}
\includegraphics[width=40mm]{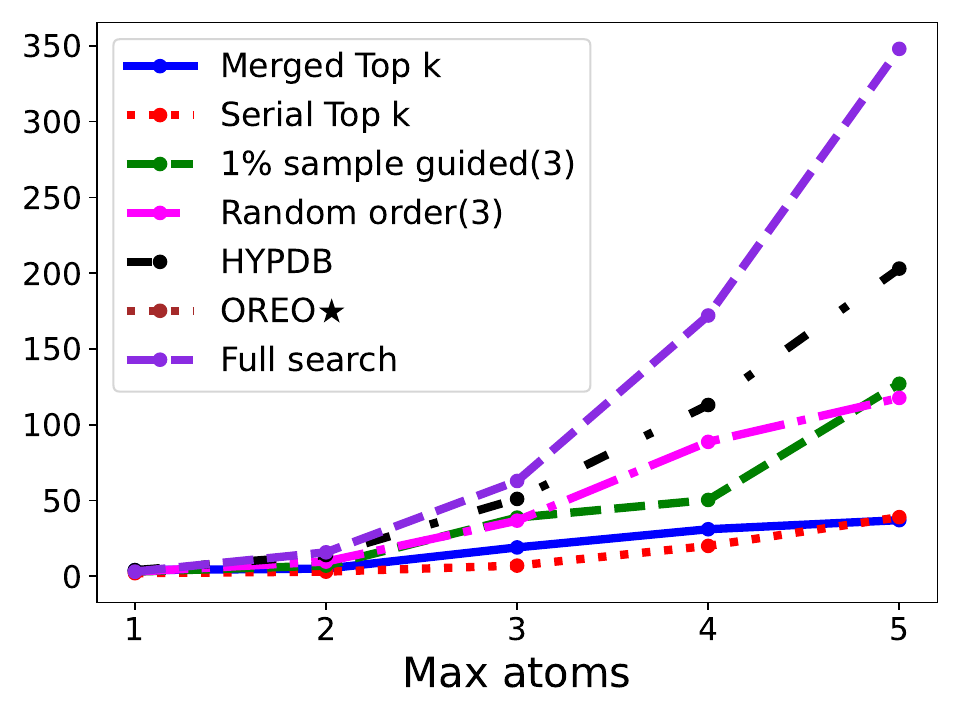}
%\caption{\revb{Sensitivity to maximal number of atoms: time until 95\% avg. naturalness score recall with increasing number of atoms over stack overflow with 10 attributes.}}
}
% \end{minipage}
% \vspace{-3mm}
\vspace{-1em}
\caption{\revb{Time for 95\% score recall of avg.~naturalness over \textit{ACS} with at most $m=2$ atoms, with varying number of (a) tuples, (b) columns, and (c) varying values of $k$. (d) Sensitivity to maximal number of atoms ($m$): time for 95\% recall of average naturalness with increasing number of atoms over \emph{Stack Overflow} with 10 attributes.
%\OREOS results were omitted from \Cref{fig:num_tuples_ACS} (and \ref{fig:num_columns_ACS}), because the time until 95\% recall was on average 7.8X (84X) longer than Random Order. Therefore, we did not include \OREOS in the $m$-sensitivity experiment (\Cref{fig:num_atoms_SO}) either.}
}}
\vspace{-1em}
\end{figure*}

We measured the time until 95\% top-$k$ score recall of the average naturalness measures in the five prioritizations with $k$ values from 100 to 1000 on ACS (\Cref{fig:k_sensitivity_ACS}). Similar trends were observed on the other datasets. %order of other methods is different, but serial and merged top k are always lower than the rest.
For all methods except \oursolutionserial, the choice of $k$ does not affect the search algorithm, only the recall computation. For \oursolutionserial, $k$ refinements are fetched from each naturalness ranking at the beginning of the search. Therefore, for \oursolutionserial each point in \Cref{fig:k_sensitivity_ACS} represents a different algorithm run.

As expected, for all methods
%, \revb{including \OREOS and HypDB}, 
the time to 95\% recall grows with $k$.
%i.e., it takes a longer time to find more top refinements. 
% Do we want to discuss this anomaly?
%Exceptional is that original order prioritization took the same time for $k=100$ and $k=200$. This is because some attribute combinations can yield many refinements at once. %In our case - (OCCP, WKHP) yielded 124 refinements.
However, \oursolutionk exhibited the most subtle rise in time to 95\% recall out of all methods (ranging from 1.3 to 3.6 minutes), showing the merit of this combined prioritization method.
%followed by \oursolutionserial (8.7 to 26.6 minutes). 

\subsubsection*{Sensitivity to $m$ (maximal number of attributes per combination)}\label{sec:atom_sensitivity}
% comment R2-D2
\revb{Increasing $m$ may lead to complex statements, which are less natural. For example, $(\textsf{OpSysProfessional use}{=}\textsf{Windows}) \land (\textsf{YearsCode}=0{-}10) \land (\textsf{OrgSize}{=}20{-}99) \land (\textsf{DatabaseHaveWorkedWith}{=}\textsf{MariaDB}) \land (\textsf{AccessibilityDifficulties}{=}\textsf{None})$. Most of the refinements with a large $m$ are specializations of a more general refinement (e.g.,  for stack overflow with 10 attributes, 99.2\% of the 44,890 refinements with 5 atoms specialize 3-atom refinements).
%However, we evaluated the applicability of our methods when increasing the maximal number of atoms $m$ over stack overflow.

While the run times of our algorithms are relatively short, the precomputation times for computing $\anova$ and $\mi$ in advance were very long (about 24 hours for $m{=}3$). Therefore, we limited this experiment to the 10 attributes responsible for the largest number of refinements retrieved in the 2-atom search. Due to \OREOS's long run times in the previous experiments, it was not included in this experiment. The results are shown in Figure~\ref{fig:num_atoms_SO}.
}

\revb{
As previously, \oursolutionk and \oursolutionserial were the fastest to reach 95\% recall. For comparison, we added the time to conduct a full claim endorsement search (regardless of the recall level), and as expected, it grows exponentially with $m$. It can be seen that \oursolutionk and \oursolutionserial were much faster than that.
}

\section{Related Work}\label{sec:related}
% \shunit{Benny wanted this to be in the beginning.}
% We next survey related work in fields of cherrypicking and explanations for query results. 
% {\em To the best of our knowledge, this is the first work that generates natural views that satisfy a given user claim.}

% \brit{papers to cite:}
% \begin{itemize}
%     \item \cite{zhou2020survey} A survey on fake news detection 
%     \item \cite{hassan2017claimbuster} - Fact checking (published in VLDB)
%     \item \cite{guo2022survey} A survey on fact-checking methods in ML, NLP, and DB
%     \item  \cite{murphy2019harking} cherry-picking psychological aspect
%     \item \cite{jo2019aggchecker} : A fact-checking system for text summaries of relational data sets
%     \item \cite{karagiannis2020scrutinizer} fact checking statistical claims
%     \item \cite{wu2014toward} computational fact checking
%     \item \cite{farinha2018towards} computational fact-checking NLP techniques
%     \item \cite{wu2017computational} fact-checking through query perturbations
% \end{itemize}

 %We next survey related work on cherry-picked claims and explanations of query answers. 
%{\em To the best of our knowledge, this is the first work that generates natural refinements for endorsing user claims.}

\subsubsection*{Fact-checking and cherry-picking}
% The task of finding a subset of the dataset where the claim holds is similar to counterargument identification from~\cite{lin2021detecting}. However, the focus there is on the generality of the counterargument, which is based on a user-defined hierarchy of group-by attributes, and can be seen as a proxy to coverage \ag{Can we give a more convincing difference? This work seems very similar}.
% \brit{the discussion about OREO predicate level search vs ours attributes level search should go to the related work section, since we are not comparing runtime here}
% cherry picking detection and mitigation in different settings

Automatic methods to identify cherry-picked results and fake news have been widely studied~\cite{zhou2020survey,guo2022survey,hassan2017claimbuster}. 
Traditional fact-checking methods rely on domain knowledge~\cite{hassan2017toward,hassan2015detecting} and lack scalability and persuasiveness without supporting datasets.
Other methods employ machine learning and NLP techniques for efficient computational fact-checking~\cite{farinha2018towards,guo2022survey}.
A considerable body of research focuses on automating fact-checking using structured data~\cite{wu2014toward,wu2017computational,asudeh2020detecting,lin2021detecting,jo2019aggchecker}. 
In one line of work~\cite{wu2014toward,wu2017computational}, parameterized queries are defined, and the robustness of a fact to parameter perturbations is analyzed. Perturbations are assessed for their \emph{relevance and naturalness}, depending on the attributes and domain knowledge.
In another line of work~\cite{asudeh2020detecting, asudeh2021perturbation}, trendlines are examined with respect to their robustness to changes in their start and end points. Cherry-picking has also been examined in rankings~\cite{asudeh2021perturbation}, where items are ranked according to a linear combination of numeric attributes, and the weights can be cherry-picked.

\paragraph*{Explanations of query answers}
Claim endorsement can be viewed as the opposite of the intervention or responsibility approaches in query result explanations~\cite{wu2013scorpion, roy2014formal, roy2015explaining,MeliouGNS11,MeliouGMS11,MeliouRS14}. In the latter, the tuples conforming to a predicate are removed from the database so that the result on the smaller database satisfies a user assertion. Conversely, a predicate in our setting qualifies the tuples that 
\emph{remain} in the database to satisfy the user claim. 
A commonly used type of explanation for query results is a set of predicates that differentiate between tuples in the answer of aggregate queries \cite{el2014interpretable,li2021putting,TaoGMR22,roy2015explaining,roy2014formal,ten2015high}. 
Our framework similarly provides a ranked list of predicates that define the refinements where the claim holds. However, our primary goal differs in that we seek the most natural refinements to endorse a user's claim, rather than predicates explaining unexpected query answers. 
Query refinements have also been used for improving query results according to desired properties~\cite{chapman2009not}, including the number of results~\cite{MishraK09,muslea2005online,KoudasLTV06,VelezWSG97,Ortega-BinderbergerCM02,mishra2009interactive,chu1994structured}, diversity~\cite{LiMSJ23}, or bias removal~\cite{SalimiGS18}. We adopt the query refinement approach for 
the new goal of claim endorsement.

% However, we do observe some similarities when it comes to examining the influence of a predicate and when considering coverage as one the parameters thereof~\cite{wu2013scorpion,roy2014formal,TaoGMR22}. 

\paragraph*{Multidimensional data aggregation}
%Another closely related line of work explores multidimensional data aggregation for various downstream tasks. 
Previous work on multidimensional data aggregation developed methods that extend the traditional drill-down and roll-up operators to find the most interesting data parts for exploration~\cite{agarwal1999cube,sathe2001intelligent,joglekar2017interactive,DBLP:journals/pvldb/YoungmannAP22}.
\reva{Others works have focused on assessing the similarity between two data cubes~\cite{baikousi2011similarity}).} % comment R1-D6, R1-D7
%Other works use data aggregations to discover intriguing data visualizations and explanations~\cite{vartak2015seedb,deutch2022fedex}. 
None of these methods corroborate a user's claim in the aggregate view, or checks whether it represents a meaningful definition of a subpopulation for the claim. 
Some employ the CUBE operator in SQL~\cite{agarwal1999cube}, which may seem relevant here to materialize all possible combinations of attributes for refinements. Yet, this approach is not scalable; database systems usually limit the number of attributes in the cube operator to 12~\cite{salimi2018}, due to its size being exponential in the number of attributes. It has been shown~\cite{lin2021detecting} that over 6 attributes, CUBE is impractical for its time and memory consumption.

\paragraph*{View Recommendation.}
\reva{ %comment R1-D6, R1-D7
Another relevant area of research has used data aggregations to discover intriguing data visualizations~\cite{vartak2015seedb,deutch2022fedex}, with various techniques proposed to recommend the top-$k$ most interesting views~\cite{ehsan2016muve,mafrur2018dive,ding2019quickinsights}. Some studies recommend visualizations based on visual quality utility functions, such as visual expressiveness~\cite{wongsuphasawat2015voyager}. Others  use deviation-based utility functions, where the interestingness of a view is measured by its distance from a reference view~\cite{vartak2015seedb,ding2019quickinsights,deutch2022fedex}. Additional studies has explored multi-objective utility functions that capture different aspects of view interestingness (e.g., visual quality, deviation) or aimed to discover the most appropriate utility function for the current analysis context~\cite{ehsan2016muve,mafrur2018dive,zhang2021viewseeker}.
The key distinction from the current work is that interestingness differs significantly from naturalness; in fact, they can be seen as almost opposite concepts. While they seek to identify interesting views that reveal anomalies in the data, our objective is to find the most natural views---those with the fewest special characteristics that do not attract particular attention. The goal is to support claims that hold true across natural subpopulations.

Nonetheless, we share common techniques with this line of work in finding the top-$k$ views. We employ sharing-based optimizations to minimize the number of query executions, similarly to \cite{vartak2015seedb}, and pruning-based optimizations (like filtering attribute combinations when their maximum group size is too small) to improve running times, as done in \cite{wongsuphasawat2015voyager,vartak2015seedb}. However, given our different objective, we have developed novel optimizations tailored to our specific context.
}

\section{Conclusions and Future Directions}\label{sec:conc}
We presented a framework for endorsing user claims through query refinements over the data. Our framework quantifies the quality of a refinement using naturalness measures that have been adapted from the literature but is able to support any measure of naturalness. 
We proposed an efficient approach for computing refinements using an anytime algorithm with a two-step approach that first produces a ranked list of attribute combinations and then computes value assignments to produce the refinements.

% comment R3-W3
There are multiple avenues of future work that build on this framework. 
% Choices in the problem definition
First, it is possible to extend our work to a wider problem definition.
\revc{Our work currently considers a single relation and can be extended to multi-relation databases, by creating a single view in advance, or by modifying the query for each attribute combination to enable multiple relations.} 
\reva{
While we focused on refinements (assuming that the claim does not hold in general), ideas in this paper could also be used for relaxations. Attributes combinations can be created from the attributes that appear in the where clause of a given query. The attribute combinations can be prioritized as in Section~\ref{sec:methods:prioritization} and searched as in Section~\ref{sec:methods:find_preds}. 
Relaxation for conjunctions appears to be a simpler problem, with the main challenge being prioritization, where our methods also apply.}
\reva{Additionally, an important direction is to enrich the predicate language to inequalities and disjunctions.} \revc{Inequality predicates can easily be supported in the algorithmic framework, while disjunctions would require changes to the algorithm and different optimizations.}

% displaying the results
Second, one may consider methods to make the results more accessible to users. For example, one can incorporate desiderata for the ranking of refinements that considers the entire list, such as diversification which would introduce variation in the refinements. \revc{This would require designing metrics that combine both naturalness and diversity, and re-ranking the returned refinements while taking the new metric into account.}
Alternatively, one can consider a summarization strategy for the display of results, grouping similar refinements together and thus facilitating the analysis process.

% change the search
Third, to make the search process interactive, it would be helpful to develop decision rules that help the user decide when to end the search and could be based on bounds on the quality of potential results.
\reva{Another intriguing future work is to discover the most suitable naturalness measure in the current analysis context based on user interaction (inspired by \cite{zhang2021viewseeker}).} %comment R1-D7

Finally, our work considers user claims that compare two groups. We intend to consider more elaborate claims such as trends in the query results. \revc{This may require considerable adjustment of the problem definition, the predicate space and the algorithm.}

% \begin{acks}
%  This work was supported by the [...] Research Fund of [...] (Number [...]). Additional funding was provided by [...] and [...]. We also thank [...] for contributing [...].
% \end{acks}

\clearpage
\bibliographystyle{ACM-Reference-Format}

\bibliography{cherrypick}

%\clearpage
%\appendix
%\section{Queries with Additional Information for Naturalness Measures}
%\label{appendix:full_queries}
%\input{full_mean_q.tex}
%\input{full_median_q.tex}
\clearpage
\appendix
\section{Sample size for sampling-based search}\label{sec:sample_size_exp}
\begin{figure*}[t]
\centering
% now sampling
\subfloat[$\anova$]{
\label{fig:quality_sampling_anova_ACS}
  \includegraphics[width=27mm]{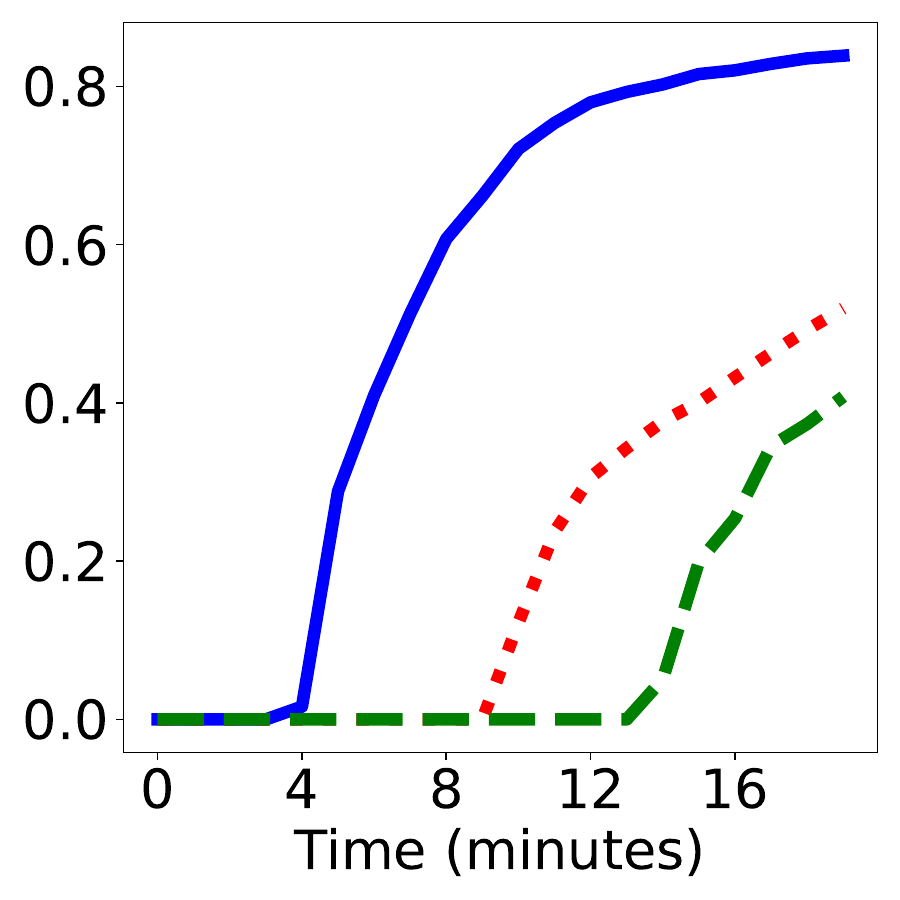}
}
\subfloat[$\mi$]{
\label{fig:quality_sampling_mi_ACS}
  \includegraphics[width=27mm]{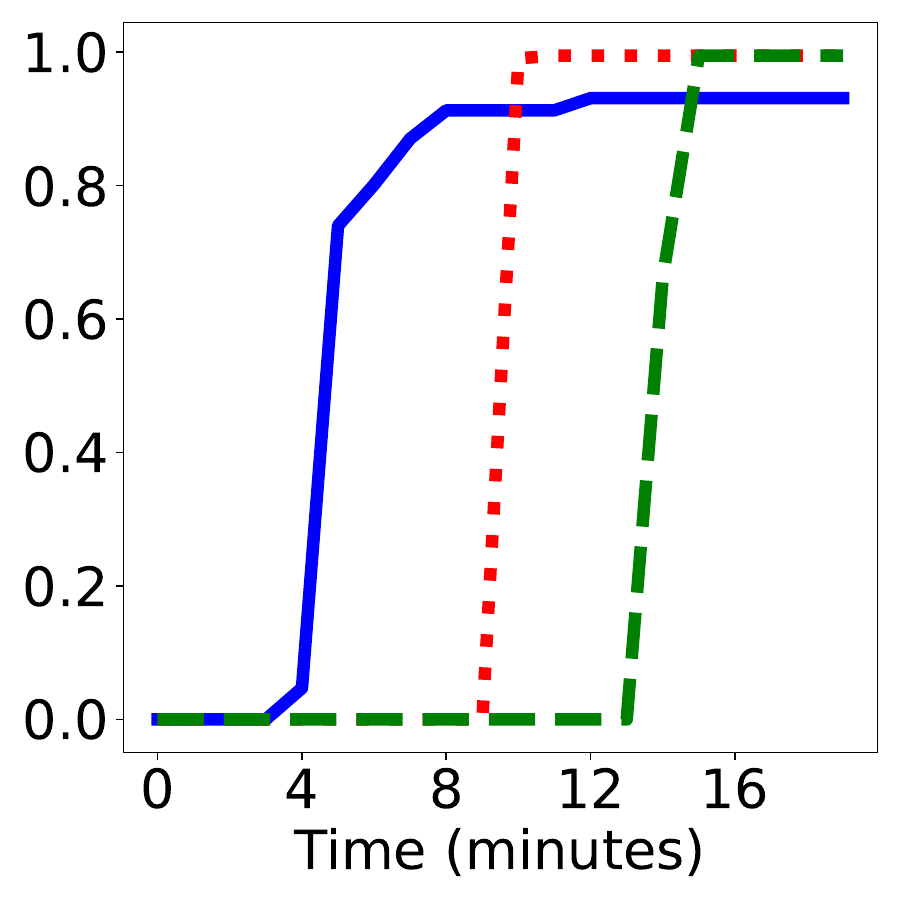}
}
\subfloat[$\embsim$]{
\label{fig:quality_sampling_embsim_ACS}
  \includegraphics[width=27mm]{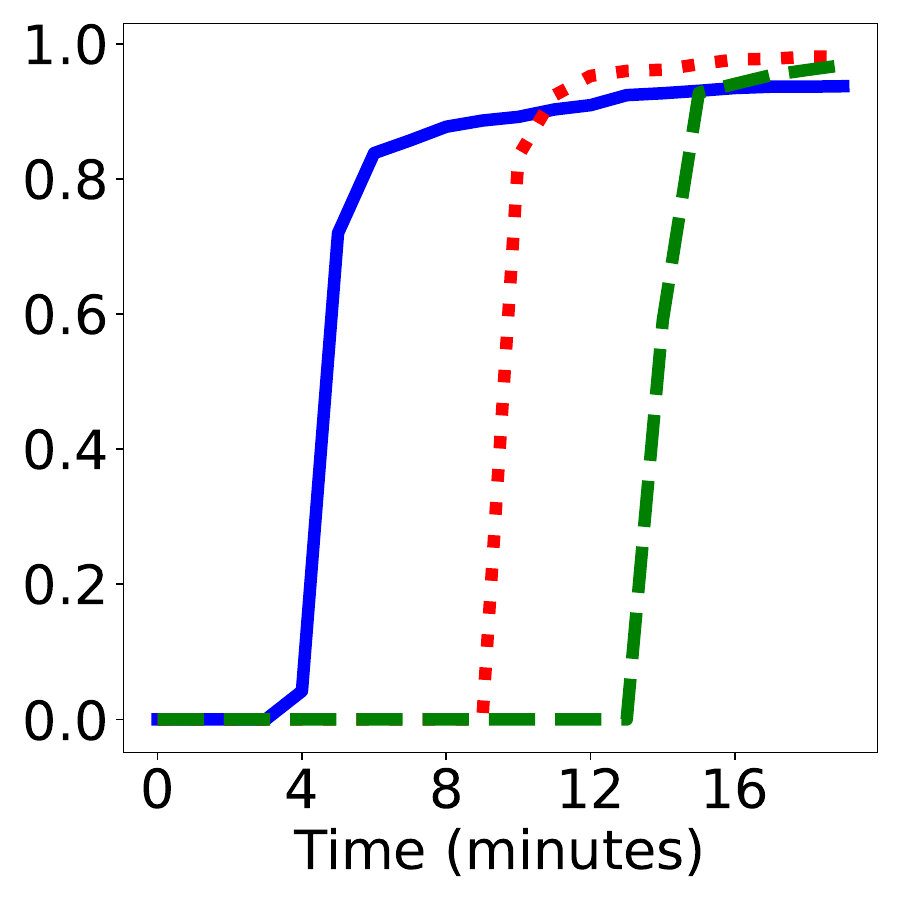}
}
\subfloat[$\coverage$]{
\label{fig:quality_sampling_coverage_ACS}
  \includegraphics[width=27mm]{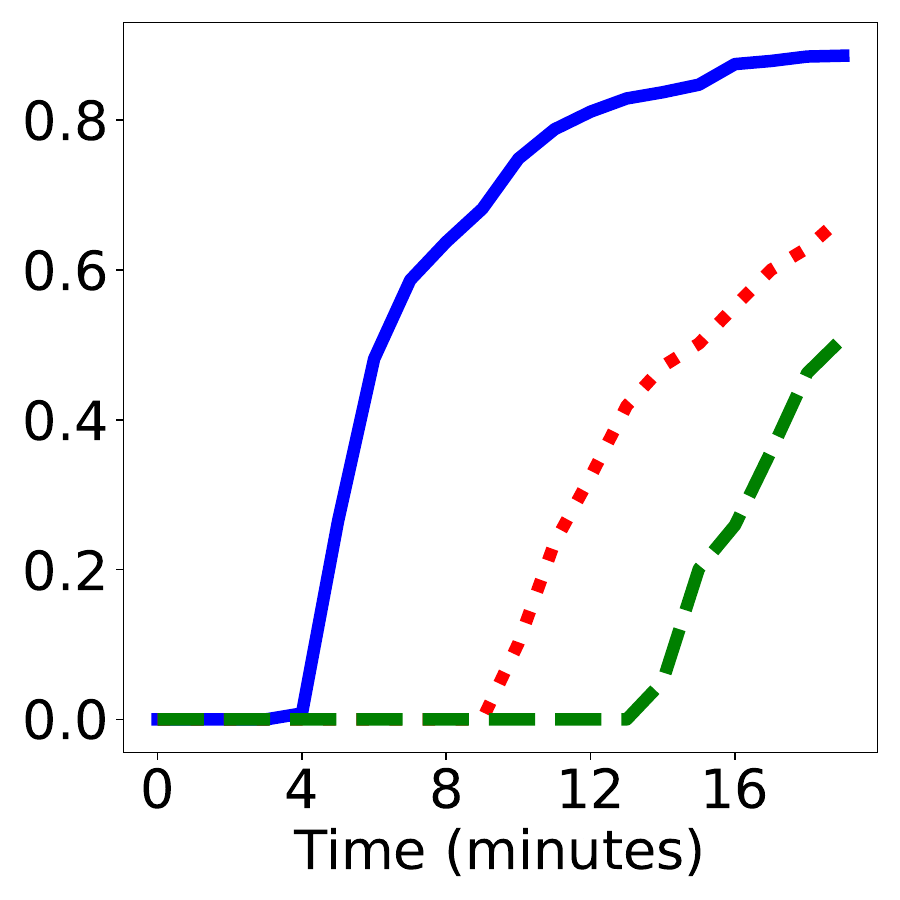}
}
\subfloat[$\statsig$]{
\label{fig:quality_sampling_statsig_ACS}
  \includegraphics[width=27mm]{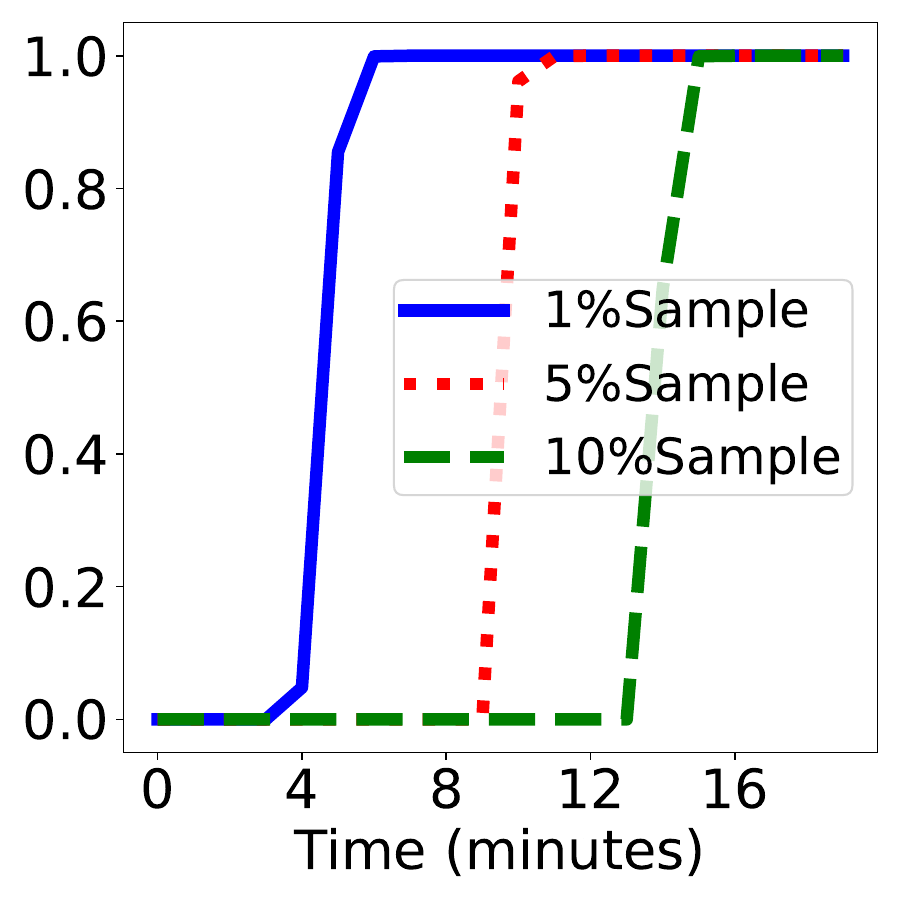}
}
\subfloat[Average of all]{
\label{fig:quality_sampling_avg_ACS}
  \includegraphics[width=27mm]{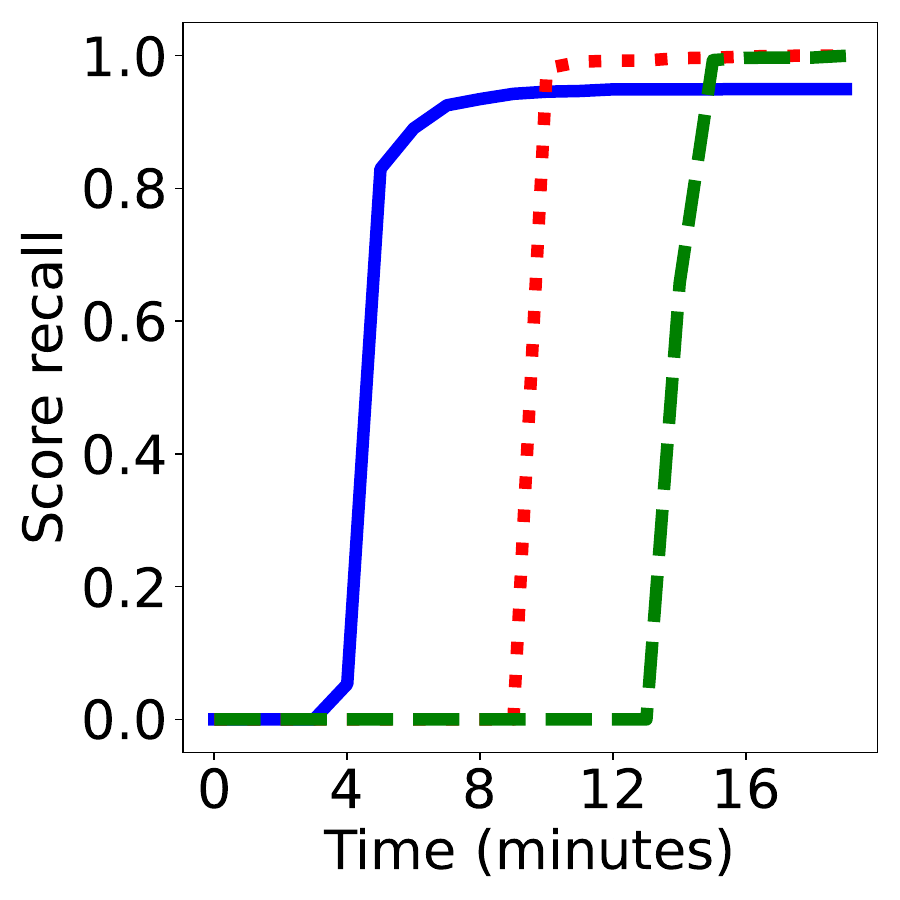}
}
\caption{\revb{Top 100 score recall for each naturalness measure over time for various sampling sizes in sampling guided search over the ACS dataset. 
}}
\label{fig:quality_our_methods_ACS_sampling}
\vspace{-0.8em}
\end{figure*}

As mentioned in Section~\ref{sec:methods:prioritization}, we first randomly sample a subset of the dataset and run a fast full claim endorsement search on it. We experimented with 1\%, 5\% and 10\% sample sizes, and we repeat each experiment three times with each sample size. 
We measure the time until 95\% score recall for each naturalness measure in all datasets (shown in the first 3 rows for each dataset in \Cref{tab:time_to_recall}), and the score recall over time in all naturalness measures for ACS (\Cref{fig:quality_our_methods_ACS_sampling}).
%(\Cref{fig:quality_sampling_anova_ACS,fig:quality_sampling_mi_ACS,fig:quality_sampling_embsim_ACS,fig:quality_sampling_coverage_ACS,fig:quality_sampling_statsig_ACS,fig:quality_sampling_avg_ACS}).

\common{We observe a trade-off between the time until first result (which grows with the sample size, due to the preliminary search on the sample) and the time until a high score recall (which is shorter for larger samples, as the preliminary search yields more accurate results).}
%We observe a delay for all measures and all sample sizes before the score recall starts rising - caused by the preliminary search on the sample, that takes several minutes, depending on the sample size. As the sample size grows, so does the delay. On the other hand, for some naturalness measures, a small sample size (1\%) takes a longer time to reach a score recall as high as larger sample sizes. Meaning, there is an observed trade-off between the time until first result and the time until a high score recall. 
However, for most naturalness measures, \oursolutionsampleone does not fall short behind 5\% and 10\%. Moreover, for $\anova$ and $\coverage$, in the first 20 minutes of the experiment, \oursolutionsampleone reaches the highest score recall. This is likely because for both of these measures, larger sample sizes do not provide additional information that helps predict the score. %Therefore, the sooner the search begins, the sooner the score recall will rise.
We conclude that 1\% is a good sample size for the tested naturalness measures. Similar trends were seen for the other datasets, as can be seen in \Cref{tab:time_to_recall}.

\section{Ablation Study}\label{sec:ablation}
We next isolate several properties of our queries (\cref{sec:methods:find_preds}) and compare them to alternative plausible implementations. 

\paratitle{Attribute level vs. predicate level queries.}
\label{sec:pred_level}
In our algorithm, we use an attribute level group-by query to search for refinements that endorse the user claim (\cref{sec:methods:find_preds}). It is possible to use a predicate-level query instead and iterate over all possible predicates, as done in the OREO framework~\cite{lin2021detecting} when searching for counterexamples. In the following experiment, we show that this process is much more time-consuming. The times for a full run of our algorithm with at most one atom ($m=1$) and with a random order of attributes are shown in \cref{tab:pred_level_ablation} for average and median aggregation functions over two of the datasets. The times in parenthesis are the total SQL times. For average aggregation, the predicate-level search was 10 times slower than the attribute level. For median aggregation, the predicate level search was 5 times slower.

\begin{table}\small
    \centering
    \caption{Time (SQL time) in seconds for full run of \oursolution with at most 1 atom, with random order of attributes.}\vspace{-0.5em}
    \label{tab:pred_level_ablation}
    \begin{tabular}{ccll}
        \toprule
        Dataset & Agg & Predicate level & Attribute level \\
        \midrule
        Stack overflow & Average & 48.54 (30.6) & 4.48 (1.59) \\
        Stack overflow & Median & 49.42 (35.22) & 9.32 (6.60) \\
        ACS7 &  Average & 3681 (3578.8) & 136 (98.5) \\
        ACS7 &  Median & 4163 (4069.0) & 405 (367.17) \\
        \bottomrule
    \end{tabular}
\end{table}

\paratitle{Comparison of case-when to join based queries.}
\label{sec:case_when}
The structure of the group-aggregate queries may largely affect the execution time, as it is run for each attribute combination. For example, instead of the CASE-WHEN query in %\cref{fig:mean_gb_query}
\Cref{sec:methods:find_preds}, we could use a naive query with a join: for each group $(g_1, g_2)$, compute the values of the aggregation $\alpha$ grouped by the combination of split attributes, and then join the two results by the split attributes.
% , as presented in \cref{fig:mean_gb_join_query}.
%\brit{this figure takes a lot of space..can you shorten it? remove the white space and maybe use a smaller font}.

% \begin{lstlisting}[language=SQL, mathescape=true]
% WITH table1 AS (SELECT $A_1,..A_l,\alpha(\Aagg)$ AS a1 \\
% FROM $R$ WHERE $\Agb=g_1$
% GROUP BY $A_1,...A_l$
% HAVING COUNT($\Aagg$) > $M$), 
% table2 AS (SELECT $A_1,..A_l,\alpha(\Aagg)$ AS a2$
% FROM $R$ WHERE $\Agb=g_2$
% GROUP BY $A_1,..,A_l$
% HAVING COUNT($\Aagg$) > $M$)
% SELECT * FROM table1 INNER JOIN table2 
% USING ($A_1,..A_l$) WHERE a1 > a2;
% \end{lstlisting}

% \begin{figure}
% \hrule
%         \centering
%         \begin{minipage}[b]{1.0\linewidth}
%             %\begin{tcolorbox}[colback=white]
% \begin{center}
% \footnotesize
%     \begin{tabular}{l}
% \textsf{WITH table1 AS (}\\
% \textsf{SELECT} $A_1,...,A_l,\alpha(\Aagg)$ \textsf{AS a1 FROM} $R$ \textsf{WHERE} $Agb=g_1$ \\
% \textsf{GROUP BY} $A_1,...,A_l$ \textsf{HAVING COUNT(}$\Aagg$\textsf{)>}$M$\textsf{),}\\

% \textsf{table2 AS (}\\
% \textsf{SELECT} $A_1,...,A_l,\alpha(\Aagg)$ \textsf{AS a2 FROM} $R$ \textsf{WHERE} $Agb=g_2$ \\
% \textsf{GROUP BY} $A_1,...,A_l$ \textsf{HAVING COUNT(}$\Aagg$\textsf{)>}$M$\textsf{)}\\

% \textsf{SELECT * FROM table1 INNER JOIN table2 USING(}$A_1,...,A_l$\textsf{)}\\
% \textsf{WHERE a1>a2;}
%     \end{tabular}
% \end{center}
%             %\end{tcolorbox}
%         \end{minipage}%%
%         \hrule
%         \caption{Aggregation query for a given attribute combination $A_1,\ldots, A_n$ using join instead of CASE WHEN.}
%         \label{fig:mean_gb_join_query}
%     \end{figure}

\begin{table}[b]
    \centering
    \small
    \caption{Average and total times in seconds for CASE-WHEN vs.~join based queries on ACS with at most $m=2$ atoms.}
    \label{tab:case_when_join}
    \vspace{-3mm}
    \begin{tabular}{llll}
        \toprule
        Aggregation & Run time & CASE-WHEN & join-based\\
        \midrule
        \multirow{2}{*}{$\alpha=\AVG$}
        % 1 atom
        % & Total time & 101.3 &404.1 \\
        % & Avg. (std.dev.) time &0.85 (0.07) & 3.4 (1.10)\\
        % 2 atoms
        & Total time & 6102 (=1.7 hr) & 24081 (=6.7 hr) \\
        & Avg. (std.dev.) time & 0.86 (0.1) & 3.40 (1.97)\\
        \multirow{2}{*}{$\alpha=\MED$}
        % 1 atom
        %& Total time & 366.8 & 560.0 \\
        %& Avg. (std.dev.) time & 3.09 (0.14) & 4.71 (0.42)\\
        % 2 atoms
        & Total time & 25908 (=7.2 hr) & 40256 (=11.2 hr) \\
        & Avg. (std.dev.) time & 3.66 (0.35) & 5.69 (0.82)\\
        \bottomrule
    \end{tabular}
\end{table}

We compare the running times for these two versions of the query on the ACS dataset, with the query about women's and men's average/median total income, for $m=2$, i.e., at most two atom predicates (\Cref{tab:case_when_join}). Similar trends were observed for other scenarios and thus omitted.
%Over 120 split attributes, the CASE WHEN query was consistently faster, for both $\MED$ and $\AVG$ aggregates.
Our findings illustrate that employing CASE WHEN queries leads to a consistent and significant improvement in runtime compared to naive join-based queries, particularly for the $\AVG$ aggregation function (4x improvement), but also for the $\MED$ aggregation function (1.5x improvement).
%for both $\MED$ and $\AVG$ aggregation functions. 
For example, the average runtime for a single CASE WHEN $\AVG$ query was 0.85 seconds, while the average runtime for the corresponding join based queries was 3.4 seconds. This highlights the advantage of utilizing this optimization.
%The average improvement was 2.54 seconds per attribute (i.e., per query), with the largest improvement being 6 seconds, the smallest 1.2 seconds, and the total - 302.3 seconds. The average runtime for the CASE WHEN query was 0.85 seconds, while the average runtime for the join based query was 3.4 seconds.
%For median, the average improvement was 1.62 seconds, ranging from 0.68-2.71 seconds, and the total - 193.2 seconds. 
% \brit{it will be better to present these findings in a table, then point to the table and say something like: "Our findings illustrate that employing CASE WHEN queries leads to a significant improvement in runtime compared to naive join-based queries For example, the average runtime for a single CASE WHEN query was 0.85 seconds, while the average runtime for join based queries was 3.4 seconds. This highlights the advantage of utilizing this optimization." I have added an example table that summarizes the results below. Please change it as you like }

%For median, on stack overflow:
%The average improvement is 0.04 seconds, maximal is 0.07, and minimal is 0.01. The average time for CASE WHEN query over all attributes was 0.15 seconds while the average time for join based query was 0.20 seconds.

\section{User Study Survey}\label{sec:survey_questions}
We bring here the instructions and the questions used in our user study\footnote{A graphic version is available at \url{https://shorturl.at/eCLk0}}.

In this study, you will be asked to rate statements (created based on data) according to how much you would recommend an author to use them.  The form takes about 10-15 minutes to fill.

There are 3 topics and in each topic several statements. For each topic, please rate at least one statement as 1 ("Least recommended") and another one as 5 ("Most recommended"), and rate the others as you see fit. To be considered as "Most recommended", the statement should be the best relative to the set of given statements.

The first few questions are meant to provide us with some information about the academic background of study participants. Other than that, no personal information will be saved.
Participation is voluntary. You may withdraw at any time, without any negative consequences.

This study is conducted by Shunit Agmon, Benny Kimelfeld, Brit Youngmann and Amir Gilad. For any questions, feel free to contact Shunit Agmon by e-mail: shunit.agmon@gmail.com.

$\boxtimes$ I have reviewed the consent form and agree to participate in this study.

\subsection*{Academic Background}
\textbf{What is you current academic status?} (Mark the first option that applies to you). \\
$\boxtimes$ Pursuing a graduate degree (master's or PhD) \\
$\boxtimes$ Pursuing a bachelor’s degree \\
$\boxtimes$I have a PhD \\
$\boxtimes$I have a Master's degree \\
$\boxtimes$I have a Bachelor's degree \\
$\boxtimes$Not pursuing a degree \\
\textbf{What is your major or field of study?} (open question)

\subsection*{Flight Delays on Mondays vs. Saturdays}
Alice is writing a traveling blog recommending her readers to prefer flights on Mondays over Saturdays to minimize delays. She has the statements below, based on US flight delays records from 2015. To what extent would you recommend to Alice to use each of them in her article?

Please rate at least one statement as ``Least recommended'' and another one as ``Most recommended'', and rate the others as you see fit (2 statements can get the same score).

\begin{itemize}
    \item Among flights scheduled to depart between 3:00-4:00 AM, there are more delays on Saturdays (43) vs. Mondays (30).
    \item Among flights with flight number 2, there are more delays on Saturdays (49) vs. Mondays (48).
    \item Among flights operated by Hawaiian Airlines Inc., there are more delays on Saturdays (1,028) vs. Mondays (986).
    \item In October, there were more flight delays on Saturdays (9,549) vs. Mondays (8,744).
    \item Among flights with flight number 3507, there were more flight delays on Saturdays (13) vs. Mondays (10).
    \item Among flights with flight number 3430, there were more flight delays on Saturdays (16) vs. Mondays (14).
    \item Among flights with flight number 4876, there were more flight delays on Saturdays (14) vs. Mondays (9).
    \item Among flights with air time between 390 and 400 minutes, there were more flight delays on Saturdays (51) vs. Mondays (41).
    \item Among flights on the last days of the month (30-31), there were more flight delays on Saturdays (3,753) vs. Mondays (2,573).
    \item Among flights with scheduled duration between 480 and 490 minutes, there were more flight delays on Saturdays (17) vs. Mondays (14).
\end{itemize}

\textbf{Before you continue to the next section...} please make sure that in the previous section, you have rated at least one of the statements as 5 (most recommended) and at least one of the statements as 1 (least recommended).

\subsection*{Master's vs. Bachelor's Salary}
Bella is writing a article about the salary benefits of a Master's degree.  She has the statements below, drawn from the 2022 StackOverflow Developer's Survey. To what extent would you recommend to Bella to use each of them in her article?  

Please rate at least one statement as "Least recommended" and another one as "Most recommended", and rate the others as you see fit (2 statements can get the same score).

\begin{itemize}
    \item Among the 174 people (out of 38K) in Iran who have been professionally coding for under 10 years, MSc graduates earn a higher salary on average (\$56K) than BSc graduates (\$39K).
    \item Among the 40 People (out of 38K) in Croatia who have been coding for under 10 years, MSc graduates earn a higher salary on average (\$46K) than BSc graduates (\$22K).
    \item Among the 4,525 people (out of 38K) who use Linux based operating systems and have been coding for up to 10 years, MSc graduates earn a higher salary on average (\$134K) than BSc graduates (\$121K).
    \item Among the 234 people (out of 38K) who want to work with Bash/Shell language and decide what new tools to buy by asking developers they know, MSc graduates earn a higher salary on average (\$313K) than BSc graduates (\$197K).
    \item Among the 280 people (out of 38K) who define themselves as Southeast Asian, MSc graduates earn a higher salary on average (\$83K) than BSc graduates (\$42K).
    \item Among the 40 people (out of 38K) in Croatia who have up to 10 years of work experience, MSc graduates earn a higher salary on average (\$46K) than BSc graduates (\$30K).
    \item Among the 10,698 people (out of 38K) who use Linux based operating systems, MSc graduates earn a higher salary on average (\$167K) than BSc graduates (\$166K).
    \item Among the 16 people (out of 38K) who want to work with PyCharm and learn to code through online courses, MSc graduates earn a higher salary on average (\$54.7K) than BSc graduates (\$13.5K).
    \item Among the 687 people (out of 38K) who are independent or freelancers and have been coding for 10 to 20 years, MSc graduates earn a higher salary on average (\$145K) than BSc graduates (\$100K).
    \item Among the 836 people (out of 38K) who have 20-30 years of work experience, MSc graduates earn a higher salary on average (\$256K) than BSc graduates (\$225K).
    \item Among the 2049 people (out of 38K) who work in India, MSc graduates earn a higher salary on average (\$52.5K) than BSc graduates (\$51.6K).
    \item Among the 1442 people (out of 38K) who decide what new tools to buy by asking developers they know, MSc graduates earn a higher salary on average (\$182K) than BSc graduates (\$166K).
    \item Among the 1872 people (out of 38K) who work in Germany, MSc graduates earn a higher salary on average (\$135K) than BSc graduates (\$114K).
    \item Among the 22 people (out of 38K) who have worked with OpenStack platform and would like to work with Microsoft Teams, MSc graduates earn a higher salary on average (\$98K) than BSc graduates (\$48K).
    \item Among the 32 people (out of 38K) who have worked with Notepad++ and do not know the size of their organization, MSc graduates earn a higher salary on average (\$231K) than BSc graduates (\$62K).
    \item Among the 27 people (out of 38K) who have worked with Jira and want to work with Rocketchat, MSc graduates earn a higher salary on average (\$160K) than BSc graduates (\$75K).
    \item Among the 60 people (out of 38K) who define themselves as Indian and want to work with Vim, MSc graduates earn a higher salary on average (\$160K) than BSc graduates (\$75K).
    \item Among the 308 people (out of 38K) who use Windows operating system and have worked with the Express web framework, MSc graduates earn a higher salary on average (\$180K) than BSc graduates (\$103K).
\end{itemize}

\textbf{Before you continue to the next section...} please make sure that in the previous section, you have rated at least one of the statements as 5 (most recommended) and at least one of the statements as 1 (least recommended).

\subsection*{Women's vs. Men's Income}
Charlotte is writing an article about women's income being higher than is commonly expected. Note the difference between income and salary: it is possible to get an income without being employed (e.g., from various support sources or passive income from investments). She has the statements below, drawn from the US Census survey of 2018 in the largest seven states in the US. To what extent would you recommend to Charlotte to use each of them in her article?

Please rate at least one statement as "Least recommended" and another one as "Most recommended", and rate the others as you see fit (2 statements can get the same score).

\begin{itemize}
    \item Among the 930 (out of 1.2M) 12th grade students who have last worked 1-5 years ago, women have a higher income on average (\$2,768) than men (\$1,656).
    \item Among the 28,366 people (out of 1.2M) with insurance through an employer or union, who have a child, women have a higher income on average (\$1,044) than men (\$1,017).
    \item Among the 49,011 (out of 1.2M) African Americans who were never married, women have a higher income on average (\$21.5K) than men (\$16.7K).
    \item Among the 582 (out of 1.2M) army veterans who have not reported a Veteran Service Disability Rating and who do not have cognitive difficulty, women have a higher income on average (\$57K) than men (\$46K).
    \item Among the 730 (out of 1.2M) army veterans who have not reported a Veteran Service Disability Rating, women have a higher income on average (\$50K) than men (\$43K).
    \item Among the 129,462 people (out of 1.2M) who are not in the labor force and were never married, women have a higher income on average (\$7,325) than men (\$7,201).
    \item Among the 7,595 people (out of 1.2M) who work for a private company and have a child, women have a higher income on average (\$3,474) than men (\$3,436).
    \item Among the 1,599 people (out of 1.2M) of age 20-30 who work in the armed forces, women have a higher income on average (\$39K) than men (\$37K).
    \item Among the 91K people (out of 1.2M) who were never married and have not worked in the last 5 years, women have a higher income on average (\$6,155) than men (\$5,649).
    \item Among the 118 people (out of 1.2M) who depart for work between 6:45-7:35 and work as producers and directors, women have a higher income on average (\$105K) than men (\$98K).
    \item Among the 9,892 people (out of 1.2M) who do not have health insurance and work in construction, women have a higher income on average (\$29K) than men (\$28K).
    \item Among the 583 people (out of 1.2M) who work as paralegals and legal assistants in west US, women have a higher income on average (\$60K) than men (\$58K).
    \item Among the 174 African American people (out of 1.2M) in Polk county (Florida), women have a higher income on average (\$22K) than men (\$15K).
    \item Among the 86 people who define themselves as White (out of 1.2M) who were born in Vietnam, women have a higher income on average (\$51K) than men (\$43K).
\end{itemize}

\textbf{Before you finish...} please make sure that in the previous section, you have rated at least one of the statements as 5 (most recommended) and at least one of the statements as 1 (least recommended).

\end{document}